\documentclass[]{raa} 

\usepackage{graphicx,times}             
\usepackage{natbib}
\usepackage{amssymb,amsmath}
\bibpunct{(}{)}{;}{a}{}{,}

\usepackage{enumerate}
\usepackage{amsmath}
\usepackage{float}
 \usepackage{multirow}
  \usepackage{longtable}
  \usepackage{lipsum}
  \usepackage{pdflscape}
\usepackage{hyperref}
\hypersetup{colorlinks = true, linkcolor = green, anchorcolor = red, citecolor = blue, filecolor = red, pagecolor = red, urlcolor = red}

\newcommand{\obs}{^\textrm{obs}}

\newcommand\rmxaa{Revista Mexicana de Astronom\'ia y Astrof\'isica}
   
  \DeclareMathOperator\erf{erf}

\begin{document}

\title{A Quark-Nova in the wake of a core-collapse Supernova: a unifying model for long duration Gamma-Ray Bursts and 
  Fast Radio Bursts}
  
   \volnopage{Vol.0 (200x) No.0, 000--000}      
   \setcounter{page}{1}          

\author{Rachid Ouyed
\and Denis Leahy
\and Nico Koning
}

 \institute{Department of Physics and Astronomy, University of Calgary, 2500 University Drive NW, Calgary, AB, T2N 1N4, Canada; {\it rouyed@ucalgary.ca}}

\abstract{By appealing to a Quark-Nova (QN; the explosive transition of a neutron star to a quark star) in the wake of a core-collapse Supernova (CCSN)
explosion  of a massive star, we develop a unified model for long duration Gamma-ray Bursts (LGRBs) and  Fast Radio Bursts (FRBs). The time delay (years to decades) between the SN and the QN and, the fragmented nature (i.e. millions of chunks) of the relativistic QN ejecta are  key to yielding a
 robust LGRB engine. In our model, a  LGRB light-curve exhibits the interaction of the fragmented QN ejecta with a turbulent 
  (i.e. filamentary and magnetically saturated) SN ejecta  which is shaped by  its interaction with an underlying pulsar wind  nebula (PWN).
  The afterglow is due to the interaction of the QN chunks, exiting the SN ejecta, with the surrounding medium.
 Our model can fit  BAT/XRT prompt and afterglow light-curves,
  simultaneously with their spectra, thus yielding  the observed properties of LGRBs (e.g. the Band function and the X-ray flares). 
  We find that the peak luminosity-peak photon energy relationship (i.e. the Yonetoku law), and the isotropic energy-peak photon energy  relationship (i.e. the Amati law) are not fundamental but phenomenological.
  FRB-like emission in our model result from coherent synchrotron emission (CSE) when  the QN chunks 
interact with non-turbulent weakly magnetized PWN-SN ejecta, where conditions are prone to the Weibel instability. 
Magnetic field amplification induced by the Weibel instability in the shocked chunk frame sets the bunching length for electrons and pairs to radiate
coherently. The resulting emission frequency, luminosity and duration 
in our model are consistent with FRB data. We find a natural unification of  high-energy burst phenomena from
  FRBs  (i.e. those connected to CCSNe) to LGRBs including X-ray Flashes (XRFs) and X-ray rich GRBs (XRR-GRBs) as well as Super-Luminous SNe (SLSNe).   We find a possible  connection between Ultra-High Energy Cosmic Rays and FRBs
  and propose that a QN following a binary neutron star merger can yield a short GRB (SGRB) with fits to BAT/XRT light-curves. 
}

\authorrunning{Ouyed, R. et al.}            
   \titlerunning{A QN in the wake of a core-collapse SN: a model for LGRBs and FRBs}  

   \maketitle

\section{Introduction}
 \label{sec:intro}

 \subsection{Gamma Ray Bursts (GRBs)}
  \label{sec:intro-GRBs}

Ever since their discovery \citep{klebesadel_1973} and the confirmation of their cosmological origin (\citealt{meegan_1992,vanparadijs_1997}), GRBs have challenged physicists and astrophysicists 
who have yet to understand fully the driving mechanism and the nature of the underlying engine. 
 The intense, intermittent prompt emission  in  hard X-rays and gamma-rays lasts from milliseconds to hundreds of seconds 
with short-duration GRBs (SGRBS)  peaking at $\sim 0.3$ s and long-duration GRBs (LGRBs) peaking at $\sim 30$ seconds (\citealt{mazets_1981,norris_1984,kouveliotou_1993,horvath_1998}; see \citealt{mukherjee_1998}
 for a possible intermediate group).  Their emission  in the afterglow phase (i.e. X-ray, optical and radio) can last from hours to weeks (\citealt{costa_1997,vanparadijs_1997,meszaros_1997}).  The measured  redshift distributions of the two groups show a median of $\sim 2.4$ for LGRBs (e.g. \citealt{bagoly_2006})
and $\sim 0.4$ for SGRBs (e.g. \citealt{oshaughnessy_2008}; in \citet{berger_2007} it is
suggested that between 1/3 to 2/3 of SGRBs are at a redshift $\sim 1$). 

The spectra of SGRBs and LGRBs  are  non-thermal and well described  by the phenomenological Band-function (\citealt{band_1993,preece_2000})
 which has yet  to be explained fully (see however e.g. \citealt{peer_2006,beloborodov_2010}). 
  Recent analysis supports the synchrotron origin (\citealt{li_2019a,li_2019b}). 
  In some GRBs
   a thermal component in addition to the Band-function (\citealt{band_2004}) seems necessary  to reproduce the observed spectrum
 (\citealt{ghirlanda_2003,ryde_2005,basak_2015}).

 There is a rich literature on the topic of GRBs covering the history, the observations and the physics 
 of these intriguing bursts. We refer the interested reader to past, and  recent, reviews and references therein for details 
 (e.g. \citealt{fishman_1995,piran_1999,piran_2000,vanparadijs_2000,meszaros_2002,lu_2004,piran_2005,meszaros_2006,bisnovatyi_2006,zhang_2007,nakar_2007,gehrels_2009,costa_2011,berger_2014,peer_2015,davanzo_2015,iyyani_2018,zhang_2018}). While our model applies
 to LGRBs, in this introduction, we briefly discuss general properties of SGRBS and LGRBs.

\subsubsection{Standard models}

In the standard and widely accepted picture, a catastrophic event  yields a relativistic fireball  which consists of  ejecta with a wide range of Lorentz factors whose
energy is harnessed by internal shocks (\citealt{cavallo_1978,goodman_1986,paczynski_1986,kobayashi_1997,piran_1999}; 
see also \citealt{zhang_2011}). LGRBs are believed to originate from collapsars
 (i.e. collapsing massive Wolf-Rayet type stars; \citealt{woosley_1993,macfadyen_1999}).
  Models involving collapsars utilize a hyper-accreting stellar mass BH as a central engine
wich drives a jet (e.g. \citealt{popham_1999, li_2000, lee_2000,  dimatteo_2002, gu_2006, chen_2007, janiuk_2007,
lei_2009, liu_2015, li_2018a, lei_2013a,lei_2013b}).

 SGRB are from the merging of two compact objects in binary systems (two neutron stars or a neutron star and a stellar-mass black hole; \citealt{blinnikov_1984,paczynski_1986,eichler_1989,narayan_1992})\footnote{The detection of a kilonova in GRB 130603B (\citealt{tanvir_2013}) and the recent gravitational wave event GW170817 (\citealt{abbott_2017a}) and
its associated SGRB (\citealt{abbott_2017b}) gave support for the binary-merger origin
of at least some SGRBs.}. These two phenomena produce  highly collimated ultra-relativistic jets and appeal to colliding shells with different Lorentz factors to harness the jet's kinetic and
  internal energy  yielding the highly intermittent prompt emission (\citealt{rees_1994,kobayashi_1997}). The afterglow emission is from the interaction of the jet with the surrounding  ambient medium farther away from the engine involving jet deceleration 
    (e.g. \citealt{wijers_1997,meszaros_1997}). The observation of jet breaks is often used as evidence
for collimation (\citealt{rhoads_1997,rhoads_1999,frail_2001}) and while it seems generally capable of accounting  for some features of LGRBs and SGRBs, it nevertheless requires fine-tuning in some cases (e.g. \citealt{grupe_2009,covino_2010}).  
Recent studies show that the achromatic break expected to be associated with the jets is absent
in some GRBs (\citealt{willingale_2007}). This can only be explained with models
involving impulsive jets  or multiple jets (see e.g. \citealt{granot_2005,vaneerten_2011}).
Alternative scenarios such as the cannonball model
of \citet[and references therein]{dar_2004}  and the  ``ElectroMagnetic Black Hole (EMBH)" model (\citealt{christodoulou_1971,damour_1975,preparata_1998}) may 
account for some features of some seemingly non-standard  GRBs.

\subsubsection{The galaxy, the metallicity and the supernova association}
\label{sec:intro-SN-association}

LGRBs are  often associated with  star forming environments (e.g. \citealt{bloom_2002,fruchter_2006}  and references therein).
  Specifically,  LGRBs are  associated with low-mass, gas-rich and low-metallicity star-forming  galaxies (like the Large Magellanic Cloud; \citealt{bloom_2002,fruchter_2006,wang_2014}) that are fainter and more irregular 
than core-collapse SNe host galaxies.

The SN-LGRB association (\citealt{woosley_1993,galama_1998,bloom_1999,hjorth_2003,stanek_2003}) 
together with the association of LGRBs with star forming environments 
 link LGRBs  to the deaths of
  massive stars (suggestive of the collapsar model; e.g. \citealt{macfadyen_1999}). 
 Specifically, all SNe, spatially and temporally, associated with LGRBs are classified as 
broad-line (BL) Type Ic (Type Ic-BL; see \citealt{hjorth_2012}). 
 However, some LGRBs show no underlying Type Ic core-collapse SNe  
 (\citealt{fynbo_2006,niino_2012}) as expected in the collapsar model.  These are
 found in metal-rich environments with little to no star formation (e.g. \citealt{tanga_2018}).
  It is suggested that  a non-negligible  fraction of LGRB hosts have a metallicity around the solar value (e.g. \citealt{prochaska_2009,savaglio_2012,elliot_2013,schady_2015}). The collapsar model requires the progenitor to be metal-poor 
 in order to maintain the massively rotating cores required to launch a LGRB
 (e.g.  \citealt{woosley_2006}).
 
SGRBs tend to reside in environment with relatively reduced star formation (e.g. \citealt{gehrels_2005,barthelmy_2005,davanzo_2009,zhang_2009,levesque_2010} and
references therein). However, as demonstrated in \citet{berger_2014} 
SGRBs lacking SN associations are predominantly associated with star-forming galaxies. While SGRBs have not been
associated with any SNe so far, they have been associated with a variety of galaxies ranging
from LGRB-like galaxies to elliptical ones (e.g. \citealt{gehrels_2005,davanzo_2009})
and in some cases SGRBs are found to be in isolation (e.g. \citealt{berger_2010}) as expected if they originate from
binary mergers.

\subsubsection{The extended emission (EE) and the late-time X-ray plateaus}
\label{sec:intro-EE}

Some GRBs show re-brightening (the extended emission; EE)
which occurs tens of seconds after the prompt emission
 and can last for hundreds of seconds (e.g. \citealt{norris_2006,norris_2010}).
These bursts seem to show properties characteristic of  both SGRBs and
LGRBs and may require a complex engine activity (e.g. \citealt{thompson_2004,rosswog_2007,metzger_2008,metzger_2010a,barkov_2011,bucciantini_2012}).

A canonical GRB afterglow light-curve emerged from the  Swift XRT observations (\citealt{nousek_2006}). 
 Spanning a very wide time-interval  of $10^{-1}$-$10^5$ s, the observed
light-curves show phases of a rapid decline in the early X-ray afterglow (i.e. a steep decay component; e.g. \citealt{tagliaferri_2005}) followed by a plateau  (also referred  to  as the shallow decay component which lasts $10^4$-$10^5$ seconds; e.g. \citealt{zhang_2006}) and then a normal decay component.  The plateaus   
 are common to both SGRBs (\citealt{rowlinson_2013})  and LGRBs with 
spectral properties  similar to those of the prompt emission (\citealt{chincarini_2010}). 

Some of these canonical light-curves show occasional flaring during the late X-ray afterglow emission (e.g. \citealt{obrien_2006}),
in particular for LGRBs and in some SGRBs (e.g. \citealt{barthelmy_2005,campana_2006,laparola_2006}).
  These, sometimes repetitive, X-ray flares superimposed on the  X-ray light-curve 
have been observed in about half  of the afterglows with a fluence which is 
on average a few percents of the GRB  prompt emission (e.g. \citealt{burrows_2005}). 
 In some cases, giant flares have been observed with fluence equaling that of the prompt emission (e.g. \citealt{falcone_2007}). 
These flares are not expected in the standard model  and are suggestive of energy injection into the jet hundreds of seconds following the prompt emission
or a very late re-start of the engine   (e.g. \citealt{king_2005,zhang_2006}). 
Recent  analyses concluded that the flares may be linked to the prompt emission
and are not an afterglow effect (\citealt{falcone_2007,dainotti_2008,chincarini_2010}). 
I.e. they seem to involve a mechanism that is similar to the one behind the prompt
emission but acting at lower energies and at later times (e.g. \citealt{peng_2014}).

Keeping the central engine active for much longer than the duration of the prompt emission
(hours to days of extended activity) is difficult for  the collapsar model of LGRBs, because  accretion disc
viscous timescale are  short (see however \citealt{rosswog_2007}).  
Magnetars and their spin-down power (\citealt{duncan_1992,thompson_1993}) may 
  explain the $>10^4$ s  engine activity in the X-ray afterglow 
(e.g. \citealt{gompertz_2014,lu_2015}) but not necessarily 
the flares.   Merging of two neutron stars into a hyper-massive quark star (QS) and then collapse into a black hole (BH),  could be responsible for plateaus and following bump in the X-ray light curves of some GRBs (\citealt{li_2016,hou_2018}). In the context of SGRBs, it is pointed out that NS-NS mergers may not  lead to magnetars and one has to 
 deal with the limited energy input from the rotational energy (see however \citealt{gompertz_2014}).  
Others appeal to curvature effect (e.g. \citealt{kumar_2000}), magnetic
dissipation processes (e.g. \citealt{giannios_2006}) or light scattering in the jet to induce rebrightenning (e.g. \citealt{panaitescu_2008}). At this stage, 
 it is not unreasonable to state that the origin of the extended activity as well as the flares are debatable in the standard models (see  \citealt{dar_2006} for alternative explanations).

 \subsubsection{GRB prompt phase two-component relationships}
 \label{sec:intro-correlations}

 Several two-component relationships have been proposed (\citealt{fenimore_2000,norris_2000,schaefer_2001,amati_2002,yonetoku_2004,ghirlanda_2004,liang_2005,firmani_2006,li_2007,butler_2007,tsutsui_2008}; see
 also \citealt{schafer_2007} for a review). In particular, 
  \citet{amati_2002}  found a correlation between the  cosmological rest-frame  spectrum peak photon energy, 
 $E_{\rm peak}$, and the isotropic-equivalent radiated energy, $E_{\rm iso}$ (the Amati relation). 
 \citet{yonetoku_2004,yonetoku_2010} found a tight correlation between  $E_{\rm peak}$ and the 1-second peak
luminosity ($L_{\rm iso, peak}$) in GRBs (the Yonetoku relation). These relationships are 
debated in the literature with pro- and con- camps (e.g. \citealt{nakar_2005,butler_2007,collazzi_2012,heussaff_2013,dainotti_2018a}).  
Other correlations not considered here are reviewed in details in \citet{dainotti_2018b}. 

\subsubsection{Quark stars (QSs) and GRBs}
\label{sec:intro-uds}

The strange matter hypothesis states that matter made of {\it up}, {\it down} and {\it strange} quarks 
(i.e. {\it (uds)} matter) could be the most stable state of matter (\citealt{itoh_1970,bodmer_1971,terazawa_1979,witten_1984}; see also \citealt{weber_2005} and reference therein). If true, then {\it strange}-quark seeding in the deconfined core (where the quarks are not confined inside neutrons) of some NSs 
would imply that   the whole
system could lower its energy by converting to the more stable  {\it (uds)} matter.  
There is an extensive literature devoted to the existence and properties of
quark stars and the conversion of a NS to a quark star (e.g. \citealt{olinto_1987,lugones_1994,dai_1995,cheng_1996,horvath_1988,ouyed_2002a,keranen_2005,niebergal_2010, herzog_2013,pagliara_2013,furusawa_2015a,furusawa_2015b,drago_2015a,drago_2015b,ouyeda_2018a,ouyeda_2018b}).
The strange-quark seeding needed to trigger the conversion
 has also been investigated with different seeding
 mechanisms and  timescales suggested in the literature (e.g. \citealt{olesen_1994,iida_1998,drago_2004,bombaci_2004,mintz_2010,perez_2010,logoteta_2012}).
These studies together find different paths to the formation of a quark star from a strange-quark seeded core of a NS.

  Early investigations of QSs  as GRB engines use general arguments to argue that the energy
  release during the conversion
  of a NS to a QS (of order $10^{53}$ ergs) combined with properties of the
  resulting QS (e.g. its spin-down power, the exotic phases of quark matter)  may yield
  a GRB engine (\citealt{usov_1992,dai_1998,wang_2000,ouyed_2002a,ouyed_2002b,berezhiani_2003,drago_2004,ouyed_2005,paczynski_2005,xu_2009,dai_2011,perez_2013,drago_2016}). Other models involve  the conversion of a NS to a strange star by accretion in a low-mass X-ray binary (\citealt{cheng_1996,ouyed_2011a,ouyed_2011b,ouyed_staff_2013}). In  the post-QN phase 
     highly variable hyper-accretion onto the QS, 
  which appeals to the exotic phase of quark matter,  ejects intermittent relativistic shells, reminiscent of the energetics and variability seen
   in GRBs (\citealt{ouyed_2005}).
However, most of these
 models fail to account for the many unique features of GRBs mentionned in this introduction 
  (e.g. the spectrum, variability, etc...).

 \subsection{Fast Radio Bursts (FRBs)}
   \label{sec:intro-FRBs}

 The discovery of intense, millisecond, highly dispersed radio bursts in the GHz range (\citealt{lorimer_2007})
 opened a new era in radio astronomy and a window into one of the most enigmatic phenomena in
 modern astronomy, Fast Radio Burst (FRBs).  Dozens of FRBs are known (see http://frbcat.org/) with
 two repeating ones (\citealt{spitler_2016,sholz_2016,chime_2019}). 
 Their dispersion measures (DM; of hundreds of  pc cm$^{-3}$) put them at extra-Galactic to cosmological distances 
which makes them extremely bright ($> 10^{41}$ erg s$^{-1}$). 
While a typical GRB prompt emission is made of many sub-second pulses yielding an intermittent
 emission, FRBs consists of a single pulse of milliseconds duration, except for the multiple pulses in
 repeating FRBs. The story of FRBs so far seems to resemble that of GRBs (e.g. \citealt{kulkarni_2014,kulkarni_2018}). 
 A  full  account  of  the  discoveries, observations  and properties of these 
 FRBs can be found in \citet{lorimer_2007,thornton_2013,spitler_2014,petroff_2016,ravi_2016,gajjar_2018,michilli_2018a}
 with a recent analysis given in \citet{lorimer_2018}.

Because of the large beam width at Parkes    and Areciob, FRBs are weakly localized 
which makes it difficult to isolate their host galaxies or associate them with any astrophysical objects.
With no discernible source and with no counterparts at other frequencies FRBs are 
hard to model. One can infer that FRBs are associated with  high brightness temperatures 
requiring a coherent emission mechanism (\citealt{katz_2014}).
 A discussion of current theoretical models can be found in the literature (e.g. \citealt{katz_2016a,platts_2018,popov_2018}). 
 Many of these models involve single or double compact stars undergoing catastrophic processes 
 such as merging, comet impact or bursting.   Specifically, models involving  intense pulses from pulsars or magnetars
  (\citealt{connor_2016,cordes_2016,katz_2016b,metzger_2017,margalit_2018}) have been proposed. Other models appealing to standard compact objects  include NS-NS mergers (\citealt{yamasaki_2018}), impact of
asteroids with NSs (e.g. \citealt{geng_2015,dai_2016}), 
as well as  WD-WD, WD-NS and WD-BH interactions (e.g. \citealt{kashiyama_2013,gu_2016,li_2018b}).
\citet{montez_2014} make use of the interaction of planets, large asteroids, or white dwarfs with a pulsar wind. Repeating FRBs may be used as an argument to disfavor catastrophic scenarios preferring instead models involving magnetar-like bursting activity.  
 Because FRBs are relatively new compared to GRBs, so far there have been only
     a handful attempts at explaining them using QSs (e.g. \citealt{shand_et_al_2016}).

   \subsection{The Quark-Nova model for GRBs}
   \label{sec:intro-QN}

Our working hypothesis is that a QN can occurs whenever the underlying NS's core density reaches the quark-decofinement limit
$\rho_{\rm NS, cr.}$ where quarks roam freely and are no longer confined to hadrons. For static configurations, and for a
 given Equation-of-State of neutron matter, we define a critical NS mass $M_{\rm NS, cr.}$ when the
 density in the NS core is $\rho_{\rm NS, cr.}$.
 If a NS is born with a mass above
  this critical value but is rapidly rotating then $\rho_{\rm NS, cr.}$ is only reached after spin-down and/or by accreting more mass
   (see discussion \S ~ 2.1 in \citealt{ouyed_staff_2013} for example). An increase in mass can occur 
 following a SN if fallback is important  or in a binary system
where the NS can gain mass from a companion (\citealt{ouyed_2011a,ouyed_2011b,ouyed_staff_2013}) or during a Common Envelope phase (e.g. \citealt{ouyed_2015a,ouyed_2015b,ouyed_2016} and references therein).  In this paper, we consider
deconfinement, immediately followed by the QN, triggered by spin-down.

If the QN occurs early in the wake of a SN, meaning that the NS explodes
   weeks to months following its birth in the SN, the kinetic energy of the QN ejecta
   (the outermost layers of the NS crust ejected during the explosion)\footnote{As shown in \citet{ouyed_leahy_2009}, the QN ejecta 
   fragments into millions of chunks (see also \S ~ \ref{sec:QN-ejecta} here).} 
     is efficiently converted to radiation (\citealt{leahy_2008,ouyed_2009a}).  
   Crucially,  the extended envelope means that PdV losses
are negligible when it is shocked by the QN ejecta, yielding
a Super-Luminous SN (SLSN; \citealt{ouyed_2009a}). 
Effectively,  the QN re-energizes and re-brightens the extended SN ejecta 
 giving  light-curves very similar to  those of  SLSNe (\citealt{ouyed_2012, ouyed_2013a,kostka_2014a,ouyed_2016}).  
  A number of  SLSNe and double-humped SNe  have been modelled in this framework 
  (see {\it http://www.quarknova.ca/LCGallery.html} for a picture
gallery of the fits). The  QN model  predicts that the interaction of the neutron-rich QN ejecta with the SN ejecta would 
   lead to unique nuclear spallation products, in particular an excess of $^{44}$Ti at the expense
   of $^{56}$Ni (\citealt{ouyed_2011c,ouyeda_2014,ouyeda_2015a}), 
   which may  have been observed  in Cas A SN  (e.g. \citealt{laming_2014,ouyed_2015a}).

         Including a QN event in the collapsar model (e.g. \citealt{staff_2007,staff_2008a,staff_2008b,ouyed_2009b}) or in binaries
   (\citealt{ouyed_2011a,ouyed_2011b,ouyed_2015c}) provides an intermediary stage (between the
   NS and the Black Hole (BH) phases; the BH forms from the collapse of the QS)   that extends
    the engine's activity and provides an extra source of energy.   In \citet{staff_2008a,staff_2008b}  it was found that 
    a three stage model within the context of a core-collapse supernova involving a NS, converting to a QS followed by a BH phase
 from the collapsing QS allowed some  interpretation of the observations of early  and late X-ray afterglows of GRBs.   
      Basically, 
   this model harnesses the QN energy (\citealt{leahy_2009}) in addition to the QS spin-down power (\citealt{staff_2007}). 
    However, these models  did not capture  important features of GRBs
   such as the variable light-curve and the spectrum.

  \subsubsection{Our current model for LGRBs and FRBs}     
  \label{sec:intro-QN-new}

  For time-delays between the SN and the QN exceeding a few years, the SN ejecta is too large and diffuse to experience any substantial re-brightening 
  (i.e. no SLSNe can result). However, the density in the inner layers of the SN ejecta is still high enough 
 to induce shock heating of the QN chunks yielding either a LGRB or an FRB as we show in this paper.

Specifically, we demonstrate that a  QN event which occurs years to 
 decades following the core-collapse SN explosion of a  massive star  (hereafter we assume to be a Type Ic SN) 
   can explain the photometric and spectroscopic activity of LGRBs.  In addition, we find a regime
  where the interaction between the QN ejecta and the PWN-SN ejecta (i.e. the shell born from the interaction
  between the SN and the PWN) allows for the development
  of the Weibel instability which induces coherent synchrotron emission (CSE) with power, duration and
   DM consistent with FRBs.

 The storyboard in our model, elaborated in this paper, can be very briefly summarized as follows:

\begin{enumerate}

\item A normal Type Ic (no broad lines) SN occurs following the collapse of a Wolf-Rayet star stripped 
of its Hydrogen and Helium envelopes. The resulting SN compact remnant is a  massive NS (either
born with mass exceeding $M_{\rm NS, cr.}$ or  can exceed it via mass accretion) but born rapidly rotating so 
to keep the core density below the quark deconfinement limit $\rho_{\rm NS, cr.}$;

\item Concurrently a Pulsar Wind Nebula (PWN) is powered by the spinning down pulsar born from the SN. The interaction
of the PWN with the SN ejecta generates a PWN-SN shell (we refer to as the ``wall" in this paper);

\item NS spin-down  drives the NS core density  above $\rho_{\rm NS, cr.}$ and triggers the QN.

\item The explosion 
releases $\sim 10^{53}$ ergs in kinetic energy imparted to the NS's outermost crust  layers which expands and 
fragments into millions of pieces (i.e. $\sim ~ 10^6$ chunks of  $\sim 10^{22}$-$10^{23}$ gm each). This QN ejecta moves out radially from the
explosion site with a Lorentz factor $\Gamma_{\rm QN}$ of a few thousand;

\item The chunks crash into the PWN-SN shell (i.e. the wall) to create a GRB\footnote{The QN chunks may be reminiscent of previous LGRB models involving a shower of ``Bullets'' (\citealt{heinz_1999}) and  a trail of ``cannonballs" (\citealt{dar_2004}) but ours
 is  fundamentally different. For example:   (i) The QN is an instantaneous event and occurs years to decades after the SN explosion;
 (ii) The QN explosion is isotropic yielding millions of chunks in a thin expanding spherical front; (iii) The GRB duration 
 in our model is due to the radial distribution of matter in the PWN-SN shell, the QN chunks interact with.}  
 (presented in details in \S ~ \ref{sec:grbs-non-filamentary} and \S ~ \ref{sec:grbs-filamentary}) or an FRB  (presented in details in \S ~ \ref{sec:frbs}).

\begin{enumerate}

\item The LGRB is synchrotron emission induced by  the interaction of a  dominant QN chunk (i.e. the one closest to the observer's
line-of-sight) with the turbulent PWN-SN shell. The afterglow 
is from the chunk's interaction  with the surrounding medium past the SN ejecta (\S ~ \ref{sec:the-wall-phase} and \S ~ \ref{sec:SN2-prompt});

\item A flare is from secondary chunks (surrounding the primary chunk) going through the same filaments as the
  primary chunk effectively echoing  the prompt emission (i.e. with emission occurring later in time and
  at lower frequencies; \S ~ \ref{subsec:flare-I} and \S ~ \ref{subsec:flare-II}); 

  \item The prompt emission from a chunk interacting with a filament in the turbulent PWN-SN shell
  yields a fast cooling synchrotron spectrum. A chunk passing through multiple
  filaments in the PWN-SN shell gives a convolution of many  fast cooling synchrotron spectra
     resulting in a Band-like spectrum (see \S ~ \ref{subsec:spectrum-multiple-filaments});
    
    \item The interaction of the primary and secondary chunks with the PWN-SN shell together
     yield the light-curve (including the prompt, flares and afterglow) and spectrum of  an LGRB (see \S ~ \ref{subsec:fits});
     
     \item The Yonetoku and Amati relationships are found to be phenomenological and
     not fundamental properties of LGRBs (see \S ~ \ref{sec:yonetoku-amati-theory} and \S ~ \ref{sec:yonetoku-amati-simulations});

   \end{enumerate}
     
     \item If the QN occurs in a non-turbulent or weakly turbulent PWN-SN shell
      with a weak magnetic field, 
     the Weibel instability  develops in the shocked chunk's frame when colliding with the
     PWN-SN shell. The instability  allows for particle bunching and the switch from incoherent to coherent
     synchrotron emission. FRBs result in this case with properties consistent with data (see \S ~ \ref{sec:frbs});
     
     \item  We propose a unification   of LGRBs and FRBs,  based on the degree of magnetization of the PWN-SN ejecta   (see \S ~ \ref{sec:unification}).

\end{enumerate}


  The paper is organized as follows: In \S ~ \ref{sec:QN} we give a brief overview of the physics and astrophysics
  of the QN. We describe the characteristics of the QN ejecta which is  ultra-relativistic and heavy-ion-rich 
   which fragments as it expands away from the explosion site. In \S ~ \ref{sec:QN-SN} we analyze the interaction of the
   SN ejecta with the underlying PWN. This section describes the PWN-SN shell (i.e. the wall) with which  the
   QN chunks, ejected years to tens of years after the SN, 
   interact. We first consider, in \S ~ \ref{sec:grbs-non-filamentary}, an analytical model based on a non-turbulent
    PWN-SN shell, and as a proof-of-principle, we show how the interactions of the QN chunks with such a wall and
    later with its surroundings can yield key properties of LGRBs. Improvement of the analytical model is
    given in \S ~ \ref{sec:grbs-filamentary} where a turbulent and filamentary PWN-SN ejecta is considered. This captures
     many more properties of LGRBs including the complex light-curves and the Band  spectrum, 
     while demonstrating that the Yonetoku and Amati laws are phenomenological in nature.
     Here we test our model against 48 
     observed LGRBs and show its success at fitting simultaneously their 
        light-curves (including the afterglow and flares) and spectra. 
      We end the GRB part of the paper by listing specific
      predictions of our model.
       
       FRBs (i.e. those  related to star-forming regions) are discussed in \S ~ \ref{sec:frbs} where we demonstrate that a QN
      occurring in a non-turbulent and weakly magnetized  PWN-SN (i.e. when the SN
      ejecta is not blown out by the NS spin-down power),
       allows the development of the Weibel instability in the shocked QN chunks. Coherent synchrotron emission (CSE) is triggered, 
        yielding luminosity, frequency and 
       duration consistent with observed FRBs. Some predictions 
      are listed at the end of this part of the paper. Other astrophysical implications (e.g. Ultra-High Energy
       Cosmic Rays; magnetar formation and SLSNe) of
       our model are explored in \S ~ \ref{sec:other-implications}. In particular, in \S ~ \ref{sec:SGRBs} we investigate  how
       a QN in the wake of a binary NS  merger can yield a SGRB.  In \S ~ \ref{sec:general-discussion} we present a general discussion of our model
       and list its limitations. We also suggest a scheme  which unifies 
        LGRBs and FRBs including XRFS, XRR-GRBs and SLSNe simply by varying the degree of 
        magnetization of the PWN-SN shell when it is hit by the QN chunks.  We conclude in \S ~ \ref{sec:conclusion}.

\section{The Quark Nova: key ingredients}
 \label{sec:QN}

In the QN model, quark deconfinement (i.e. when quarks  are no longer confined to hadrons) in the core of a massive NS  can be initiated by 
an increase of the core density above the deconfinement value $\rho_{\rm NS, c.}$.
 This can occur via spin-down as assumed in our paper here 
(e.g. \citealt{staff_2006}) and/or mass accretion (\citealt{ouyed_2011a,ouyed_2011b,ouyed_2015c})  triggering  
a hadronic-to-quark-matter conversion front.
Recent  analytical  (e.g. \citealt{vogt_2004,keranen_2005}) and  numerical  
(\citealt{niebergal_2010,ouyed_2013b,ouyeda_2015b,ouyeda_2018a,ouyeda_2018b}; see also references listed in \S ~ \ref{sec:intro-uds}) analyses of the microphysics and macrophysics of the transition, suggest that  the
transition could be of an explosive type which might occur via a Deflagration-Detonation-Transition (DDT) and/or
 quark-core collapse QN where the ``halted" quark core is prone to collapse in a mechanism similar to
 a core-collapse SN (see \citealt{niebergal_2011,ouyeda_2018}).  While our working hypothesis (i.e. the QN explosion) 
  remains to be confirmed in multi-dimension simulations which is currently the main focus
 of the QN group\footnote{See \url{www.quarknova.ca}}, our findings in this paper (and our work on SLSNe in the
 context of QNe in the wake of SNe; e.g. \citealt{ouyed_2016} and references therein)   lend it
 some support.

 \subsection{Quark deconfinement}
 \label{sec:QN-deconfinement}

As shown in \citet{staff_2006}, a QN is most likely
to occur when a NS is born with a mass just above  $M_{\rm NS, cr.}$.  If non-rotating, the QN will occur promptly. If the NS is 
rapidly rotating (i.e. a birth period $P_{\rm NS}$ of the order of a few milliseconds)  the core density at birth is below  $\rho_{\rm NS, cr.}$.
As the NS spin-downs, the core density eventually increases above $\rho_{\rm NS, cr.}$, triggering quark deconfinement in the core 
thus initiating the hadronic-to-quark-matter transition (see also \citealt{mellinger_2017}).  We assume the expanding conversion
front induces an explosive conversion (by means of a DDT or quark-core-collapse)  to a QS
(\citealt{niebergal_2010,ouyeda_2018a,ouyeda_2018b}).

 Hereafter we take $M_{\rm NS, c.}=2M_{\odot}$ in order to take into account  the  $\sim 2M_{\odot}$ NS observed by \citet{demorest_2010}.
 The precise value of $N_{NS,c.}$ does not affect the results of the current study. 
 A $2M_{\odot}$ NS does not rule out the existence of quark stars, since quark matter can be stiff enough to allow massive QSs (e.g. \citealt{alford_2007}; see also \S ~ 2.1  in \cite{ouyed_staff_2013} for a discussion).

\subsection{Energetics}
\label{sec:QN-energetics}

A QN can   release $(M_{\rm NS, cr.}/m_{\rm n}) E_{\rm conv.}\sim  3.8\times 10^{53}\ {\rm ergs}\times (M_{\rm NS, cr.}/2M_{\odot})$  from the direct conversion of its hadrons   to quarks with  an average of 
  $E_{\rm conv.}\sim 100$ MeV of energy released per hadron  (e.g. \citealt{weber_2005}); $m_{\rm n}$ is the neutron mass. Accounting for  gravitational  energy and additional  energy release during phase transitions within the quark phase, the total energy may easily exceed 
    $5\times 10^{53}$ ergs (e.g. \citealt{jaikumar_2004,vogt_2004,yasutake_2005}). Taking into account
     that a sizeable percentage of energy,   about 1/3 of the total conversion energy, 
    is transferred to the kinetic energy of the QN ejecta; $E_{\rm QN}\sim 10^{53}$ erg is adopted as 
   the fiducial value for the kinetic energy  of the QN ejecta. The  fiducial 
    Lorentz factor is taken as $\Gamma_{\rm QN}=10^{3.5}$  which translates to a QN ejecta mass $M_{\rm QN}=10^{-4.75}M_{\odot}$,  effectively, the outermost crust region of the NS (e.g. \citealt{keranen_2005,ouyed_leahy_2009,marranghello_2010}). Hereafter we write\footnote{We adopt a nomenclature for the dimensionless quantities as $f_{\rm x}= f/10^{x}$ with quantities in cgs units.}
$E_{\rm QN, 53}= E_{\rm QN}/10^{53}$ erg  and $\Gamma_{\rm QN, 3.5}=\Gamma_{\rm QN}/10^{3.5}$).

\subsection{Fragmentation of the Quark Nova ejecta: {\it the QN chunks}}
 \label{sec:QN-ejecta}

 The expanding relativistic QN ejecta cools rapidly enough to solidify or liquify  and break up
 into of order one million fragments (\citealt{ouyed_leahy_2009}).

\subsubsection{Chunk's mass and statistics}
\label{sec:c-stats}

The mass of a typical QN chunk for typical QN parameters is
 $10^{19}\ {\rm gm} < m_{\rm c}< 10^{23}\ {\rm gm}$ depending on whether the QN ejecta breaks up
in the solid or liquid phase. 
In  reality, the fragmentation (i.e. the mass of a typical chunk and the
resulting number of fragments; whose parameters are hereafter assigned the subscript ``c") is more complicated than
 the estimates presented in \citet{ouyed_leahy_2009}. For simplicity, we 
 set the number of chunks fixed  to $N_{\rm c}=10^6$ which implies a typical chunk mass
 $m_{\rm c}=M_{\rm QN}/N_{\rm c}\simeq 10^{22.5}$ gm; we assume that they all have the same mass
 (the implications of having a mass distribution
are mentioned at the end of \S~ \ref{sec:limitations}).  The distribution of QN chunks is equally spaced in solid angle and centered on the explosion site
 (see Figure \ref{fig:stencil} and Appendix \ref{appendix:fireworks}).

\subsubsection{Chunk's maximum size}
\label{subsec:optically-thin-regime}
  
 The very early stages of the evolution of the QN ejecta are dominated by adiabatic losses  inducing an almost instantaneous loss of the ejecta's internal energy; mainly due to the rapid expansion in the degenerate relativistic regime
of the  ejecta (\citealt{ouyed_leahy_2009}). However, re-heating from $\beta$-decay\footnote{Being neutron-rich, the QN ejecta was shown to be a favorable environment for r-process nucleosynthesis (\citealt{jaikumar_2007,kostka_2014b,kostka_2014c,kostka_2014}; see also Appendix \ref{appendix:rprocess} here).
  Heating from $\beta$-decay by the r-process yield may temporarily delay the cooling and fragmentation process but the outcome
   remains the same.} and  from sweeping of ambient material (see Appendix \ref{appendix:sweeping}) keeps the chunk's
temperature high enough that it will continue to expand until it  reaches the transparency radius
(when the chunk becomes optically thin). The chunk's size at transparency, 
 in the co-moving frame where quantities are primed, is given by 
 \begin{equation}
 \kappa_{\rm c}^\prime \rho_{\rm c, T}^\prime R_{\rm c, T}^\prime= 1\ .
 \end{equation}
 Here $\kappa_{\rm c}^\prime$ is the chunk's opacity and the subscript ``T" stands for ``Transparent".
 Writing the density as $\rho_{\rm c, T}^\prime= m_{\rm c}/(A_{\rm c, T}^\prime R_{\rm c, T}^\prime)$
  yields a  critical cross-section area

 \begin{equation}
 \label{eq:AcT}
A_{\rm c, T}= A_{\rm c, T}^{\prime}\sim 2.4\times 10^{21}\ {\rm cm}^2\times m_{\rm c, 22.5} \kappa_{\rm c, -1}\ ,
 \end{equation}
 or a radius of $R_{\rm c, T}^{\prime}\sim 2.8\times 10^{10}\ {\rm cm}^2\times m_{\rm c, 22.5}^{1/2} \kappa_{\rm c, -1}^{1/2}$. 
 When the chunk hits this critical size it stops expanding.
 The un-primed cross-section area is in the NS frame of reference (i.e. the GRB cosmological rest frame; see Appendix \ref{appendix:frames} for the different reference frames involved here). We take 
  a typical chunk's opacity to be $\kappa_{\rm c}= 0.1$ cm$^2$ g$^{-1}$ (see  Appendix \ref{appendix:rprocess});
  $\kappa_{\rm c}^\prime= \kappa_{\rm c}$ since opacity is frame independent.

The corresponding {\it baryon number density}, $\frac{m_{\rm c}/m_{\rm H}}{(4\pi/3){R_{\rm c, T}^{\prime}}^3}$, in the chunk's
frame  is 

 \begin{equation}
 \label{eq:nc}
 n_{\rm c, T}^\prime\sim 2.2\times 10^{14}\ {\rm cm}^{-3}\times m_{\rm c, 22.5}^{-1/2}\times \kappa_{\rm c, -1}^{-3/2}\ .
 \end{equation}

  \subsubsection{The chunk's sweeping luminosity}
  \label{sec:Lcsweep}

   As it sweeps up protons and electrons from the ambient medium of baryon number density $n_{\rm amb.}$, the chunk
   gains energy which it emits as radiation.
 The evolution of the chunk's sweeping luminosity $L_{\rm c, sw.}^\prime(t^\prime)$ and its 
 Lorentz factor $\Gamma_{\rm c}(t^\prime)$  are given by Eqs. (\ref{eq:Lcsw}) and (\ref{eq:Gammac})  in Appendix \ref{appendix:sweeping}. 
    When the  chunk is coasting at its maximum, constant size, given by  $A_{\rm c, T}\propto m_{\rm c}$, then 
    the chunk's mass cancels out of  Eq. (\ref{eq:Gammac}). This implies that the time evolution of 
   the chunk's Lorentz factor $\Gamma_{\rm c}(t^\prime)$ is determined by
the ambient density alone  $\rho_{\rm amb.} = \mu_{\rm H} m_{\rm H} n_{\rm amb.}$;  we take a mean molecular weight of $\mu_{\rm H}=1.25$
with $m_{\rm H}$  the hydrogen mass.
 Equations (\ref{eq:Lcsw})  and  (\ref{eq:Gammac}) can be integrated to yield
   the evolution of the chunk's Lorentz factor and the resulting,  promptly radiated, sweeping luminosity:
   
   \begin{align}
   \label{eq:GammacT}
   \Gamma_{\rm c}(t^\prime) &= \frac{\Gamma_{\rm QN}}{\left(1 + \frac{t^\prime}{t_{\Gamma}^\prime} \right)^{1/2}}\\
    \label{eq:LcswT}
   L_{\rm c, sw.}^\prime(t^\prime) &=   1.1\times 10^{36}\ {\rm erg\ s}^{-1}\times\\\nonumber
    &\times  m_{\rm c, 22.5}\kappa_{\rm c, -1} n_{\rm amb., 0} \Gamma_{\rm c, 3.5}(t^\prime)^2 \ ,
   \end{align}
   with $\Gamma_{\rm c}(0)=\Gamma_{\rm QN}$.
     The chunk's Lorentz factor  decreases after a characteristic hydrodynamical timescale (taking $\beta_{\rm c}=v_{\rm c}/c=1$
   where $c$ is the speed of light):

   \begin{equation}
  \label{eq:tGamma}
 t_{\Gamma}^\prime\simeq      \frac{9.9\times 10^6\ {\rm s}}{n_{\rm amb., 0} \Gamma_{\rm QN, 3.5}^2 \kappa_{\rm c, -1}} \ ,        
\end{equation}
       which is effectively set by the ambient density for a given QN explosion; i.e. for a given $\Gamma_{\rm QN}$.
        The equations above  assumes a constant ambient density. 
       In the case of varying ambient density, Eq.(\ref{eq:LcswT}) still holds but Eq.(\ref{eq:GammacT}) must be replaced by re-integrating Eq.(\ref{eq:Gammac}) accordingly.

\section{A QN in the wake of a SN} 
\label{sec:QN-SN}

We now consider a QN occurring after a SN explosion in which a rapidly rotating NS was 
born with a mass  above the critical mass $M_{\rm NS, cr.}$.
 For  such NSs,  the increase
   in core density is most dramatic at $t_{\rm SpD}$, the spin-down characteristic timescale (see \citealt{staff_2006}).  
   In other words, it is natural to assume that the time-delay $t_{\rm QN}$ between the SN proper  and the
   exploding neutron star is  the spin-down timescale; i.e. $t_{\rm QN}\simeq t_{\rm SpD}$ when
   $M_{\rm NS}\ge M_{\rm NS, c.}$.  Hereafter, we keep fixed
the radius and mass of the NS and 
set them to $R_{\rm NS}=10$ km and $M_{\rm NS}= M_{\rm NS, cr.}=2M_{\odot}$, respectively. 
    The NS moment of inertia we take to be $I_{\rm NS}=10^{45}$ g cm$^2$. 

 The decline of the pulsar spin-down power,
   assuming a magnetic dipole, depends on time as (\citealt{deutsch_1955,manchester_1977}),
   \begin{equation}
   L_{\rm SpD}(t) = L_{\rm SpD, 0} \left( 1+ \frac{t}{t_{\rm SpD}}\right)^{-2}\ ,
   \end{equation}
   with 
   \begin{equation}
   L_{\rm SpD, 0}\simeq 3.9\times 10^{41}\ {\rm erg\ s}^{-1}\times P_{\rm NS, -2.4}^{-4}B_{\rm NS, 12.5}^{2}\ ,
   \end{equation}
    \begin{equation}
    t_{\rm SpD}\simeq 103.7\ {\rm years}\times P_{\rm NS, -2.4}^2B_{NS, 12.5}^{-2}\ ,
    \end{equation}
    and a rotational energy, $E_{\rm SpD}= (1/2)I_{\rm NS} (2\pi/P_{\rm NS})^2$, of
    \begin{equation}
    E_{\rm SpD}  \simeq 1.2\times 10^{51}\ {\rm erg}\times P_{\rm NS, -2.4}^{-2}\ .
  \end{equation}
  The subscript ``SpD" stands for spin-down. 
     The NS's birth period and magnetic field are given in units of 4 milliseconds ($P_{\rm NS, -2.4}=P_{\rm NS}/4$ ms)
   and $10^{12.5}$ G ($B_{\rm NS, 12.5}=B_{\rm NS}/10^{12.5}$ G), respectively (hereafter our fiducial values).

  \subsection{Summary of model's  parameters}
  \label{subsec:model-parameters}

The parameters described below are in chronological order starting with the 
SN, followed by the Pulsar Wind Nebula (PWN) phase 
describing the interaction between the SN ejecta and the PW and, the QN proper which occurs at time $t_{\rm QN}=t_{\rm SpD}$ 
following the SN.

\begin{itemize}

  \item  {\bf SN parameters}:   There are 5 parameters. The first 3 are the SN ejecta's kinetic energy $E_{\rm SN}$, the
 SN ejecta's mass $M_{\rm SN}$ and $n$ which is the power-law index of the
 SN's steep density part overlaying the density plateau. We set the SN fiducial values as $E_{\rm SN}=10^{51}\ {\rm erg}$, $M_{\rm SN} = 10^{34}$ grams (i.e. $5M_{\odot}$)  and $n=9$.
  
   The SN ambient medium the SN explodes into is defined by its baryon number density  $n_{\rm amb.}$ and magnetic
  field $B_{\rm amb.}$. We  list them as part of the SN parameter set with fiducial values of
  1 cm$^{-3}$ and  $10^{-5}$ G.

 \item {\bf NS parameters}:  With the radius and mass of the NS
set to $10$ km and $2M_{\odot}$, respectively, there are  only 2 free parameters which are the NS's birth period $P_{\rm NS}$ and
   birth magnetic field $B_{\rm NS}$.  The  birth period varies only by a factor of
  a few from one SN to another since only massive NSs with  spin period $P_{\rm NS}$ of a few milliseconds 
  can  experience substantial increase in their core's
  density and explode as QNe (\citealt{staff_2006}). For  $B_{\rm NS}$, we take a log-normal distribution 
  with mean $12.5$ and variance $\sigma_{\rm \log B}=0.5$ (e.g. \citealt{faucher_2006} and references therein).

  The range in $t_{\rm SpD}$ is controlled by the NS birth magnetic field $B_{\rm NS}$ which varies
  by orders of magnitude from one source to another unlike $P_{\rm NS}$ which varies by a factor of a few at most.

   \item {\bf QN parameters}:  There are 4 parameters. The first 2 parameters are the ejecta's kinetic energy $E_{\rm QN}=10^{53}$ ergs (kept fixed; effectively
    set by the NS mass $M_{\rm NS, c.}$), and its Lorentz factor $\Gamma_{\rm QN}=10^{3.5}$ also kept fixed.
     The ejecta's mass is then given by $M_{\rm QN}=E_{\rm QN}/\Gamma_{\rm QN}c^2\sim 10^{-4.75}M_{\odot}$ (representative of  the NS's outermost crust).
  The third parameter we take to be the total number of chunks $N_{\rm c}=10^6$ which
  yields a typical QN chunk's mass $m_{\rm c}=10^{22.5}$ gm.
  Other parameters/properties of the QN ejecta such as the chunk's critical cross-section
 area $A_{\rm c, T}$ and the corresponding rest-frame baryon number density $n_{\rm c, T}^\prime$ (given by Eqs. (\ref{eq:AcT}) and (\ref{eq:nc}), respectively)  
 are all known  once the mass of the chunk is known.

 \end{itemize}

 Table \ref{table:parameters} lists the fiducial values for our parameters. 
The SN and NS parameters combined  yield the properties of the  PWN-SN turbulent shell 
(resulting from the interaction between the PWN and the
SN ejecta) which we refer to  as the ``wall".

\subsection{The  SN ejecta}
\label{subsec:SN-ejecta}

We  describe the SN  ejecta and how it is shaped by its interaction with the underlying
PWN and  with  the overlaying ambient medium (with constant baryon number density $n_{\rm amb.}$
and magnetic field $B_{\rm amb.}$) prior to the QN event.

The analytical SN and PWN models we adopt here are the self-similar solutions   given  in \citet{chevalier_1977,chevalier_1982,chevalier_1984,blondin_1998,blondin_2001}; see also  \citealt{reynolds_1984,vanderswaluw_2001,vanderswaluw_2003,vanderswaluw_2004,chevalier_2005}. 
For a constant $n_{\rm amb.}$, the size of the SN ejecta
 is  given by $R_{\rm SN} = \alpha_{\rm n} (Av_{\rm t}^n/\rho_{\rm amb.})^{1/n} t^{(n-3)/n}$ 
 with $A = ((5n-25)/(2\pi n))\times E_{\rm SN} v_{\rm t}^{-5}$ and
$v_{\rm t}= ((10n-50)/(3n-9)\times E_{\rm SN}/M_{\rm SN})^{1/2}$ is the velocity at the
intersection between the density plateau of the SN ejecta. The SN ejecta's  power law steep density gradient is set by the parameter $n$.
For $n=9$, $\alpha_{9}=1.048$, $A=1.5\times 10^7$ g cm$^{-3}$ s$^3 \times {E_{\rm SN, 51}}^{-3/2} M_{\rm SN, 34}^{5/2}$
and $v_{\rm t}=  4.7\times 10^3$ km s$^{-1}\times {E_{\rm SN, 51}}^{1/2} M_{\rm SN, 34}^{-1/2}$. I.e.

 \begin{equation}
 \label{eq:Rsn}
 R_{\rm SN}(t)  \simeq 3.0\times 10^{18}\ {\rm cm}  \times
 \left({E_{\rm SN, 51}}^{1/3}M_{\rm SN, 34}^{-2/9}n_{\rm amb., 0}^{-1/9}\right)\times t_{9.5}^{2/3}\ .
 \end{equation}
 Since $t_{\rm QN}=t_{\rm SpD}\sim 100$ years for our fiducial values, the time-dependency of the SN 
properties are given in units of $100$ years; i.e. $t_{9.5}=t/(10^{9.5}\ {\rm s})=t/(100\ {\rm years})$.
 
 The  SN ejecta's density profile is 
 
 \begin{equation}
\label{eq:rhoenv}
\rho_{\rm SN}(r,t)=\begin{cases}
\rho_{\rm Plat.}(t), &   \text{for\ } r <  R_{\rm Plat.}\\
\rho_{\rm Plat.}(t)\times \left( \frac{R_{\rm Plat.}}{r} \right)^{-n}, &   \text{for\ } r > R_{\rm Plat.}
\end{cases}
\end{equation}
with $R_{\rm Plat.}(t)= v_{\rm t} t$ 
defines the edge of the density plateau in the inner
SN ejecta and $\rho_{\rm Plat.}(t)=A t^{-3}$ is the time-evolving plateau's density.

For our fiducial SN parameters,  we get

\begin{equation}
\label{eq:RPlat}
R_{\rm Plat.}(t)\simeq 1.5\times 10^{18}\ {\rm cm}\times  \left({E_{\rm SN, 51}}^{1/2} M_{\rm SN, 34}^{-1/2}\right) \times t_{9.5}\ ,
\end{equation}

with the plateau's baryon number density, $n_{\rm Plat.}(t)=\rho_{\rm Plat.}(t)/m_{\rm H}$, being

\begin{equation}
\label{eq:nPlat}
n_{\rm Plat.}(t)\simeq 2.9\times 10^2\ {\rm  cm}^{-3}\times  \left({E_{\rm SN, 51}}^{-3/2} M_{\rm SN, 34}^{5/2}\right) \times t_{9.5}^{-3}\ .
\end{equation}

\subsection{The PWN-SN shell: the ``wall"}

The collision between the PWN and the inner SN ejecta
 (the plateau) leads to the formation of an PWN-SN dense shell (i.e.  
  the ``wall" with its parameters denoted with subscript ``w").
      The wall is at a radius $R_{\rm w} = 1.5 \left( \frac{n}{n-5} \right)^{1/5} \left( \frac{n-5}{n-3} \right)^{1/2} \left( \frac{{E_{\rm SN}}^3 L_{\rm SpD}^2}{M_{\rm SN}^5} \right)^{1/10} t^{6/5}$ or, 
   
   \begin{align}
    \label{eq:Rwall}
   R_{\rm w}(t) &\sim 1.5\times 10^{18}\ {\rm cm}~\times \\\nonumber
   &\times \left( \frac{{E_{\rm SN, 51}}^3}{M_{\rm SN, 34}^5} \right)^{1/10} \left(\frac{P_{\rm NS, -2.4}^{-2}}{t_{\rm SpD, 100}} \right)^{1/5} \times t_{9.5}^{6/5}\ ,
   \end{align}
  which assumes a constant pulsar luminosity $L_{\rm SpD}= E_{\rm SpD}/t_{\rm SpD}$.
     The corresponding wall's speed, $V_{\rm w}= dR_{\rm w}/dt = R_{\rm w}/t\propto t^{1/5}$, is
     
    \begin{align}
    \label{eq:Vwall}
   V_{\rm w}(t)&\sim 5.7\times 10^{3}\ {\rm km\ s}^{-1}\times \\\nonumber
  &\times  \left( \frac{{E_{\rm SN, 51}}^3}{M_{\rm SN, 34}^5} \right)^{1/10} \left(\frac{P_{\rm NS, -2.4}^{-2}}{t_{\rm SpD, 100}} \right)^{1/5} \times t_{9.5}^{1/5}\ .
   \end{align}
   For $t=t_{\rm QN}=t_{\rm SpD}$, we have $V_{\rm w}\propto P_{\rm NS}^{-2/5}$. 
   {\it I.e. the wall's speed varies very little in time and is roughly constant from one source to another in our model}.

   The wall's baryon number density is $n_{\rm w}=   f_{\gamma_{\rm ad.}}\times (\rho_{\rm SN}/m_{\rm H})$  with $\rho_{\rm SN}/m_{\rm H}$   is the SN ejecta's baryon number density
 and $f_{\gamma_{\rm ad.}} =(\gamma_{\rm ad.}+1)/(\gamma_{\rm ad.}-1)$ is the shock compression factor
 set by  the adiabatic index $\gamma_{\rm ad.}$. This gives, for $\gamma_{\rm ad.}=5/3$, 

\begin{equation}
\label{eq:nwall}
n_{\rm w}(t)\sim  1.2\times 10^3\ {\rm  cm}^{-3}\times   \left({E_{\rm SN, 51}}^{-3/2} M_{\rm SN, 34}^{5/2}\right)\times t_{9.5}^{-3}\ .
\end{equation}

The maximum wall's thickness can roughly 
be estimated from mass conservation to be $\Delta R_{\rm w}/R_{\rm w} = 1/(3 f_{\gamma_{\rm ad.}})$ so that $\Delta R_{\rm w}/R_{\rm w} = 1/12$. In principle, the wall can be thinner  if cooling is taken into account.

 \subsubsection{The wall's magnetic field}
 \label{subsec:Bwall}

 The development of the Rayleigh-Taylor (RT) instability within the
  PWN-SN interface (e.g. \citealt{chevalier_1978}) means that the wall's magnetic field is prone to
  turbulent amplification (e.g. \citealt{jun_1995,jun_1998,bucciantini_2004,duffell_2017}; see also \citealt{stone_2007,porth_2014}). These studies find that 
  the amplified magnetic field can be estimated using equipartition arguments  which allows us to assume for simplicity that 
$B_{\rm w}^2/8\pi= \epsilon_{\rm w}\times n_{\rm w} k_{\rm B} T_{\rm w, sh.}$
where the shocked wall temperature is given by $k_{\rm B} T_{\rm w, sh.}= 3/16\times \mu_{\rm e} m_{\rm H} V_{\rm w}^2$ and $\epsilon_{\rm w}$ being
  the ratio of magnetic to thermal energy at the PWN-SN shock.  We take the mean molecular weight per electron  to be $\mu_{\rm e}=2$ representative
  of the type-Ic SN ejecta.
This yields

\begin{equation}
\label{eq:Bwall}
B_{\rm w}(t)\simeq 0.8\ {\rm mG}\ \times \epsilon_{\rm w, -5}^{1/2} n_{\rm w, 3}(t)^{1/2} V_{\rm w, 8.7}(t)\ ,
\end{equation}
where we set $\epsilon_{\rm w}=10^{-5}$ as our fiducial value since it gives milli-Gauss values in line with simulations
 and measurements of the magnetic field strength in SNRs (e.g. \citealt{reynolds_2012} and references therein).
 This parameter enters 
 when calculating the spectrum (i.e. the synchrotron emission)  and is thus listed in Table \ref{table:parameters}.
Since $V_{\rm w}$ (given in units of $10^{8.7}$ cm s$^{-1}$= 5000 km s$^{-1}$) varies
 little from one source to another (see Eq. (\ref{eq:Vwall})), 
the wall's magnetic field depends essentially on the wall's
 density and the strength of the turbulent amplification parameter $\epsilon_{\rm w}$.  
 However, since $\epsilon_{\rm w}$ is expected to be constant once turbulence saturation is reached
 in the PWN-SN ejecta this  leaves $n_{\rm w}$ as the controlling parameter.

 \begin{table}[t!]
\begin{center}
\caption{Fiducial parameters  in our model (see \S ~ \ref{subsec:model-parameters}).}
 \label{table:parameters}
\resizebox{1.0\textwidth}{!}{
\begin{tabular}{|c|c|c|c|c||c|c||c|c|c||c|c|c|}\hline
 \multicolumn{5}{|c||}{SN} &    \multicolumn{2}{|c||}{NS}  & \multicolumn{3}{|c||}{QN} &  \multicolumn{3}{|c|}{Spectrum (see \S  ~\ref{subsec:spectrum-single-filament})}\\\hline
   $E_{\rm SN}$    &  $M_{\rm SN}$ & $n$  & $n_{\rm amb.}$ & $B_{\rm amb.}$ & $P_{\rm NS}$ & $B_{\rm NS}^\mathbf{a}$ & $E_{\rm QN}$   & $\Gamma_{\rm QN}$  & $N_{\rm c}$  &  $\epsilon_{\rm w}^\mathbf{b}$   & $p$ & $n_{\rm pairs}$\\\hline
  $10^{51}$ erg   &  $10^{34}\ {\rm gm}$ ($= 5M_{\odot}$) & 9  &1 cm$^{-3}$ & $10^{-5}$ G & $10^{-2.4}$ s (= 4 ms) & $10^{12.5}$ G & $10^{53}$ erg  & $10^{3.5}$  & $10^{6}$ & $10^{-5}$&   2.4 & 10\\\hline
\end{tabular}
}
 \end{center}
$^\mathbf{a}$ We adopt a log-normal distribution for the magnetic field with mean $10^{12.5}$ G and variance $\sigma_{\rm \log B}=0.5$ (see \S ~  \ref{subsec:model-parameters}).\\
  $^\mathbf{b}$ The wall's magnetization factor as defined in Eq. (\ref{eq:Bwall}).\\
\end{table}

  \subsection{Characteristic timescales}
  \label{sec:timescales}

There are two critical timescales  that define the interaction between the PW and the SN ejecta (e.g. \citealt{blondin_2001,chevalier_2005}), prior
to the QN explosion:

\begin{itemize}

\item {\bf The SN density plateau}: The wall would reach the edge of the SN ejecta plateau at time
 (obtained by equating Eqs. (\ref{eq:Rwall}) and (\ref{eq:RPlat}))
\begin{eqnarray}
\label{eq:tPlat}
t_{\rm Plat.} \sim \left(\frac{E_{\rm SN}}{E_{\rm SpD}}\right) t_{\rm SpD} \ ,
\end{eqnarray}
in the  constant spin-down luminosity case. 
 The condition $t_{\rm QN}=t_{\rm SpD}>  t_{\rm Plat.}$  is satisfied whenever $E_{\rm SN} \le  E_{\rm SpD}$
  and is equivalent to 

\begin{equation}
\label{eq:PNScr}
P_{\rm NS}  \le   P_{\rm NS, cr.} \sim \frac{4\ {\rm ms}}{{E_{\rm SN, 51}}^{1/2}}\ \cdot
\end{equation}

For $P_{\rm NS}  <   P_{\rm NS, cr.}$ (i.e. when $t_{\rm Plat.}<t_{\rm SpD}$),
   the QN occurs after the wall has reached the edge of the density plateau.
I.e., the SN ejecta is already blown-out by the PWN (see Figure 6 in \citet{blondin_2017}) and  can no longer be described
by a self-similar solution\footnote{The blow-out regime is simulated 
in \citet{blondin_2017}  by extending spin-down power beyond $t_{\rm Plat.}$.}. This case is more relevant 
in our model since it gives best fits to light-curves and spectra of observed LGRBs as we show in \S ~ \ref{subsec:fits}.
 Thus for our fiducial values, the blow-out regime corresponds
to  NSs born with a period in the range  $1.5\ {\rm ms} < P_{\rm NS} \le P_{\rm NS, cr.}$ 
  with the  lower limit set by the r-mode instability on rapidly rotating accreting NS (\citealt{andersson_1999,andersson_2000});

 \item {\bf The SN reverse shock:}   When the SN reverse shock (RS) propagates inward to the edge of the SN plateau 
  it triggers its inward motion 
   and eventually ``crushes" the PWN. The relevant timescale for $n=9$ is
  
  \begin{equation}
  \label{eq:trs}
  t_{\rm SN, RS}\simeq 459\ {\rm years}\times {E_{\rm SN, 51}}^{-1/2} M_{\rm SN, 34}^{5/6} n_{\rm amb., 0}^{-1/3}\ .
  \end{equation}

    For the  constant pulsar luminosity case, the ratio between the pressure in the PWN
  and behind the RS can be estimated (e.g.  Eq. (9) in \citealt{blondin_2001};
  see also \citealt{vanderswaluw_2001}) to be $P_{\rm PWN}/P_{\rm RS}\sim 1.5$ for our fiducial values. Thus, no crushing is more likely. Nevertheless,  we impose
  $t_{\rm SpD} <  t_{\rm SN, RS}$ (which guarantees $t_{\rm Plat.} <  t_{\rm SN, RS}$ because $t_{\rm Plat.}<t_{\rm SpD}$). This  means we do not need to consider  the effect of the SN reverse shock on the PWN.  
  We cannot rule out the scenario where the QN occurs while
   the wall has been crushed to smaller radii.  However, 
    the evolution of the crushed PWN changes so that the current model is not
   applicable. This is a complication beyond the scope of this paper and  may be worth exploring elsewhere.
   
   \end{itemize}
   
   Other  timescales  relevant to our model:
   
   \begin{itemize}

\item {\bf The SN optical depth $\tau_{\rm SN}$}:    The conditions $\tau_{\rm SN} <1$  (i.e.
an optically thin SN; see Appendix \ref{appendix:tauSN}),  
    translates to 
   
   \begin{equation}
     \label{eq:tSLSN}
   t_{\rm QN} > t_{\rm SLSN} \sim 1.8\ {\rm years}\times {E_{\rm SN, 51}}^{-1/2}M_{\rm SN, 34}\ .
   \end{equation}
  
    For $\tau_{\rm SN} >1$,
    the QN kinetic energy is ``absorbed" in the SN envelope re-brightening the SN
    and yielding a SLSN (see  \S ~ \ref{sec:QN-Type-Ic-BL});

  \end{itemize}

 In summary,  the range in time delay between the
   SN and the QN  applicable to GRBs is  $t_{\rm SLSN} <  t_{\rm QN}=t_{\rm SpD}  <t_{\rm SN, RS}$
   which for our fiducial values gives

   \begin{equation}
   \label{eq:tQNrange}
  1.8\ {\rm yr} <  t_{\rm QN}  <  460\ {\rm yr} \ .
   \end{equation}
         
   The corresponding range in wall's density (which controls the GRB luminosity in our model)  can be derived by incorporating
    the range in $t_{\rm QN}$  given in Eq. (\ref{eq:tQNrange}) into Eq. (\ref{eq:nwall}) to get
    
     \begin{equation}
   \label{eq:nwrange}
   12.6\ {\rm cm}^{-3} <  n_{\rm w}  < 2.1\times 10^8\ {\rm cm}^{-3} \ .
   \end{equation}
   
   The corresponding  wall's size (which controls GRB timescales in our model) is 
   derived by incorporating
    the range in $t_{\rm QN}$ in Eq. (\ref{eq:tQNrange}) into Eq. (\ref{eq:Rwall}) to get
    
     \begin{equation}
   \label{eq:rwrange}
   1.2\times 10^{16}\ {\rm cm} <  R_{\rm w}  < 9.3\times 10^{18}\ {\rm cm} \ .
   \end{equation}
     
       The first 3 panels in Figure \ref{fig:Wall-distributions} show the $n_{\rm w}, R_{\rm w}$ and
       $B_{\rm w}$ distributions, for our fiducial values of parameters, applicable to GRBs. The
       time delay, $t_{\rm QN}=t_{\rm SpD}$, is set
         by drawing $B_{\rm NS}$  from 
         a log-normal distribution with mean of 12.5 and a variance
     $\sigma_{\rm \log B_{\rm NS} }=0.5$ (see Table \ref{table:parameters}).

\section{Application to long duration GRBs I:  {\it A non-turbulent PWN-SN ejecta}}
\label{sec:grbs-non-filamentary}

  In this proof-of-principle section, we present the simple but analytically tractable case of: (i) the QN chunks
   colliding with a non-turbulent self-similar PWN-SN shell (i.e. the wall
    as described above)  located at $R_{\rm w}$ (with thickness $\Delta R_{\rm w}=R_{\rm w}/12$), 
 density $n_{\rm w}$ and magnetic field $B_{\rm w}$; 
  (ii) $t_{\rm QN}=t_{\rm SpD}$ which gives, for the fiducial values of our parameters, $t_{\rm QN}\sim 100$ years,
   $n_{\rm w}\sim 10^3$ cm$^{-3}$, $R_{\rm w}\sim 10^{18}$ cm and $B_{\rm w}\sim$ mG.
   
   Once the NS explodes,  the QN ejecta is ultra-relativistic. It catches up with the wall in less than a year
   during which time we assume the wall properties did not evolve.
   There are 3 distinct regions the QN ejecta interacts with:
    (i) the pre-wall phase (the inside of the PWN) before they collide
   with the wall; (ii)  the wall phase (giving us the prompt emission and the GRB proper); (iii) the post-wall phase when the
   chunks interact with the  ambient medium (giving us the  afterglow).
   
  This simple case will be used later as a reference when applying our model 
  to the fully turbulent PWN-SN shell in the blow-out regime (i.e.
  for $E_{\rm SN}<E_{\rm SpD}$; see Eq. (\ref{eq:tPlat})) which is relevant to most LGRBs  (see \S ~ \ref{subsec:fits}). 
   
   \subsection{The pre-wall phase: QN chunks inside the PWN}

    Inside the PW bubble (see Appendix \ref{appendix:PWN})  the density is low enough that a chunk's sweeping luminosity (Eq. (\ref{eq:Lcsw})) is
    dwarfed by heating from the $\beta$-decay of r-process elements in the chunk; i.e. $L_{\rm c, sw.}^\prime(t^\prime) << L_{\rm c, \beta}^\prime(t^\prime)$ with the $\beta$-decay power
    given by Eq. (\ref{eq:L-beta}). 
   The time evolution of the chunk's temperature $T_{\rm c}^\prime(t^\prime)$ and  cross-section area $A_{\rm c}^{\prime}(t^\prime)$  during the optically thick regime 
   (i.e. before transparency) is found from  Eqs. (\ref{eq:Tc}) and (\ref{eq:Ac}), respectively, in Appendix \ref{appendix:PWN}.

  The distance travelled by a chunk in the NS  frame before it becomes optically thin is $R_{\rm T}$ given
 by Eq. (\ref{eq:RT}) which, within a factor of a few,   is close to $R_{\rm w}$.
  The chunk's temperature  in the pre-wall phase at  transparency 
 (i.e. the corresponding blackbody at $ t_{\rm T}^\prime$ found from $A_{\rm c, T}\sigma_{\rm SB}{T_{\rm c, T}^\prime} ^4= L_{\rm c, \beta}(t^\prime)$) is 
  
  \begin{equation}
  \label{eq:BBprime}
 k_{\rm B}T_{\rm c, T}^\prime \simeq
       0.23\ {\rm eV} \times m_{\rm c, 22.5}^{-0.194}\kappa_{c, -1}^{-3.3/6.7} \ .
\end{equation}

Thus,  in the non-turbulent PWN-SN shell (i.e. the single 
  wall) scenario, we have the simple picture of the chunks
   being cool and optically thin when they 
 start colliding with the wall.   

   \subsection{The wall phase: QN chunks inside the wall}
   \label{sec:the-wall-phase}

   Important properties:
  
  \begin{itemize}
  
  \item {\bf Doppler effects}: Appendix \ref{appendix:frames} lists the references frames in our model: 
   the chunk's frame (primed quantities), the NS frame
  (unprimed quantities) and the observer's frame (quantities with superscript ``obs."). 
  Since  $\Gamma_{\rm c}(t^\prime)^2>>1$ and $\theta_{\rm c}<<1$ applies 
  we can write the Doppler factor as $D_{\rm c}(\Gamma_{\rm c}(t^\prime),\theta_{\rm c})\simeq 2\Gamma_{\rm c}(t^\prime)/(1+\Gamma_{\rm c}(t^\prime)^2\theta_{\rm c}^2)$;

  \item {\bf The primary chunk at $\theta_{\rm P}$ (closest to the line-of-sight) causes the prompt and afterglow emission}:
     Figure \ref{fig:stencil}
  shows the spacing between the QN chunks as presented in Appendix \ref{appendix:fireworks}. 
  The distribution of QN chunks is equally spaced in solid angle and centered on the explosion site.
  Because the angular spacing between chunks is several times
   larger than $1/\Gamma_{\rm c}\sim 3.2\times 10^{-4}/\Gamma_{\rm c, 3.5}$, there
will almost always be a single chunk dominating the observed prompt emission. This
   chunk  we refer to as the 
  ``{\it primary}" chunk and is depicted with subscript ``P". The primary's viewing angle is 
 $0 < \theta_{\rm P} < 2\times 10^{-3}/N_{\rm c, 6}^{1/2}$ with an average
 value $\bar{\theta}_{\rm P}\sim (4/3)/N_{\rm c}^{1/2}\simeq 1.3\times 10^{-3}/N_{\rm c, 6}^{1/2}$;

 \item {\bf The secondary chunk at $\theta_{\rm S}$ causes the flares}: Each primary chunk is surrounded by about 6 peripheral chunks (the secondaries)
 as described in  Figure \ref{fig:stencil} with $\theta_{\rm S}=\theta_{\rm sep.}-\theta_{\rm P}=4/N_{\rm c}^{1/2}-\theta_{\rm P}$;
 $\theta_{\rm sep.}=4/N_{\rm c}^{1/2}$ is the separation between adjacent chunks. Hereafter we
 use the simplification that these secondary chunks are combined into
 a single chunk whose  viewing angle is in the range 
 $2\times 10^{-3}/N_{\rm c, 6}^{1/2}< \theta_{\rm S} < 4\times 10^{-3}/N_{\rm c, 6}^{1/2}$ with an average
 value $\bar{\theta}_{\rm S}= (28/9)/N_{\rm c}^{1/2}\simeq 3.1\times 10^{-3}/N_{\rm c, 6}^{1/2}$. 
  The secondary chunk defines the flaring activity in our model and acts as a repeat, or echo, 
  of the prompt GRB induced by the primary chunk;
  
  \item{\bf The chunk's forward shock (FS) and reverse shock (RS)}:   The QN chunk collision with the wall yields a FS and a 
RS. 
      The RS is relativistic when $n_{\rm c}^\prime/n_{\rm w} << \Gamma_{\rm QN}^2$,  (e.g.
   \citealt{landau_1959,blandford_1976,meszaros_1992,sari_1995}). This case implies that  most of the chunk's kinetic energy
   is converted to internal energy, slowing down the chunk in a fraction of a second (the time
   it takes the RS to cross the chunk). 
 Using Eq. (\ref{eq:nc}) for $n_{\rm c}^\prime$, this occurs when

\begin{equation}
\label{eq:chunk-RS}
n_{\rm w}>  n_{\rm w,  RS}=2.2\times 10^7\ {\rm cm}^{-3}\times \left(m_{\rm c, 22.5}^{-1/2}\kappa_{\rm c, -1}^{-3/2}\Gamma_{\rm c, 3.5}^{-2}\right)\ .
\end{equation}
Using Eq. (\ref{eq:nwall}) this happens when

\begin{align}
\label{eq:tQN-RS}
t_{\rm QN}  < t_{\rm QN, RS} &=3.8\ {\rm yrs}\times \left(E_{\rm SN, 51}^{-1/2}M_{\rm SN, 34}^{5/6}\right)\times \\\nonumber 
&\times \left(m_{\rm c, 22.5}^{1/6}\kappa_{\rm c, -1}^{1/2}\Gamma_{\rm c, 3.5}^{2/3} \right)\ .
\end{align}
 The above is for $n_{\rm w}= f_{\gamma_{\rm ad.}}n_{\rm Plat.}=4n_{\rm Plat.}$. For higher compression factor $f_{\gamma_{\rm ad.}}$, 
$t_{\rm QN, RS}$ is higher by a multiplicative factor $(f_{\gamma_{\rm ad.}}/4)^{1/3}$.

For $t_{\rm QN, RS} < t_{\rm QN} < t_{\rm SN, RS}$, the chunk's RS is Newtonian. In this case, the dynamics and the emission
is dominated by the  FS which moves with a Lorentz factor $\Gamma_{\rm FS}(t^\prime)$;

  \item  {\bf The wall's (i.e. PWN-SN shell) geometry}: We assume that the  wall 
  is perfectly aligned along a spherical shell centered on the QN explosion. 
   In addition we assume that the wall is continuous spatially,  
    and has   a uniform density $n_{\rm w}$; 
   
    \item {\bf The relevant timescales}: There are two contributions:
    (i) a  radial time delay which arises 
   as the primary chunk crosses the wall and; (ii)  an angular  time
   delay between the primary chunk hitting at $\theta_{\rm P}$ and the secondary chunk hitting the wall at a higher viewing
   angle $\theta_{\rm S}$.   
 The angular time delay\footnote{We recall that unprimed quantities are given in the NS (i.e.  GRB cosmological rest) frame  (see Appendix \ref{appendix:frames}).} between them   is 
  
  \begin{equation}
  \Delta t_{\rm S-P}= \frac{R_{\rm w}\theta_{\rm S}^2}{2c} - \frac{R_{\rm w}\theta_{\rm P}^2}{2c} \ .
  \end{equation}

The component which dominates the GRB duration enters later when we consider a turbulent PWN-SN ejecta 
 is the radial time delay which  takes into account the radial distribution and extent of multiple filaments from the ``shredded" wall (see \S ~ \ref{subsec:fits});

  \item  {\bf The thin and thick wall scenarios}: Let us define  $t_{\rm w}^\prime  = \Delta R_{\rm w}/\Gamma_{\rm FS}c$
  as a measure of the wall's crossing time in the chunk's frame
   with $ \Delta R_{\rm w} =  R_{\rm w}/12$ the wall's thickness in the NS frame.    
    The distribution of the thickness parameter $t_{\rm w}^\prime/t_{\Gamma}^\prime = t_{\rm w}/t_{\Gamma}$ is
   shown in the lower right  panel  in Figure \ref{fig:Wall-distributions} for fiducial values of our parameters.
   If $t_{\rm w}^\prime < t_{\Gamma}^\prime$ then the chunks will 
   wall experience no deceleration (thin wall case) while in the thick wall case ($t_{\rm w}^\prime > t_{\Gamma}^\prime$ ) there is significant deceleration
   on timescales of a few times $t_{\Gamma}^\prime$.

  \end{itemize}
  
   In the remainder of this section,  we use the thin wall case (i.e. $t_{\rm w}^\prime <  3 t_{\Gamma}^\prime$) where
  the chunk's Lorentz factor remains roughly constant when crossing the wall so we can write  $\Gamma_{\rm FS}(t^\prime)\simeq \Gamma_{\rm QN}=10^{3.5}$. 
  The thick wall case (with $t_{\rm w}^\prime > 3t_{\Gamma}^\prime$) is presented in Appendix \ref{appendix:thick-wall} and is compared to the thin wall case
    at the end of this section.

    It is useful to differentiate between the 3 sets of parameters: (i) the wall (i.e. PWN-SN shell) parameters;
     (ii) the chunk/QN parameters;  (iii) the observer's parameters mainly defined by the viewing angle $\theta_{\rm c}$.
  For the solutions 
    presented in what follows we only vary the viewing angles $\theta_{\rm P}$ and $\theta_{\rm S}$  and the time delay between
     the QN and SN, $t_{\rm QN}=t_{\rm SpD}$. In the thin wall case, the Doppler
    factor depends only on the viewing angle so that that $D(\Gamma_{\rm c}(t^\prime),\theta_{\rm c})\simeq D(\Gamma_{\rm QN},\theta_{\rm c})= 2\Gamma_{\rm QN}/f(\theta_{\rm c})$ with
    
    \begin{equation}
     f(\theta_{\rm c})= 1+(\Gamma_{\rm QN}\theta_{\rm c})^2\ .
     \end{equation}
  For $0 < \theta_{\rm c}= \theta_{\rm P} < 2\times 10^{-3}/N_{\rm c, 6}^{1/2}$ this implies  $0 \le f(\theta_{\rm P})\le 41$
     and $f(\bar{\theta}_{\rm P})\simeq 17.9$ for the primary chunk. For the secondary chunk we have $2\times 10^{-3}/N_{\rm c, 6}^{1/2}< \theta_{\rm c}=\theta_{\rm S} < 4\times 10^{-3}/N_{\rm c, 6}^{1/2}$ with a corresponding $41 \le f(\theta_{\rm S})\le 161$ and $f(\bar{\theta}_{\rm S})\simeq 97.1$.

     Hereafter, we will   refer to the prompt emission (induced by the
   primary chunk) by the subscript ``GRB", the flaring  (induced by the secondary chunk) by the subscript ``Flare" and the afterglow (induced
   by the primary chunk)  by the subscript  ``AG", respectively.

 \subsubsection{The luminosity}
\label{sec:chunk-luminosity}

  When $t_{\rm QN, RS} < t_{\rm QN} < t_{\rm SN, RS}$, the RS into the chunk is purely Newtonian.
     The  emission is dominated by the chunk's FS 
     moving at a  Lorentz factor $\Gamma_{\rm FS}\simeq \Gamma_{\rm QN}$.
        The observed luminosity from a single chunk seen at an angle $\theta_{\rm c}=\theta_{\rm P}$ 
         from the line-of-sight hitting the wall   of density $n_{\rm w}$ 
 is $L_{\rm GRB}=D(\Gamma_{\rm QN},\theta_{\rm P})^4 L_{\rm c, sw.}^\prime$ 
 where the chunk's sweeping luminosity $L_{\rm c, sw.}^\prime(t^\prime)$ is given by Eq. (\ref{eq:LcswT});
 emitted as synchrotron radiation (see \S ~ \ref{subsec:spectrum-single-filament}). This
  gives
  
  \begin{align}
  \label{eq:LGRB}
  L_{\rm GRB} & \simeq \left( \frac{1.7\times 10^{54}\ {\rm erg\ s}^{-1}}{ f(\theta_{\rm p})^4}\right) \times \\\nonumber  
     &\times (m_{\rm c, 22.5}\kappa_{\rm c, -1} \Gamma_{\rm QN, 3.5}^6)\times \left( n_{\rm w, 3}\right)\ .
      \end{align}
 
        With $0 < \theta_{\rm P} < 2/N_{\rm c}^{1/2}$ and for the range in $n_{\rm w}$ given in Eq. (\ref{eq:nwrange}) we get
 
 \begin{equation}
 \label{eq:LGRB-range}
  7.5\times 10^{45}\ {\rm erg\ s}^{-1} <  L_{\rm GRB} < 3.5\times 10^{59}\ {\rm erg\ s}^{-1}\ ,
 \end{equation}
 with an average value of $8.8\times 10^{48}\ {\rm erg\ s}^{-1}$.

\subsubsection{The duration}
\label{sec:duration}

 The observed duration of emission from a single chunk going through the wall 
of thickness $\Delta R_{\rm w}$ is
$\Delta t_{\rm GRB}= \Delta R_{\rm w}/(D(\Gamma_{\rm QN}, \theta_{\rm P})\Gamma_{\rm QN}c) = f(\theta_{\rm P})\times \Delta R_{\rm w} /(2\Gamma_{\rm QN}^2c)$.
 For $ \Delta R_{\rm w}/R_{\rm w}=1/12$, we get 

\begin{equation}
\label{eq:tGRB}
 \Delta t_{\rm GRB} \sim \left( \frac{1}{6}\ {\rm s}\times f(\theta_{\rm P}) \right)\times  
  \frac{(R_{\rm w, 18})}{(\Gamma_{\rm QN, 3.5}^{2})}\ .
   \end{equation}
    For $0 < \theta_{\rm P} < 2/N_{\rm c}^{1/2}$  and for the
   range of $R_{\rm w}$ given in Eq. (\ref{eq:rwrange}) we arrive at 
   
   \begin{equation}
     4.8\ {\rm ms} < \Delta t_{\rm GRB} < 63.6\ {\rm s}\ ,
    \end{equation}
    with an average value of 3 seconds.

    \subsubsection{The isotropic energy}  
\label{subsec:Eiso-I}

The isotropic energy ($E_{\rm GRB}=  L_{\rm GRB} \Delta t_{\rm GRB}$) is 

\begin{align}
\label{eq:EGRB}
  &E_{\rm GRB}     \simeq \left( \frac{2.8\times 10^{53}\ {\rm erg}}{ f(\theta_{\rm p})^3}\right)\times \\\nonumber  
      &\times \left(m_{\rm c, 22.5}\kappa_{\rm c, -1} \Gamma_{\rm QN, 3.5}^4\right) \times \left( n_{\rm w, 3} R_{\rm w, 18}\right)\ .
 \end{align}
 
With $n_{\rm w, 3} R_{\rm w, 18}\simeq 2.9\times t_{9.5}^{-9/5}$ (i.e.
$0.19 <  n_{\rm w, 3} R_{\rm w, 18} < 183$), the  range in isotropic energy is 
\begin{equation}
  2.1\times 10^{48} \ {\rm ergs} < E_{\rm GRB} <  2.7\times 10^{57}\ {\rm ergs}  \ ,
\end{equation}
with an average value of $\sim 5\times 10^{49} \ {\rm ergs}$.

    \subsubsection{The afterglow}
    \label{sec:GRB-afterglow}

    Exiting the wall and the SN with a Lorentz factor of $\sim \Gamma_{\rm QN}$,  the primary chunk interacts with the surrounding ambient medium (subscript ``amb.") and radiates at a rate of

 \begin{align}
\label{eq:LAG}
 L_{\rm AG} &\simeq \left(\frac{1.7\times 10^{51}\ {\rm erg\ s}^{-1}}{f(\theta_{\rm P})^4}\right)\times \\\nonumber
&\times (m_{\rm c, 22.5}\kappa_{\rm c, -1} \Gamma_{\rm QN, 3.5}^6)\times (n_{\rm amb., 0}) \ ,
\end{align}
with a corresponding range, due to $\theta_{\rm P}$, of 
  \begin{equation}
  6\times 10^{44}\ {\rm erg\ s}^{-1} <  L_{\rm AG}   < 1.7\times 10^{51}\ {\rm erg\ s}^{-1}\ ,
 \end{equation}
 and an average value of $\sim 1.7\times 10^{46} \ {\rm erg}$ s$^{-1}$.
 
 The luminosity ratio
 between the prompt and afterglow emission is given by the density ratio $n_{\rm w}/n_{\rm amb.}\sim 10^3$
  in the single wall scenario.
   However, 
  in order to simultaneously fit the prompt, afterglow and flare emission of observed LGRB light-curves, the density jump alone is
  not sufficient and a decrease in $\Gamma_{\rm FS}$ prior to exiting the GRB phase is necessary (see \S ~ \ref{subsec:fits}), 
  which is suggestive of a thick wall. A thick wall is  also needed to recover the Band-like spectrum (see \S ~ \ref{subsec:spectrum-multiple-filaments}).

 The duration of the afterglow is $t_{\rm AG}= t_{\rm \Gamma, amb.}^\prime/D(\Gamma_{\rm QN}, \theta_{\rm P})$
 where $t_{\rm \Gamma, amb.}^\prime$ is the dynamical timescale (see Eq. (\ref{eq:tGamma})) in the ambient medium:
 
 \begin{equation}
 t_{\rm AG}= 1.5\times 10^3\ {\rm s}\times \frac{f(\theta_{\rm P})}{n_{\rm amb., 0}\Gamma_{\rm QN, 3.5}^3\kappa_{\rm c, -1}}\ ,
 \end{equation}
 with a range of $1.3\times 10^3\ {\rm s}\le t_{\rm AG}^{\rm obs.}\le 6.4\times 10^4\ {\rm s}$
and an average  value of $\sim 2.7\times 10^4\ {\rm s}$.

    \subsubsection{The spectrum}
    \label{subsec:spectrum-single-filament}
    
        There are 3 more parameters that define the spectrum. The electron energy
        distribution with the  power-law index $p$, the number of pairs $n_{\rm pairs}$ 
        generated in the chunk's FS per proton swept-up  and, $\epsilon_{\rm w}$
        the ratio of magnetic to thermal energy defining the wall's magnetization (see Eq. (\ref{eq:Bwall})).
        Important effects include:
        
 \begin{itemize}
        
        \item   {\bf Acceleration in the FS}:   A typical electron (or positron) accelerated by the FS acquires the average  Lorentz factor  of the electrons distribution 
  $\gamma_{\rm e, av.}=(\zeta_{\rm p} \Gamma_{\rm FS} m_{\rm p}/m_{\rm e})/2 n_{\rm pairs}$ (e.g. \citealt{piran_1999}; recall that $\zeta_{\rm p} =1$ in our case as
  explained in  Appendix \ref{appendix:sweeping}.  We define  $n_{\rm pairs}$ as the  number
of pairs created per proton by dissipative
processes in the FS (e.g. \citealt{thompson_2000,beloborodov_2002} and references therein) with  10 pairs
created per  proton swept-up as our fiducial value. The minimum Lorentz factor of the distribution is
   $\gamma_{\rm e, m}= \gamma_{\rm e, av.}\times (p-2)/(p-1)$ where $p>2$ is the power-law index
   describing the  distribution of Lorentz factors of the electrons. We get

   \begin{align}
   \label{eq:gamma_e_m}
    \gamma_{\rm e, m} &\simeq 8.3\times 10^4\times \frac{g(p)/g(2.4)}{n_{\rm pairs, 1}} \times \Gamma_{\rm QN, 3.5} \ ,
   \end{align}
   with $g(p)=(p-2)/(p-1)$.  The no-pairs case is recovered by setting $n_{\rm pairs}=1/2$ in all equations involving $n_{\rm pairs}$;
   
   \item {\bf Synchrotron emission}: We consider synchrotron emission from the chunk's FS.  
There are two relevant timescales in the chunk's co-moving frame. The  first is the synchrotron cooling time ($t_{\rm Syn.}^\prime\simeq 7.7\times 10^8$ s$/{B_{\rm w}^{\prime}}^2 \gamma_{\rm e}$; e.g. \citealt{rybicki_1986,lang_1999}). Here $B_{\rm w}^\prime=B_{\rm w, sh.}= \Gamma_{\rm FS} B_{\rm w}$
 is the shock compressed wall's magnetic field (in the shocked chunk's frame), which yields 

\begin{equation}
t_{\rm Syn.}^\prime\simeq \frac{7.7\times 10^7\ {\rm s}}{\Gamma_{\rm QN, 3.5}^2 B_{\rm w, -3}^2 \gamma_{\rm e}}\ ,
\end{equation}
with $B_{\rm w}$ given by Eq. (\ref{eq:Bwall}).

The above can be compared to  the chunk's  
  hydrodynamic time $t_{\Gamma}^\prime\simeq 9.9\times 10^3$ s$/(n_{\rm w, 3} \Gamma_{\rm QN, 3.5}^2\kappa_{\rm c, -1})$; see Eq.(\ref{eq:tGamma}). The  ratio is
  \begin{equation}
  \label{eq:tratio}
 \frac{t_{\rm Syn.}^\prime}{t_{\Gamma}^\prime}\sim \frac{1.2\times 10^{4}}{\gamma_{\rm e}}\times \frac{\kappa_{\rm c, -1}}{\epsilon_{\rm w, -5} V_{\rm w, 8.7}^2}\ ,
 \end{equation}
 where  $n_{\rm w}$ cancels out of the equation above since $B_{\rm w}^2\propto n_{\rm w}$.
  A critical electron Lorentz factor  is found by setting  $t_{\rm Syn.}^\prime =  t_{\Gamma}^\prime$ to get
   
   \begin{equation}
     \label{eq:gamma_e_c}
   \gamma_{\rm e, c}\simeq  1.2\times 10^4\times  \frac{\kappa_{\rm c, -1}}{\epsilon_{\rm w, -5}V_{\rm w, 8.7}^2} \ ,
   \end{equation}
   which is the Lorentz factor of an electron that cools on a hydrodynamic timescale.  The injected high-energy
    electrons will be cooled to this value in the fast-cooling regime;
 
  \item {\bf The peak photon energy}: For an electron of Lorentz factor $\gamma_{\rm e}$, the observed synchrotron photon energy is $E_{\rm \gamma}= D(\Gamma_{\rm QN},\theta_{\rm c}) E_{\rm \gamma}^\prime$ (with  $E_{\rm \gamma}^\prime=  (\hbar e/m_{\rm e}c) B_{\rm w, sh.} \gamma_{\rm e,}^2$; e.g. \citealt{lang_1999}):
  
  \begin{equation}
  \label{eq:Egamma}
  E_{\rm \gamma}\simeq \frac{2.3\times 10^{-4}\ {\rm eV}}{f(\theta_{\rm P})}\times \Gamma_{\rm QN, 3.5}^2 B_{\rm w, -3} \gamma_{\rm e}^2\ .
  \end{equation}

    \end{itemize}

        The  fast cooling regime occurs when $ \gamma_{\rm e, m} >  \gamma_{\rm e, c}$
        which is equivalent to
        
        \begin{equation}
        \label{eq:npairs}
        n_{\rm pairs} < 69 \frac{\Gamma_{\rm QN, 3.5}\epsilon_{\rm w, -5}V_{\rm w, 8.7}^2 \left(g(p)/g(2.4)\right)}{\kappa_{\rm c, -1}}\ .
        \end{equation}

       To derive the spectrum from a single chunk we first estimate 
          the   cooling photon energy (setting $\gamma_{\rm e}=\gamma_{\rm e, c}$ in Eq. (\ref{eq:Egamma}))    to be
   \begin{equation}
   \label{eq:Ecool}
   E_{\rm \gamma, c}\simeq \left(\frac{25.9 \ {\rm keV}}{f(\theta_{\rm P})}\right)\times \left( \Gamma_{\rm QN, 3.5}^2 \kappa_{\rm c, -1}^2\right)\times
    \left( \frac{n_{\rm w, 3}^{1/2}}{\epsilon_{\rm w, -5}^{3/2}V_{\rm w, 8.7}^3} \right) \ ,
   \end{equation}
   where we replaced $B_{\rm w}$, given by Eq. (\ref{eq:Bwall}), in Eq. (\ref{eq:Egamma}). Similarly, 
 The observed characteristic photon energy   (setting $\gamma_{\rm e}=\gamma_{\rm e, m}$ in Eq. (\ref{eq:Egamma}))  is

   \begin{align}
    \label{eq:Epeak}
  E_{\rm \gamma, p}&\simeq \left(\frac{1.2\ {\rm MeV}}{f(\theta_{\rm P})}\right)\times \left(\Gamma_{\rm QN, 3.5}^4\right)\times \left(\frac{g(p)/g(2.4)}{n_{\rm pairs, 1}}\right)^2 \times \\\nonumber 
   &\times \left(n_{\rm w, 3}^{1/2} \epsilon_{\rm w, -5}^{1/2}V_{\rm w, 8.7}\right) \ .
   \end{align}
   
      For $0 < \theta_{\rm P} < 2/N_{\rm c}^{1/2}$  and for the range in $n_{\rm w}$ given in
   Eq. (\ref{eq:nwrange}) we get 
   \begin{equation}
   3.4\ {\rm keV} <  E_{\rm \gamma, p} < 42.7\ {\rm MeV} \ ,
   \end{equation}
  with an average value of 218 keV.
   
     In the single thin wall case, the spectrum is a fast cooling synchrotron spectrum (since $E_{\rm \gamma, p} >E_{\rm \gamma, c}$) 
      which is different from the Band function (\citealt{band_2004}).  However, as we show in  \S ~ \ref{subsec:spectrum-multiple-filaments}, slowing down of the chunk in the case of a single thick
    wall  (i.e. a time-varying Lorentz factor $\Gamma_{\rm FS}=\Gamma_{\rm FS}(t^\prime)$) and/or when
    considering a primary chunk interacting with multiple filaments yields a Band function.

 \subsubsection{The flare}  
\label{subsec:flare-I}

A flare in our model is from the chunk  (at $\theta_{\rm c}=\theta_{\rm S}$) colliding with the wall.
 In this case, flares
can be seen as a repetition of the prompt emission with a smaller Doppler factor (i.e. stretched in time
but reduced in intensity). 
The luminosity ratio between a flare and a burst is thus

\begin{equation}
\label{eq:L-Flare}
\frac{L_{\rm Flare}}{L_{\rm GRB}} =  \frac{f(\theta_{\rm P})^4}{f(\theta_{\rm S})^4}=  \left(\frac{1+(\Gamma_{\rm QN}\theta_{\rm P})^2}{1+(\Gamma_{\rm QN}\theta_{\rm S})^2} \right)^4 \ .
\end{equation}
With $\theta_{\rm S}= 4/N_{\rm c}^{1/2}-\theta_{\rm P}$ this yields a range of 
  $1.5\times 10^{-9} \le L_{\rm Flare}^{\rm obs.}/L_{\rm GRB}^{\rm obs.}\le 1$ which is a very wide range. On average
 for  $\bar{\theta}_{\rm P}= (4/3)/N_{\rm c}^{1/2}$ and $\bar{\theta}_{\rm S}=(28/9)/N_{\rm c}^{1/2}$ 
  we get $L_{\rm f}/L_{\rm b}\simeq  10^{-3}$.

We assumed  all chunks have the same mass and Lorentz factor and pass through a wall with uniform density $n_{\rm w}$. 
 As we show in our fits to data (see \S ~ \ref{subsec:fits}), this assumption  has 
  to be relaxed to explain flares in some LGRBs.
  
  The ratio between the Flare  and the LGRB duration is 
  
   \begin{equation}
   \label{eq:t-Flare}
\frac{\Delta t_{\rm Flare}}{\Delta t_{\rm GRB}} =  \frac{f(\theta_{\rm S})}{f(\theta_{\rm P})}=\left(\frac{1+(\Gamma_{\rm QN}\theta_{\rm S})^2}{1+(\Gamma_{\rm QN}\theta_{\rm P})^2} \right) \ .
\end{equation}
  With $41 < f(\theta_{\rm S})< 161$, this gives a range in Flare duration of  $1< \Delta t_{\rm Flare}/\Delta t_{\rm GRB}< 161$.
  
  The ratio of photon peak energy between the Flare and the GRB is 
  
  \begin{equation}
  \label{eq:E-Flare}
\frac{E_{\rm \gamma, p, Flare}}{E_{\rm \gamma, p, GRB}} =  \frac{f(\theta_{\rm P})}{f(\theta_{\rm S})}=\left(\frac{1+(\Gamma_{\rm QN}\theta_{\rm P})^2}{1+(\Gamma_{\rm QN}\theta_{\rm S})^2} \right) \ ,
\end{equation}
  with a range of 
  $6.2\times 10^{-3} \le E_{\rm p, Flare}/E_{\rm p, GRB}\le 1$ and
  an average of $f(\bar{\theta}_{\rm P})/f(\bar{\theta}_{\rm S})\simeq 17.9/97.1\sim 0.2$. 
  
  The angular time delay between the secondary  and the primary, effectively the time
  of occurrence of the flare in the light-curve,  is 
  
  \begin{equation}
  \label{eq:ts-Flare}
  t_{\rm Flare}=\Delta t_{\rm S-P}= \frac{R_{\rm w}\theta_{\rm S}^2}{2c} - \frac{R_{\rm w}\theta_{\rm P}^2}{2c} \ ,
  \end{equation}
  
   which varies from 0 when $\theta_{\rm S}=\theta_{\rm P}$ to a maximum
   of $(R_{\rm w}/2c)\times (16/N_{\rm c})$ when ($\theta_{\rm P},\theta_{\rm S})= (0,4/N_{\rm c}^{1/2})$. This gives a range
   
    \begin{equation}
     0\ {\rm s} < t_{\rm Flare} <  2.5\times 10^3\ {\rm s}\ .
  \end{equation}
  
 A Flare is ``a mirror image" of the prompt emission stretched in
time, with a softer spectrum, and occurring at later time.

  \subsection{Comparison to data}
  
  Here we compare our analytical single wall model to LGRB data from \citet{ghirlanda_2009} 
  which consists of the rest frame peak luminosity $L_{\rm iso, peak}$, isotropic energy $E_{\rm iso}$
  and photon peak energy $E_{\rm peak}$.   In the single wall model  we have $L_{\rm iso, peak}=L_{\rm GRB}$ given by Eq. (\ref{eq:LGRB}), 
   $E_{\rm iso}= E_{\rm GRB}$ given by Eq. (\ref{eq:EGRB}) and the photon peak energy $E_{\textrm{peak}}=E_{\gamma, p}$ given by
 Eq. (\ref{eq:Epeak}).
   All of the model's physical 
     quantities are in the NS frame meaning the GRB cosmological rest frame.
      The duration $\Delta t_{\rm GRB}$ in our model is given by Eq. (\ref{eq:tGRB}) while 
       the observed  $t_{90}$ data (where T90 is the time to detect 90\% of the observed fluence) is  from \url{https://swift.gsfc.nasa.gov/archive/grb_table}.
 Here we include the thick wall case described in Appendix \ref{appendix:thick-wall};
   in the thick wall case, we set the GRB duration to be $3 t_{\Gamma}$. We find
   that both thin and thick wall cases are required to match data. 
  
  \subsubsection{The NS magnetic field distribution}

  To compare our analytical single wall case to GRB data, we run  models keeping
  most of our parameters fixed as given in Table \ref{table:parameters}. We only vary the viewing angle
  $\theta_{\rm P}$ and the time delay between the QN and SN, $t_{\rm QN}=t_{\rm SpD}$ (recall also that $t_{\rm SpD}\sim t_{\rm Plat.}$
 for $E_{\rm SN}=E_{\rm SpD}$; i.e. for $P_{\rm NS}=4$ ms).   The  range in time delay given by Eq. (\ref{eq:tQNrange}) translates to 
 
  \begin{align}
   \label{eq:Brange}
   & 1.5\times 10^{12}\ {\rm G}\times P_{\rm NS, -2.4}^{-1}E_{\rm SN, 51}^{-1/4}M_{\rm SN, 34}^{5/12}n_{\rm amb., 0}^{-1/6} <  B_{\rm NS} \\\nonumber
   &  <  2.4\times 10^{13}\ {\rm G}\times P_{\rm NS, -2.4} E_{\rm SN, 51}^{1/4}M_{\rm SN, 34}^{-1/2}\ .
   \end{align}

   For a $B_{\rm NS}$ randomly drawn from a log-normal distribution of the pulsars' birth magnetic field with mean $\mu_{\log{B}}=12.5$ and 
 standard deviation $\sigma_{\log{B}}=0.5$, the magnetic field distribution relevant to GRBs is a subset of the observed one
 since it is subject to
      the limits given by Eq. (\ref{eq:Brange}) above. The resulting distribution is narrower with $\sigma_{\rm \log B_{\rm NS}}=0.2$ and a mean of 12.5.
      
      We run 500 simulations (the dots in Figures \ref{fig:amati-single-theory1}
  and \ref{fig:amati-single-theory2}) of our analytical model  each representing a single chunk
 passing through a single thin or thick wall  (see Appendix \ref{appendix:thick-wall} for key differences between the
 thin and thick wall cases). 
 The randomized variables are:

\begin{itemize}
	\item $\theta_P = $acos(UniformDistribution[cos($10^{-3}$), 1])
	\item B$_{\textrm{NS}} =$ LogNormalDistribution(12.5 log(10), .2 log(10))
	\item z: Randomly choose a LGRB from a list of over 300  (retrieved from \url{https://swift.gsfc.nasa.gov/archive/grb_table/}) and use its z.
\end{itemize}

For each run, $B_{\rm NS}$ gives us  $R_{\rm w}\propto B_{\rm NS}^{-2}, n_{\rm w}\propto B_{\rm NS}^6$ and
$\Delta R_{\rm w}=R_{\rm w}/12$.
  The other parameters  were kept constant
 to their fiducial values (see Table \ref{table:parameters}).
 Our runs are compared to LGRB data (the pluses in Figures \ref{fig:amati-single-theory1}
  and \ref{fig:amati-single-theory2}). 
   The top left panel in Figure \ref{fig:amati-single-theory1} shows $E_{\rm peak}$ ($=E_{\rm \gamma, p}$) versus  redshift
  which is consistent with data.
 The  upper right panel
  shows $E_{\rm peak}$ versus the GRB duration. The duration is not expected to match 
   the data since the single wall model includes only a single pulse. 
   The slope  in the $E_{\rm \gamma, p}$-$\Delta t_{\rm GRB}$ models is due to the fact that $E_{\rm \gamma, p}\propto n_{\rm w}^{1/2}\propto B_{\rm NS}^{3}$ and $\Delta t_{\rm GRB}\propto R_{\rm w}\propto B_{\rm NS}^{-2}$
    which yields $E_{\rm \gamma, p}\propto {\Delta t_{\rm GRB}}^{-3/2}$.

 We now discuss the Yonetoku and Amati laws resulting from our analytical single
 wall  model.  The Yonetoku law is shown in the right panels in Figure \ref{fig:amati-single-theory2}
while the Amati law is in the left panels. Best overall    fits were obtained by adjusting the number of pairs from 10 to $n_{\textrm{pairs}} = 20$.

  The top panels show the case of a constant $B_{\rm NS}$ and
  varying viewing angle $\theta_{\rm P}$. The slope in our model agrees  better with Amati law than 
  with Yonetoku's.
  In the middle panels where the viewing angle is
 kept constant while varying $B_{\rm NS}$, there is a better agreement with
 Yonetoku's but a clear  deviation from Amati's 
 for the high $t_{\rm w}/t_{\Gamma}>10$ sources; we refer to this as the ``hook".
  Both laws appear to be restored when varying both $B_{\rm NS}$ and the viewing angle
  as shown in the bottom panels.

  In general for the very thick wall case (i.e. $t_{\rm w}/t_{\Gamma}>10$),  the Amati relationship
is not preserved unless the chunk's viewing angle $\theta_{\rm P}$
is varied from source to source. However even when varying $\theta_{\rm P}$ between sources there are still
some leftover effect of the ``hook" in the bottom left  panel for the thickest filaments.

 \subsubsection{The phenomenological Yonetoku and Amati laws}
 \label{sec:yonetoku-amati-theory}

These two-components relationships  are in fact phenomenological
 and are an  artifact of limited parameter space (i.e. a limited scatter effect)  
 describing a GRB in our model.
For example, $L_{\rm iso, peak}$ ($=L_{\rm GRB}$) for a single chunk is given by Eq. (\ref{eq:LGRB}) and 
depends on $\theta_{\rm P}, m_{\rm c}, \Gamma_{\rm QN}$, and $n_{\rm w}\propto t_{\rm QN}^3\propto B_{\rm NS}^{-6}$.
Most parameters vary only by a small amount, so we set them to their fiducial values, as we did above, in the following analysis.
The two parameters that have significant variation are $\theta_{\rm P}$ and $B_{\rm NS}$.

Expressing $L_{\rm iso}$ ($=L_{\rm GRB}), E_{\rm peak}\ (=E_{\rm \gamma, p})$ and $E_{\rm iso}$ ($=E_{\rm GRB}$) in terms of their dependence on $\theta_{\rm P}$ and $B_{\rm NS}$, we obtain
for the thin wall case ($t_{\rm w}^\prime/t_{\Gamma}^\prime < \sim 3$):

\begin{align}
E_{\rm peak}&=C1 \times \frac{n_{\rm w}^{1/2}}{f(\theta_{\rm P})}=C1^\prime\times \frac{B_{\rm NS}^3}{f(\theta_{\rm P})}\\\nonumber
L_{\rm iso, peak}&=C2 \times \frac{n_{\rm w}}{f(\theta_{\rm P})^4}=C2^\prime\times \frac{B_{\rm NS}^6}{f(\theta_{\rm P})^4}\\\nonumber
E_{\rm iso}&=C3 \times \frac{n_{\rm w}R_{\rm w}}{f(\theta_{\rm P})^3}=C3^\prime\times \frac{B_{\rm NS}^4}{f(\theta_{\rm P})^3} \ ,
\end{align}
with $C1 (C1^\prime), C2 (C2^\prime)$ and $C3 (C3^\prime)$ constants. The expressions in the middle are for the general
case of $t_{\rm QN}\ne t_{\rm SpD}$ while the expressions
to the right are for $t_{\rm QN}=t_{\rm SpD}$. 
Here we focus on the $t_{\rm QN}=t_{\rm SpD}$ case to demonstrate the phenomenological nature
of the Yonetoku and Amati laws but this can be easily extended to the general case of $t_{\rm QN}\ne t_{\rm SpD}$.

We see that we cannot write $L_{\rm iso, peak}= f(E_{\rm peak})$ (i.e. as a function of $E_{\rm peak}$ alone)
or  $E_{\rm iso}= f(E_{\rm peak})$ because they are two independent variables.
I.e. $L_{\rm iso, peak}$ is not a function of $E_{\rm peak}$, nor is $E_{\rm iso}$.
 Thus both Yonetoku and Amati plots will yield a scatter of points about the  relation, 
for which the scatter is determined by the range of $B_{\rm NS}$ and $f(\theta_{\rm P})$.

Let us consider two options:

\begin{itemize}

\item If we take $B_{\rm NS}= constant$, then $L_{\rm iso, peak}$ varies as $E_{\rm peak}^4$ and $E_{\rm iso}$ varies as $E_{\rm peak}^{3}$.
These slopes are recovered in the 500 analytical 
models  shown in the top panels in Figure  \ref{fig:amati-single-theory2}.  In the constant $B_{\rm NS}$ case, 
  the thickness parameter   is constant (here $t_{\rm w}^\prime/t_{\Gamma}^\prime=t_{\rm w}/t_{\Gamma}\sim 3$)
 since the wall's properties ($n_{\rm w}, R_{\rm w}$ and $\Delta R_{\rm w}=R_{\rm w}/12$)
 are all constant.

\item If we take $\theta_{\rm P} = constant$, then $L_{\rm iso, peak}$ varies as $E_{\rm peak}^2$ and $E_{\rm iso}$ varies as $E_{\rm peak}^{4/3}$.
These slopes are also recovered in the middle panels in  Figure \ref{fig:amati-single-theory2}. Note that the thick wall models  (with $t_{\rm w}^\prime > 3 t_{\Gamma}^\prime$) 
deviate slightly from these correlations and are violated for extreme cases when $t_{\rm w}^\prime > 10 t_{\Gamma}^\prime$.

\item The bottom  panels in Figure  \ref{fig:amati-single-theory2} show the 500 models when both
$\theta_{\rm P}$ and $B_{\rm NS}$ are varied.  
In our analytical model, $\log(B_{\rm NS})$ has a scatter of $\sim 0.2$, and $f(\theta_{\rm P})$ varies between 1 and  $\sim 41$.
   Using $\sigma_{\log{B_{\rm NS} } }\sim 0.5$
   gives a much larger  vertical and horizontal scatter (i.e. about $\sim 5$ times bigger) in the bottom panels.

\end{itemize}
  
  In summary, neither the Yonetoku and Amati relations are fundamental, but are phenomenological (as
also demonstrated with sumulations in \S ~ \ref{sec:yonetoku-amati-simulations}). According to our model, 
 they are both the result of GRB dependence (i.e. $L_{\rm iso, peak}, E_{\rm peak}$ and $E_{\rm iso}$) on multiple physical parameters, 
which each have a limited range of scatter.  Observationally, selection
effects (e.g.  cut-offs due to detector sensitivity as discussed
for example in \citealt{collazzi_2012}) can result in limited scatter thus yielding in principle phenomenological correlations
as described in our model.

To understand the related slopes as reported in the literature we argue the following:

  \begin{itemize}
  
  \item {\bf The slope in the Yonetoku law}: Taking different values of  $\theta_{\rm P}$ gives a succession of parallel
  lines each with a slope of 4/3.  Taking different values of  $B_{\rm NS}$  gives a succession of parallel
  lines each with a slope of 3. These series of lines in the $L_{\rm iso, peak}$-$E_{\rm peak}$ plane create a scatter which when fit
  yields a phenomenological slope in the range
  
  \begin{equation}
  \frac{4}{3} \le {\rm Slope}_{\rm Yonetoku} \le 3\ .
  \end{equation}
 The lower limit corresponds to a scatter dominated by a big range in $\theta_{\rm P}$ while the
  upper limit correspond to a wider range in $B_{\rm NS}$.
  
  \item {\bf The slope in the Amati law}: Taking different values of $\theta_{\rm P}$  gives a succession of parallel
  lines each with a slope of 2.  Taking different values of $B_{\rm NS}$ gives a succession of parallel
  lines each with a slope of 4. These series of lines in the $E_{\rm iso}$-$E_{\rm peak}$ create a scatter which when fit
  yields a phenomenological slope of
  
  \begin{equation}
  2 \le {\rm Slope}_{\rm Amati} \le 4\ .
  \end{equation}
 The lower limit corresponding to a scatter dominated by a big range in $\theta_{\rm P}$ while the
  upper limit correspond to a wider range in $B_{\rm NS}$.

\end{itemize}

We revisit the phenomenological Yonetoku and Amati laws in \S ~ \ref{sec:yonetoku-amati-simulations}.


\section{Application to long duration GRBs II: {\it A turbulent filamentary PWN-SN ejecta}}
\label{sec:grbs-filamentary}

The single filament model (i.e. considering only the analytical self-similar wall), while it helps to understand our engine and is successful
  at capturing key and general features of our model,  cannot reproduce the wider variation in duration observed in GRBs,
   the Band function for the thin wall case
   and, does not allow for variable luminosity.   
      Here we consider the  case of the QN chunks interacting with a turbulent, filamentary, PWN-SN shell
    in the blow-out regime defined by $P_{\rm NS}< 4\ {\rm ms}/{E_{\rm SN, 51}}^{1/2}$ (i.e. when $E_{\rm SN}< E_{\rm SpD}$; see Eq. (\ref{eq:tPlat})).

The top panel in Table \ref{table:GRB-regimes} is a summary of the different stages in the
 blow-out regime.  
    This regime was simulated in \citet{blondin_2017} and consists of a pre-blow-out stage
    (Figures  3 in that paper) 
    and a blow-out stage (Figure  6 in that paper). These Figures demonstrate how the self-similar solution is modified
in 2-Dimensional  simulations. 
Figure 3 in \citet{blondin_2017} shows that in the pre-blow-out stage, 
roughly 50\% of the wall is turbulent and filamentary
from the broken off Rayleigh-Taylor (RT) fingers filling
the PW bubble interior. The remaining $\sim$ 50\% of the wall  is in a
quasi-spherical self-similar layer between the filaments and the unperturbed density
plateau. 

 In the blow-out stage,  the wall and the SN ejecta are torn apart as shown in 2-Dimensional (Figure 6 in \citet{blondin_2017}) and 3-Dimensional  simulations 
 (Figures 7 and 9 in \citet{blondin_2017}).
 The  Rayleigh-Taylor fingers split into numerous smaller ``filaments" with
 density varying from much less than the wall's to that of the wall with most filaments having 
  a density of the order of the plateau's density. 
 The highly filamentary PWN-SN is  extended ($>> R_{\rm w}(t_{\rm Plat,})$) forming large  low density corridors. 
 This stage  is of particular interest to us since it  gave best fits to LGRB data in our model, as we show in section \S ~ \ref{subsec:fits}.

      \subsection{The prompt emission}
  \label{sec:SN2-prompt}
  
 To ensure that the QN occurs when the PWN-SN  is in the blown-out stage we set $P=2$ ms
  instead of $P=4$ ms as adopted earlier in the analytical model.
    Eq. (\ref{eq:tPlat}) implies that $t_{\rm Plat.}\sim  t_{\rm SpD}/4$
  or equivalently that $t_{\rm QN}=t_{\rm SpD}\sim  4t_{\rm Plat.}$ with $t_{\rm SpD}\simeq 25.9$ years for the mean magnetic field value
    of $B_{\rm NS}=10^{12.5}$ G. Figure 7  in \citet{blondin_2017}  shows the PWN-SN
   shell at $7t_{\rm Plat.}$ which helps us picture the geometry of the blown out turbulent PWN-SN ejecta.

   To simulate the filamentary shell in the blow-out regime, we:
   (i) scale the blow-out PWN-SN ejecta  with respect to $R_{\rm w}(t_{\rm Plat.})$
    which is the radius of the edge of the SN density plateau when it is reached by the wall; i.e.
     the start of blow-out when $R_{\rm w}(t_{\rm Plat.})=R_{\rm Plat.}(t_{\rm Plat.})$. For $P_{\rm NS}=2$ ms, 
     we have $t_{\rm Plat.}=t_{\rm SpD}/4\simeq 6.5$ years which gives 
     $R_{\rm Plat.}(t_{\rm Plat.})\sim 10^{16}$ cm (see Eq. (\ref{eq:RPlat})); (ii)   
    consider filaments distributed radially with filament radius $R_{\rm F}$ in the
    range $R_{\rm F, in} \le R_{\rm F}\le R_{\rm F, out}$. In general,  $10^{-3}R_{\rm w}(t_{\rm Plat.}) <  R_{\rm F, in}< 4 R_{\rm w}(t_{\rm Plat.})$ and  $R_{\rm w}(t_{\rm Plat.}) <  R_{\rm F, out} < 10^3 R_{\rm w}(t_{\rm Plat.})$; 
     (iii) set the filaments' maximum density to $n_{\rm w}(t_{\rm Plat.})$  given by Eq. (\ref{eq:nwall}); 
     (iv)  include time dependence of $\Gamma_{\rm FS}(t^\prime)$
    since the assumption
    of $\Gamma_{\rm FS}(t^\prime)\simeq
    \Gamma_{\rm QN}$ used in the previous section is no longer valid.

  Before we present detailed fits of our model to the light-curves and spectra of observed LGRBs (\S ~ \ref{subsec:fits}),
    we briefly described how the prompt emission is modified in the multiple filaments case
    when compared to the analytical results obtained in the single filament case presented in the
    previous section. We also demonstrate that a Band-like spectrum  is an outcome
     of the turbulent PWN-SN scenario.

\subsubsection{Variability}
  
  The spraying of the blown-out PWN-SN ejecta by the 
millions of QN chunks  and their tiny size (compared to the filaments' radial extent) 
together with the radial distribution of the filaments  yields highly variable
  LGRBs  in our model.  Chunks colliding with  the  very irregular structure of the turbulent PWN-SN ejecta 
   yields very different bursts (i.e. light-curve shapes) for different lines-of-sights. Key points of the picture
   we present here include:

     \begin{itemize}
      
  \item  The number of filaments the chunks
  interact with can vary from  a few to hundreds;
  
  \item For the primary  chunk (with $\theta_{\rm c}=\theta_{\rm P}$), 
  the complexity of the turbulent filaments it passes through defines the intrinsic variability and the 
   number of spikes/pulses in the resulting light-curve;
   
   \item The brightest spike correspond to when  the
   chunk first hits a high density filament, which can occur anywhere between $R_{\rm F, in}$  and
  $R_{\rm F, out}$; 
  
  \item Once the primary hits a thick filament (i.e. when the thickness parameter of filament ``F"
  is $t_{\rm F}^\prime/t_{\Gamma}^\prime >>1$; here $t_{\rm F}^\prime=\Delta R_{\rm F}/\Gamma_{\rm FS}c$), it slows
  down drastically, effectively putting an end to the prompt emission;

      \item The observed variability is a convolution between the observer's time resolution (i.e. binning which we take to be 64 ms in this work) and the
    filamentary structure of the PWN-SN ejecta.  Whenever  the radial time delay corresponding to the separation between two filaments ($\Delta R_{\rm F, sep.}^{\rm obs.}$ in the observer's frame) is less
than 64 ms, the resulting spikes will not be resolved. In general, the condition 
$\Delta t_{\rm F, sep.}^{\rm obs.}=\Delta R_{\rm F, sep.}/D(\Gamma_{\rm FS},\theta_{\rm P})\Gamma_{\rm FS} c  > 64$ ms translates to 
 a minimum observable filament separation in the NS frame of 

\begin{equation}
\label{eq:filament-separation}
\Delta R_{\rm F, sep.} > 3.8\times 10^{16}\ {\rm cm}\times \frac{\Gamma_{\rm c, 3.5}^2}{f(\theta_{\rm P})}\ .
\end{equation}

I.e. to a first order, the observed distinct spikes in GRBs implies a minimum
separation between filaments given by Eq. (\ref{eq:filament-separation}).

  \end{itemize}

 \subsubsection{The duration}
 \label{sec:SN2-duration}

The observed duration of emission is due to the radial extent of filaments so that 
$\Delta t_{\rm GRB}\sim  (R_{\rm F, out}-R_{\rm F, in})/D(\Gamma_{\rm FS},\theta_{\rm P})\Gamma_{\rm FS}c$ 
where  $R_{\rm F, in}$ and $R_{\rm F, out}$ are the radii of the innermost and outermost filaments.
For the  case of $R_{\rm F, out}>>R_{\rm F, in}$  we can write

\begin{equation}
 \Delta t_{\rm GRB}\sim  \left( \frac{1}{6}\ {\rm s}\times f(\theta_{\rm P})\right)\times  
  \frac{R_{\rm F, out, 18}}{\Gamma_{\rm FS, 3.5}^{2}}\ .
   \end{equation}
   For $0 < \theta_{\rm P} < 2/N_{\rm c}^{1/2}$  and for the
   range of $R_{\rm w}(t_{\rm Plat.})$ given in Eq. (\ref{eq:rwrange}) we arrive at 
   
   \begin{equation}
   \label{eq:duration-range}
      \frac{1}{600}\ {\rm s} < \Delta t_{\rm GRB} < 1.1\times 10^3\ {\rm s}\ .
    \end{equation}

  Longer durations than given by Eq. (\ref{eq:duration-range}) can be obtained when we take into
  account the  slowing down of the chunks from one filament to another (see \S ~ \ref{subsec:fits}) .

\subsection{The Band function}
\label{subsec:spectrum-multiple-filaments}

A primary chunk hitting a single wall yields  synchrotron emission
in the fast cooling regime in our model; see Eq. (\ref{eq:tratio}). The corresponding spectrum, given by Eq. (\ref{equation:Lfast})
in Appendix \ref{appendix:LC-algorithms}, has a  photon peak energy  at
$E_{\gamma, p}$. To explain how a Band function results in our model we consider the 
    scenarios of a single {\it primary} chunk: (i) hitting a single non-turbulent {\it thick} filament (i.e.
    a repeat of the single wall model); (ii) going through many {\it thin}
     filaments each at different density $n_{\rm F}$ in a turbulent PWN-SN ejecta.

   The spectrum from the primary chunk hitting a single thick wall is shown in the 
  top panel of Figure \ref{fig:spectrum-band} (thick red line) which agrees very well with the observed standard Band function (thick black line). 
  Also shown in this panel are spectra sampled within the thick filament starting from the moment
  the chunk enters the wall until it exits the wall. This 
 demonstrates that the individual spectra add up to the Band one as a result of 
 different Lorentz factors as the chunk slows down.
 
 The bottom panel in Figure \ref{fig:spectrum-band}  shows the spectrum
  resulting from  the same chunk going through many (here 120) thin filaments.   
  In this example, the chunk's FS Lorentz factor $\Gamma_{\rm FS}$ varies little from filament to filament 
  although the cumulative effect results in 
  decreasing from $\Gamma_{\rm QN}=10^{3.5}$ to about $2800$ at the exit
  of the last thin filament.  A band spectrum is also recovered here.

 The Band function is always recovered in our model particularly when varying other parameters (i.e. besides $\Gamma_{\rm FS}$
 and $n_{\rm F}$) from one filament to another. The convolving
effect of these parameters  results in an averaging  of the low-energy index in the fast cooling regime yielding the
        typical  low-energy slope  in a Band-like spectrum.
              Effectively, the convolution 
    ``smears out"  and smooths out the lower limit $E_{\rm \gamma, c}$  (see Eq. (\ref{eq:Ecool})) and yield a convolved low-energy slope/index
    by averaging over the 1/3 and -1/2 slopes of the fast cooling  regime (the case in our model; see Eq. (\ref{eq:tratio})).  An approximation to the 
    convolved spectrum is given by

    \begin{equation}
   \label{eq:purespec}
F(E_{\gamma}) \propto 
\begin{cases}
     \sim   E_{\gamma}^{\frac{1/3-1/2}{2}} =  E_{\gamma}^{-\frac{1}{12}}  ,& \text{if }E_{\gamma} < E_{\rm \gamma, p}\\
   \sim  E_{\gamma}^{-p/2}= E_{\gamma}^{-1.2}  ,& \text{if }   E_{\gamma} > E_{\rm \gamma, p}\
 \end{cases}
\end{equation}
We thus have $E_{\gamma}^{-1}  F(E_{\gamma})  \sim E_{\gamma}^{-13/12}$ for the low-energy index
 and $E_{\gamma}^{-1}  F(E_{\gamma})  \sim E_{\gamma}^{-2.2}$
 (for our fiducial value of $p=2.4$) for the high-energy  index. The resulting spectrum is consistent with the Band's function
 with an  observed low-energy index of $\alpha\sim -1$ and an observed high-energy  index of $\beta\sim -2.2$.

\begin{table}[t!]
\begin{center}
\caption{LGRBs and FRBs  in our model (numerical values are for fiducial parameter values (see Table \ref{table:parameters}))}
\label{table:GRB-regimes}
\resizebox{1.0\textwidth}{!}{
\begin{tabular}{|c|c|c|c|c|c|}\hline
   \multicolumn{5}{|c|}{{\bf  GRBs: The blow-out regime (i.e. $E_{\rm SN}\le E_{\rm SpD}$)$^\mathbf{a}$}}\\\hline\hline
    Stage$^\mathbf{b}$ & Time delay &  $B_{\rm NS}$ &  Burst type & Contribution to GRB rate ($r_{\rm LGRB}$)$^\mathbf{c}$\\\hline
 Post-blow-out & \multirow{2}{*}{$t_{\rm SLSN} < t_{\rm QN} <  t_{\rm QN, RS}$} & \multirow{2}{*}{$10^{13}\ {\rm G} < B_{\rm NS} <   2.4\times 10^{13}$ G} &  \multirow{2}{*}{LGRB $+$ (bright Type Ic-BL SN)$^\mathbf{d}$} & \multirow{2}{*}{$< 5$\%}  \\
 (Highly-turbulent Wall) & &  &  &  \\\hline
 Post-blow-out &  \multirow{2}{*}{$ t_{\rm QN, RS}< t_{\rm QN} < t_{\rm SN, RS}$}  &  \multirow{2}{*}{$1.5\times 10^{12}\ {\rm G} < B_{\rm NS} <   10^{13}$ G} &    \multirow{2}{*}{LGRB $+$ ($t_{\rm QN}$ old Type Ic SN)$^\mathbf{e}$} & \multirow{2}{*}{$> 95$\%} \\
 (Highly-turbulent Wall)  & &  &  &    \\\hline\hline
     \multicolumn{5}{|c|}{{\bf FRBs: The non-blow-out regime (i.e. $E_{\rm SN}> E_{\rm SpD}$)$^\mathbf{f}$}}\\\hline\hline
   Stage$^\mathbf{g}$ & Time delay$^\mathbf{h}$ &  $B_{\rm NS}$  & Burst type & $r_{\rm FRB}/r_{\rm GRB}$  \\\hline
   Non-turbulent Wall &  \multirow{2}{*}{$t_{\rm SLSN}  < t_{\rm QN} < t_{\rm SN, RS}$}  &  \multirow{2}{*}{$1.5\times 10^{12}\ {\rm G} < B_{\rm NS} <  2.4\times  10^{13}$ G} &    \multirow{2}{*}{FRB $+$ UHECRs} & \multirow{2}{*}{See \S ~ \ref{sec:FRB-rate}} \\
   (Onset of Weibel instability)  & &  &  &     \\\hline
\end{tabular}
}
 \end{center}
 $^\mathbf{a}$ This case has $t_{\rm SpD}\ge t_{\rm Plat.}$  since $t_{\rm Plat}\sim (E_{\rm SN}/E_{\rm SpD})\times t_{\rm SpD} \sim (E_{\rm SN, 51} P_{\rm NS, -2.4}^2)\times  t_{\rm SpD} $; see Eq. (\ref{eq:tPlat}). For example, $P_{\rm NS}= 2$ ms  gives $t_{\rm QN}=t_{\rm SpD}\sim 25.9$ yrs and $t_{\rm Plat.}=(1/4)t_{\rm SpD}\sim 6.5$ yrs.\\
 $^\mathbf{b}$ The Pre-blow-out  stage of the blow-out regime (i.e. $t \le t_{\rm Plat.}$) is not considered here since $t_{\rm QN}=t_{\rm SpD}> t_{\rm Plat.}$ in our model.\\
$^\mathbf{c}$ We use a lognormal distribution of $B_{\rm NS}$ with mean  $10^{12.5}$ G and variance $\sigma_{\log{B_{\rm NS}}}=0.3$ based 
on our best fits to LGRB data (see \S ~ \ref{subsec:fits}).\\
$^\mathbf{d}$ Re-brightened by the QN chunks experiencing a reverse shock (RS; see \S ~ \ref{sec:QN-Type-Ic-BL}).\\
$^\mathbf{e}$ The parent type-Ic SN seen at time  $t_{\rm QN}=t_{\rm SpD}$.\\
$^\mathbf{f}$ The PWN eventually stalls and the wall becomes ``frozen" to the SN ejecta. I.e. $t_{\rm Plat.}$ is meaningless in the non-blow-out regime.\\
   $^\mathbf{g}$ The PWN is low-power  resulting in a non-turbulent or weakly turbulent PWN-SN shell with weak
magnetic field (i.e. $\epsilon_{\rm w}< \epsilon_{\rm w, WI}$, the critical value for the onset of the Weibel instability; see \S ~ \ref{sec:frbs}).\\
   $^\mathbf{h}$ In both blow-out and non-blow-out regimes, and for  $t_{\rm QN} \le t_{\rm SLSN}\simeq  1.8\ {\rm yrs}$, the wall (i.e. PWN-SN shell) is optically thick yielding a SLSN (see Figure
    \ref{fig:unification}).\\
\end{table}

\subsection{Light-Curve and Spectral Fitting}
\label{subsec:fits}

We have fit our model to the light-curves and spectra of 48 observed and well measured  LGRBs.  In fitting the light-curve, we recall that emission is caused by the interaction of the chunks with the filaments.  Therefore, to first order, the position/width of each filament affects the variability in time whereas the density of the filaments affects the variability in flux.  
 The light-curve (the prompt and afterglow emissions) will be dominated by the chunk moving closest to our line of sight at an angle $\theta_{\rm P}$.  
 The flare is due to the secondary chunk at an angle $\theta_{\rm S}$.

\subsubsection{Data} \label{section:data}

Table \ref{table:lcfits} lists the 48 selected LGRBs. These sources were chosen because they all have an abundance of data points and their spectral parameters are available.

  The light-curve data for these sources were obtained from the The Swift Burst Analyser \citep{evans_2010} and consists of a combination of BAT and XRT data over the energy range of 0.3-10 keV (the XRT band).  The BAT data has been extrapolated to this XRT band \citep{evans_2010}.

The spectra of many LGRBs can be described by a Band function.  We compare our model spectrum to the best fit Band parameters for the sources above, obtained from  \citet{yonetoku_2010}.

\subsubsection{Chunks and Filaments} \label{section:filaments}

Our simulations consist of identical chunks distributed isotropically on the sky.  The initial Lorentz factor and mass
of each chunk are fixed to our fiducial values of $10^{3.5}$ and $10^{22.5}$ gm, respectively (see Table \ref{table:parameters}).

Each chunk travels through a succession of `filaments'.  A filament represents a region of space with a certain density
$n_{\rm F}$, thickness $\Delta R_{\rm F}$ and magnetic field $B_{\rm F}$.  The algorithm for finding the location, thickness and density of each filament is explained in \S ~ \ref{section:simulation} below.
The magnetic field is determined using 
 Eq. (\ref{eq:Bwall}) once a filament's properties are derived.

 We only consider chunks within a small angle of the observer (see above) and therefore assume the filaments these chunks encounter are identical.  Beyond the filaments is an extended region that represents the ambient medium, with $n_\textrm{amb.}$ and $B_\textrm{amb.}$.  This last region is what governs the afterglow of the GRB
  and  is represented in our simulation as a ``wide filament" with density $n_{\rm amb.}$ and 
magnetic field  $B_{\rm amb.}$.

In order to fit the LGRB light-curve, we  determine where each filament is located.  It is possible to distribute filaments randomly to produce a ``generic" light-curve, but this method is not feasible when fitting individual LGRBs (the probability of placing the filaments at the right location is essentially 0).  We therefore assume that each observed point represents the interaction of the primary chunk ($\theta_{\rm c} = \theta_{\rm P}$) with a filament.  The point with the highest flux corresponds to the filament with a maximum density of $n_w$ (Eq. (\ref{eq:nwall})). The density of the filaments corresponding to the remaining points are scaled accordingly, which means no filament has a density greater than $n_w$.

\subsubsection{Simulation}\label{section:simulation}

The simulation generates the light-curve and spectrum, simultaneously for a given set parameters, using the following algorithm:

\begin{enumerate}

\item Determine the location, width and density of each filament using the primary chunk (see Appendix \ref{section:filament_algorithm}).

\item Determine each  filament's density using the peak luminosity (see Appendix \ref{section:denstiy_algorithm}).

\item Create the light-curve using the procedure outlined in Appendix \ref{section:lightcurve_algorithm}.

\item Create the spectrum using the procedure outlined in Appendix \ref{section:spectrum_algorithm}.

\end{enumerate}

\subsubsection{Fitting}
\label{subsec:fitting}

We fit our model to observations by repeatedly generating simulations (\ref{section:simulation}) with different parameters.  The parameters we vary 
 to fit the prompt emission are:

\begin{enumerate}
\item {\boldmath $\theta_{\rm P}$}: The smallest angle to our line of sight of any chunk in the simulation.  This ``primary" chunk will have the greatest contribution to the light-curve / spectrum.   Decreasing $\theta_{\rm P}$ has the effect of increasing the luminosity of the light-curve and spectrum and shifting the peak of the spectrum to higher energies.

\item {\boldmath $B_\textrm{NS}$}: The magnetic field of the precursor neutron star (which also sets the time delay since $t_{\rm QN}=t_{\rm SpD}$).  This parameter helps determine the density of the filaments and therefore has a strong influence on the overall luminosity.  The luminosity of the afterglow is directly effected by $B_\textrm{NS}$ because a higher value implies greater filament density, which means the chunk is moving slower when it enters the ambient medium.

\item {\boldmath $n_\textrm{pairs}$}: The number of electron/positrons created per proton from pair-production.  Increasing this parameter shifts the peak of the spectrum to lower energies, while increasing the luminosity of both the light-curve and spectrum.

\item {\boldmath $\epsilon_{\rm w}$}: The ratio of magnetic to thermal energy in the turbulent PWN-SN shell.  Decreasing $\epsilon_{\rm w}$ serves to shift the peak of the spectrum to higher energies, and steepen the low energy slope of the spectrum.

\item {\boldmath $p$}: The power-law index describing the distribution of electron energies. It is used  in the synchroton equations (see Eqs. (\ref{equation:Lslow}) and (\ref{equation:Lfast})).  This value is fixed by the observed spectrum (in order to match the slope of the high energy tail).

\item {\boldmath $n_\textrm{amb.}$}: The number density of particles in the final region of our simulation (the ambient medium).  This parameter contributes to the slope and luminosity of the afterglow.  Increasing $n_\textrm{amb.}$ has the effect of steepening the slope of the afterglow decline, and increasing its overall luminosity.

\item {\boldmath $Scale_{\rm P}$}:  A scaling factor (either by chunk mass or number of chunks, or both)  of the primary chunk's luminosity necessary to fit a few LGRBs when $\theta_{\rm P}=0$. The scaling is 
 an upward shift  of the entire light-curve.

\end{enumerate}

Due to the number of parameters (7 that are adjustable), and the time required to generate one simulation, an automatic fitting of the data is not feasible.  We therefore manually vary each parameter and rerun the simulation.  We determine the ``best-fit" simulation by eye.

\subsubsection{Results} \label{section:results}

The results of fitting our model to the 48 sources described in \S ~ \ref{section:data} are given in Figure \ref{figure:lc0}  with the ``best-fit" parameters listed in Table \ref{table:lcfits}. 

~\\


\begin{footnotesize}
\begin{longtable}{|l|l|l|l|l|l|l|l|l|l|l|l|l|} 
\caption{The ``best-fit" parameters$^\mathbf{a}$ from the QN model for 48 LGRB sources.}\\
\hline
  \#&
  Source&
  $\theta_{\rm P} [\textrm{rad}]$&
  $\log{n_\textrm{amb.}}[\textrm{cm}^{-3}]$&
  $n_\textrm{pairs}$&
  $\log{\epsilon_{\rm w}}$&
  $p$&
  $\log{B_\textrm{NS} [\textrm{G}]}$&
  $F_0^\mathbf{b}$&
  $F_1^\mathbf{b}$&
  $Scale_{\rm P}^\mathbf{c}$&
  $\theta_{\rm S} [\textrm{rad}]^\mathbf{d}$&
  $Scale_{\rm S}^\mathbf{e}$
  \label{table:lcfits}
 \\\hline
1&050126&1.00E-3&2.00&13.00&-5.70&2.50&12.55&0.44&6.16&1.00&3.00E-3&0.20\\
2&050315&0.00&0.50&7.00&-5.80&2.08&12.30&3.03E-2&14.72&1.00&5.50E-4&0.50\\
3&050318&4.20E-4&1.60&26.00&-5.80&2.20&12.50&0.2&9.16&1.00&-1.00&-1.00\\
4&050319&3.00E-4&1.20&25.00&-5.80&2.70&12.40&0.2&8.08&1.00&1.50E-3&1.00\\
5&050401&1.00E-4&1.00&26.00&-6.00&3.10&12.36&0.04&18.64&1.00&-1.00&-1.00\\
6&050505&3.20E-4&1.00&10.00&-6.10&2.50&12.25&1.28E-2&8.12&1.00&-1.00&-1.00\\
7&050814&0.00&1.50&10.00&-6.30&2.50&12.28&1.28&39.40&1.00&4.00E-4&1.00E-2\\
8&050820A&2.50E-4&0.00&8.00&-5.80&2.50&12.35&2.40E-2&58.36&1.00&-1.00&-1.00\\
9&050904&2.00E-4&2.40&4.00&-6.60&2.50&12.30&2.24&85.64&5.00&1.50E-3&1.00\\
10&050908&4.80E-4&3.00&21.00&-5.80&2.50&12.35&0.08&2.80&1.00&3.00E-3&10.00\\
11&051109A&1.00E-5&0.50&24.00&-5.80&2.50&12.30&0.16&62.80&1.00&-1.00&-1.00\\
12&060115&3.50E-4&1.50&16.00&-6.20&2.50&12.45&0.16&31.52&1.00&1.30E-3&0.10\\
13&060124&3.50E-4&0.20&8.00&-5.80&2.50&12.25&0.12&23.16&1.00&-1.00&-1.00\\
14&060210&0.00&1.40&8.00&-5.80&2.50&12.32&2.66E-2&51.76&10.00&-1.00&-1.00\\
15&060223A&5.00E-4&3.00&25.00&-5.70&2.50&12.53&2.86E-2&8.16&1.00&-1.00&-1.00\\
16&060510B&2.00E-4&1.50&16.00&-5.12&2.50&12.38&3.16&36.76&1.00&7.50E-4&0.80\\
17&060522&4.00E-4&2.20&11.00&-6.20&2.50&12.78&3.12&80.24&1.00&1.40E-3&0.40\\
18&060526&1.80E-4&1.60&25.00&-6.10&2.50&12.28&0.04&33.88&1.00&-1.00&-1.00\\
19&060604&9.00E-4&1.50&15.00&-5.80&2.50&12.38&0.20&6.40&1.00&-1.00&-1.00\\
20&060707&5.00E-4&1.00&12.00&-6.20&2.50&12.43&0.08&4.20&1.00&3.20E-3&15.00\\
21&060714&3.00E-4&1.50&44.00&-6.40&2.50&12.34&0.40&40.20&1.00&-1.00&-1.00\\
22&060814&4.80E-4&2.00&16.00&-5.30&2.50&12.68&7.08E-3&40.08&1.00&3.20E-3&10.00\\
23&060908&3.50E-4&2.00&24.00&-5.90&2.50&12.45&1.40E-2&20.28&1.00&-1.00&-1.00\\
24&060927&4.00E-4&2.00&19.00&-6.00&2.50&12.51&1.49E-2&8.28&1.00&-1.00&-1.00\\
25&061007&0.00&2.00&18.00&-6.30&3.20&12.59&0.04&132.04&1.00&1.00E-3&2.00\\
26&070508&1.80E-4&2.50&20.00&-6.00&2.50&12.60&1.23E-2&40.20&1.00&1.70E-3&10.00\\
27&070521&2.00E-4&2.20&5.00&-6.50&2.50&12.30&0.04&111.20&1.00&-1.00&-1.00\\
28&070714B&2.00E-4&3.00&5.00&-6.40&2.50&12.40&3.08E-2&72.24&1.00&1.00E-3&0.05\\
29&071003&0.00&2.50&3.00&-6.60&2.50&12.30&1.77E-2&269.88&10.00&2.00E-3&2.00E3\\
30&071010B&7.00E-4&1.00&35.00&-5.00&2.50&12.68&2.63E-2&8.88&1.00&-1.00&-1.00\\
31&080319B$^\mathbf{f,g}$&0.00&-0.40&10.00&-6.40&5.80&12.62&0.16&298.32&10.00&3.30E-3&10.00\\
32&080411&0.00&0.00&22.00&-4.72&2.50&12.50&9.20E-3&200.32&1.00&-1.00&-1.00\\
33&080603B&4.00E-4&2.50&33.00&-5.80&2.50&12.50&0.12&6.16&1.00&1.20E-3&5.00\\
34&080605&0.00&2.20&6.00&-6.40&2.50&12.26&2.10E-2&258.48&8.00&-1.00&-1.00\\
35&080607&0.00&2.50&8.00&-6.20&2.50&12.40&2.92E-3&40.20&10.00&-1.00&-1.00\\
36&080721&0.00&2.20&4.00&-6.80&2.85&12.35&1.56E-2&380.48&50.00&-1.00&-1.00\\
37&080810&0.00&1.70&8.00&-6.30&2.50&12.40&0.08&29.00&1.00&-1.00&-1.00\\
38&080916A&6.50E-4&1.80&10.00&-6.30&2.50&12.50&2.56E-2&20.12&1.00&3.00E-3&0.30\\
39&081121&0.00&1.80&8.00&-6.60&3.01&12.28&0.16&127.88&10.00&-1.00&-1.00\\
40&090102&0.00&2.50&4.00&-6.60&2.50&12.32&0.04&195.84&20.00&-1.00&-1.00\\
41&090423&0.00&2.50&6.00&-6.60&2.50&12.28&0.08&134.84&10.00&-1.00&-1.00\\
42&090424&3.00E-4&0.00&30.00&-6.20&3.80&12.45&5.08E-3&60.20&1.00&1.80E-3&10.00\\
43&090618$^\mathbf{g}$&0.00&1.50&17.00&-5.80&2.99&12.50&3.64E-2&265.28&1.00&5.20E-4&1.00\\
44&090715B&0.00&2.00&12.00&-6.40&2.50&12.35&0.04&137.28&10.00&-1.00&-1.00\\
45&090812&3.50E-4&2.00&11.00&-6.40&3.00&12.58&1.06E-2&20.32&1.00&2.30E-3&10.00\\
46&090926B&3.00E-4&1.80&7.00&-6.60&2.50&12.26&2.29E-2&260.20&1.00&-1.00&-1.00\\
47&091029&4.00E-4&2.00&15.00&-5.40&2.50&12.51&1.62E-2&20.24&1.00&2.50E-3&30.00\\
48&091208B&5.50E-4&1.60&26.00&-5.10&2.65&12.44&0.04&40.16&1.00&-1.00&-1.00\\
\hline
\multicolumn{13}{l}{$^\mathbf{a}$ The chunk's opacity was adjusted to 0.05 cm$^2$ gm$^{-1}$ for best fits. Other parameters are set to their fiducial values (see Table \ref{table:parameters}).}\\
\multicolumn{13}{l}{$^\mathbf{b}$ See Appendix \ref{section:filament_algorithm} for filament location and thickness generation:}\\
\multicolumn{13}{l}{$F_0$ = location of the first filament (i.e. $R_{\rm F, in}/R_{\rm w}(t_{\rm Plat.})$). $F_1$ = location of the last filament  (i.e. $R_{\rm F, out}/R_{\rm w}(t_{\rm Plat.})$). }\\
\multicolumn{13}{l}{Blow-out occurs at $t_{\rm Plat.}$ (see Eq. (\ref{eq:tPlat})) when the wall reaches the edge of the  density plateau; i.e.  $R_{\rm w}(t_{\rm Plat.})=R_{\rm Plat.}(t_{\rm Plat.})$.}\\
\multicolumn{13}{l}{$^\mathbf{c}$  $Scale_{\rm P}$ = scale of the primary chunk's luminosity.}\\
\multicolumn{13}{l}{$^\mathbf{d}$ $\theta_{\rm S}$ = angle of the secondary fragment (-1 means no secondary needed to fit LGRB; i.e. no flare in the light-curve).}\\
\multicolumn{13}{l}{$^\mathbf{e}$  $Scale_{\rm S}$ = scale of the secondary chunk's luminosity.}\\
\multicolumn{13}{l}{$^\mathbf{f}$ This source required an extreme value for the electron power-law index,  $p=5.8$.}\\
\multicolumn{13}{l}{$^\mathbf{g}$ LGRBs with associated broad-line Type Ic SN; see \url{https://www.dark-cosmology.dk/GRBSN/GRB-SN_Table.html}.}
\end{longtable}
\end{footnotesize}

\subsubsection{Flares}
\label{subsec:flare-II}

Many LGRBs exhibit X-ray flares which are an increase in brightness by up to a factor of 1000 times from the baseline.  In our model, X-ray flares are produced by secondary chunks.  The light-curve of a chunk with a $\theta_{\rm S} > \theta_{\rm P}$ will appear shifted to longer times and lower flux.  Because this chunk goes through more or less the same filaments as the primary chunk, to a first order the X-ray flare will appear as a ``mirror" of the primary prompt emission with
 the spectrum shifted to lower energies. The parameters we vary to fit the flares are:

\begin{enumerate}
\item {\boldmath $\theta_{\rm S}$}: The secondary's angle to our line of sight with $\theta_{\rm S}= \theta_{\rm sep.}-\theta_{\rm P}= 4/N_{\rm c}^{1/2}-\theta_{\rm P}$.  
  Decreasing $\theta_{\rm S}$ has the effect of increasing the luminosity of the flare and decreasing the time between
  the prompt emission and flaring (see \S ~ \ref{subsec:flare-I}).
  
  \item {\boldmath $Scale_{\rm S}$}: A scaling factor (either by chunk mass or number of chunks, or both) of the secondary chunk's luminosity is needed 
  in order to fit flares in some LGRBs.  The scaling consists of 
  a shift  of the luminosity of the entire flaring episode.

  \end{enumerate}

In Figure \ref{figure:genericflare} (top panel) we have created a generic LGRB to show how chunks with increasing $\theta_{\rm c}$ contribute to the flares.  The ``LGRB" consists of a single filament and four chunks at $\theta_{\rm c} = 0, 0.001, 0.002, $ and $0.003$ rads.  The effect of increasing the $\theta_{\rm c}$ of the chunks is clear; the ``spike" in emission occurs later in time and at a lower flux.  If the chunks are smoothly distributed in $\theta_{\rm c}$ one would expect a smooth slope as each chunk appears to hit the filament one after another.  However, if there is a large gap between $\theta_{\rm P}$ and the next lowest $\theta_{\rm c}=\theta_{\rm S}$ we would expect a large spike in emission (from the $\theta_{\rm P}$ chunk),  and then another spike (from the $\theta_{\rm S}$ chunk).

In Figure \ref{figure:genericflare} (bottom panel) we show how this would look with a real LGRB, by using our fit to GRB 060707.   The observed data is represented by the black, open circles.  A flare is evident around $t\obs=10^{2.5}$ s.  We show, in purple, the light-curve produced by the primary chunk with $\theta_{\rm P} = 4 \times 10^{-4}$ rad.  The green represents the light-curve produced by a secondary chunk with $\theta_{\rm S} = 2.7 \times 10^{-3}$ rad.  From this figure, it is clear that a ``mirror" light-curve is produced by the $\theta_{\rm S}$ chunks, appearing lower in flux and at a later time.
 Here an upward scaling factor $Scale_{\rm S}=40$ was used. To justify this factor and in general to fit LGRB
light-curves with extreme flares (see Table \ref{table:lcfits}),  the uniform filament density and uniform  Lorentz
 factor assumptions had to be relaxed.  For example, if the secondary chunk collides with a  denser part of the  filament than the primary does,
    then the density ratio must be included in Eq. (\ref{eq:L-Flare}) allowing for brighter flares (see discussion in \S ~ \ref{sec:discuss-LC-fits}).

\subsubsection{Discussion of our fits to LGRB light-curves}
\label{sec:discuss-LC-fits}

 Figure \ref{figure:fits-histograms} shows the distributions of parameters resulting from our fits to the 48 selected LGRBs. 
    For a fixed NS period of $P_{\rm NS}=2$ ms, the distribution of  $B_{\rm NS}$ shown in  panel A 
    corresponds to $\sigma_{\log{B_{\rm NS}}}\sim 0.2$ which is narrower than the
 $\sigma_{\log{B_{\rm NS}}}\sim 0.5$ suggested by analysis of $B_{\rm NS}$ distribution of observed pulsars (e.g. \citealt{faucher_2006}).
However, a variation in $P_{\rm NS}$ could  widen the distribution. The distribution in the magnetization parameter $\epsilon_{\rm w}$ (panel B) resulting from the fits is consistent with our model's fiducial values which corresponds to a PWN-SN magnetic field in the milli-Gauss values.

  The distribution of the 
   primary chunk's viewing angle (panel C) has a mean of $\bar{\theta}_{\rm P}\sim 2\times 10^{-4}$
   which is less than the analytical (i.e. model) value of $\bar{\theta}_{\rm P}\sim 1.3\times 10^{-3} N_{\rm c, 6}^{-1/2}$. This is 
    expected since the 48 selected sources are brighter than average.  When $\theta_{\rm P} << 1/\Gamma_{\rm FS}$,
     the light-curve and spectrum are insensitive to $\theta_{\rm P}$; i.e. we cannot distinguish between
      $\theta_{\rm P}=0$ and $\theta_{\rm P}\sim 10^{-4}$. This explains the peak
     in the lowest bin in panel C.
    Similarly, the distribution of the 
   secondary chunk's viewing angle (panel D) shows a mean of $\bar{\theta}_{\rm S}\sim  1.3\times 10^{-2}  \sim 10 \bar{\theta}_{\rm P}$
   which exceeds the analytical (i.e. model) value of $\bar{\theta}_{\rm S}\sim 3.1\times 10^{-3} N_{\rm c, 6}^{-1/2}\sim 2.4 \bar{\theta}_{\rm P}$.
   
   Panel E shows the distribution of the scaling factor $Scale_{\rm P}$ (which allows us to adjust upwards the 
     entire prompt emission to fit data) which is needed  in a few (11 of 48) LGRBs.
    Scaling is also needed to fit flares (see panel F) because of the 
    larger values of $\theta_{\rm S}$ obtained from fits (mainly constrained by the location in time of flares in the LGRB 
   light-curves). There are 8 LGRBs with $Scale_{\rm S}<1$, 10 LGRBs with $Scale_{\rm S}>1$,
   1 LGRB with $Scale_{\rm S}=1$ and 27 LGRBs with no flare ($Scale_{\rm S}=-1$; see Table \ref{table:lcfits}). 
    Because we assume that each filament has a uniform density, together with uniform chunk mass and 
  Lorentz factor, flares are in general less bright than the prompt emission. 
  On average $L_{\rm Flare}/L_{\rm GRB}\sim 10^{-3}$ (see  \S ~ \ref{subsec:flare-I}). 
 Large density variations   within and between filaments are required to explain
   extreme flares which is not seen in simulations of the turbulent
   PWN-SN shell. We find that changing $\Gamma_{\rm c}$ between the primary and secondary chunks  
   to account for large flares would add an extra delay (radial) time to
   Eqs. (\ref{eq:t-Flare}) and (\ref{eq:ts-Flare}) making the fits much more complex.
   In summary, the physical origin of $Scale_{\rm P}$ and $Scale_{\rm S}$ can be partly
   due to chunk mass and/or filament density variations.

     Panel G shows the distribution of filaments'  density 
     while panel H shows an almost flat distribution in the number of filaments which can vary from a few filaments
     to hundreds of filaments from one LGRB to another.
     The location of the innermost filament (panel I) and the outermost filament (panel J) in terms of $R_{\rm w}$
     varies widely from one LGRB to another which speaks to the highly filamentary nature of the PWN-SN in the
     blow-out stage.  The distribution of the ambient medium surrounding the SN (panel K) is slightly higher 
     compared to typical
      ISM density but not unreasonable for the ambient medium immediately after the
      SN ejecta. Finally, in panel L we show the resulting $n_{\rm pairs}$ distribution with a  peak 
     at 10 consistent with our fiducial value (see Table \ref{table:parameters}). Note that
     $n_{\rm pairs}$ for a given LGRB does not exceed the limiting value given in Eq. (\ref{eq:npairs})
     which ensures that synchrotron emission occurs in the fast cooling regime.

Panel A  in Figure \ref{fig:distributions-t90} shows the distribution of $t_{\rm QN}=t_{\rm SpD}$ (the time
 delay between the SN and QN) for our best fits to light-curves and spectra. It is close to uniform, varying from 
 10 years to $< 100$ years  with a peak at $\sim 35$ years consistent
 with our fiducial value.  Since $t_{\rm QN}\propto B_{\rm NS}^{-2}$, the shorter $t_{\rm QN}$ values correspond to higher $B_{\rm NS}$ values
 (see panel A in Figure \ref{figure:fits-histograms}).
 
 Panel B  in Figure \ref{fig:distributions-t90} shows the distributions of the thickness parameter (i.e. $t_{\rm w}/t_{\Gamma}$)
 resulting from our fits to the selected LGRBs. This shows that best fits require a range of filament thickness
 for each LGRB varying from extremely thin walls to very thick walls.   As shown earlier, including thick filaments 
 is essential for obtaining a Band function and allows us
  to simultaneously fit the prompt emission and the afterglow emission. I.e. the drop in luminosity
  during the transition from prompt to afterglow emission cannot be explained as a density effect
   alone (recall that $L_{\rm GRB}\propto n_{\rm w}$ while $L_{\rm AG}\propto n_{\rm amb.}$). A simultaneous fit
   of the prompt and afterglow emission (both scaling as $\Gamma_{\rm c}^6$) require slowdown of the primary chunk which can only occur
    with the presence of a thick filament along the primary's path.

Finally, the lower panel in Figure \ref{fig:distributions-t90}  shows the distribution of the duration ($t_{90}$) of all observed GRBs.
Also shown is the $t_{90}$ of the 48 LGRBs we selected.  Our selected LGRBs are representative of the bulk
in $t_{90}$ but not  in luminosity and photon peak energy which is on average higher than the bulk.
 The duration of a typical LGRB in our model is evidence for the large the radial extent of the blown out PWN-SN
 shell when it is hit by the QN chunks. The fits suggest $10^{-1} < R_{\rm F, out}/R_{\rm w} < 10^3$ 
 as typical for LGRBs.

   \subsubsection{Revisiting the phenomenological Yonetoku and Amati laws}
   \label{sec:yonetoku-amati-simulations}
   
     
   We start by simulating a single filament case  for $P_{\rm NS}=4$ ms ($t_{\rm Plat.}\sim t_{\rm SpD}$)  in order
   to compare it to the analytical single wall case presented
   in \S ~ \ref{sec:grbs-non-filamentary}. The filament is at a radius  $R_{\rm F}= R_{\rm w}(t_{\rm SpD})$ and has a thickness
     $\Delta R_{\rm F}= R_{\rm w}(t_{\rm SpD})/12$. Instead of
   using  Eqs. (\ref{eq:LGRB}), (\ref{eq:tGRB}), (\ref{eq:EGRB}) and (\ref{eq:Epeak}), here the peak of the spectrum (E$_{\textrm{peak}}$) is obtained from the generated spectrum (see Appendix \ref{section:spectrum_algorithm}).  Similarly,  L$_{\textrm{iso},\textrm{peak}}$ is read from the generated light-curve (based
   on a random number of filaments) and the E$_{\textrm{iso}}$ value is obtained by integrating the light-curve (see Appendix \ref{section:lightcurve_algorithm}). 
   We use a 64 ms resolution  which means that the peak luminosity is the 64-ms-peak-luminosity (see Appendix \ref{appendix:LC-algorithms} for more details).
   
   We run 500 simulations each representing a single chunk
 passing through a single  filament. 
  The main parameters (see Table \ref{table:parameters})  were kept constant
 to their fiducial values for each simulation while we randomize:

\begin{itemize}
	\item $\theta_{\rm P} = $acos(UniformDistribution[cos($10^{-3}$), 1])
	\item $B_{\textrm{NS}} =$ LogNormalDistribution(12.5 log(10), .2 log(10))
	\item z: Randomly choose a LGRB  from a list of over 300  (retrieved from \url{https://swift.gsfc.nasa.gov/archive/grb_table/}) and use its z.
\end{itemize}

The resulting points are plotted against the observations (LGRB data is from \citealt{ghirlanda_2009}) in Figure \ref{fig:amati-single-simulations} .
 The simulation results are  close to the analytical models shown in Figure \ref{fig:amati-single-theory2}. 
 The binning into 64 ms time bins introduces scatter in
 Figure \ref{fig:amati-single-simulations} which is not present in 
Figure \ref{fig:amati-single-theory2}. 

 As shown in Figure \ref{fig:amati-single-simulations}, the Amati law (panels A, C and E) is preserved up to a few times $t_{\rm w}/t_{\Gamma}$.  
In particular by adjusting the number of pairs to $n_{\textrm{pairs}} = 15$ we obtain
overall better fits to the Amati data  although this results in the model's
points to fall slightly below the data in panels B, D and F (i.e. the Yonetoku law). 

The Amati law is not satisfied for the very thick wall case and a ``hook", already seen   
 in panel C  in Figure \ref{fig:amati-single-theory2}, appears in panel C (for $E_{\rm peak}>10^3$ keV)
in Figure  \ref{fig:amati-single-simulations}. The ``hook" is smeared out when  $\theta_{\rm P}$ is varied (see panels A and E).
 If we take $\sigma_{\log{B_{\rm NS} } }\sim 0.5$
instead of $\sigma_{\log{B_{\rm NS} } }\sim 0.2$, 
the vertical and horizontal scatter is much  larger in all panels and
 erases the ``hook".

   When $B_{\rm NS}$ is low, the main LGRB prompt emission and the afterglow become similar in brightness so that the
afterglow contributes to $E_{\rm iso}$. In addition, low $B_{\rm NS}$ corresponds to low $E_{\rm peak}$.
 The corresponding LGRBs are the scattered red dots at lowest $E_{\rm peak}$
 values in panel C in Figure \ref{fig:amati-single-simulations}.

  In Figure  \ref{fig:amati-multiple-simulations}, we redo the analysis  considering multiple
  filaments. Again,  we run 500 simulations  but now for each simulation there are multiple filaments with 
  radius ranging between $R_{\rm F, in}$ and $R_{\rm F, out}$. The parameters in Table \ref{table:parameters}  are kept constant
 to their fiducial values for each simulation except the number of pairs was adjusted to $n_{\textrm{pairs}} = 12$ which gave best agreement with data
 in the left panels representing the Amati law. 
The randomized variables are:

\begin{itemize}
	\item $\theta_{\rm P} = $acos(UniformDistribution[cos($10^{-3}$), 1])
	\item B$_{\textrm{NS}} =$ LogNormalDistribution(12.5 log(10), .2 log(10))
	\item $\Delta R_{\textrm{F}} = $ UniformDistribution[0, $\Delta R_{\textrm{w}}=R_{\rm w}/12$] 
	\item $n_{\textrm{F}} = $ UniformDistribution[0, $n_{\textrm{wall}}$] 
	\item z: Randomly choose a LGRB  from a list of over 300   (retrieved from \url{https://swift.gsfc.nasa.gov/archive/grb_table/}) and use its z.
\end{itemize}

We take $R_{\rm F, in}=0.2R_{\rm w}$ and $R_{\rm F, out}=2.0R_{\rm w}$ with $R_{\rm w}=R_{\rm w}(t_{\rm SpD})$.
Filament generation consists of stacking slabs of density $n_{\rm F}$,
including $n_{\rm F}=0$ (i.e. no filament), until $2.0 R_\textrm{w}$ is reached.
The random generation of the filament's thickness $\Delta R_{\textrm{F}}$ results in a random number of filaments between $0.2$-$2.0 R_\textrm{w}$ 
for each simulation run.

In  the multiple filament scenario, the phenomenological Yonetoku relationship is preserved since 
inherently each filament obeys it regardless of the  thickness.  We arrive at similar conclusions for the 
 Amati law for small and intermediate filament thickness (i.e. $1< t_{\rm w}/t_{\rm \Gamma, w}  <10$). 
 As expected for  much thicker filaments, the Amati relationship is lost.
As in the single
filament case,  for small values of $\theta_{\rm P}$, the ``hook" in $E_{\rm iso}$ in panel C (for $E_{\rm peak}>10^3$ keV)
  in Figure  \ref{fig:amati-multiple-simulations} 
shifts to higher values of $E_{\rm iso}$.   This effect can be seen in panels A
  and E where $\theta_{\rm P}$ is varied.

Figure \ref{fig:amati-everything-simulations} shows the multiple filaments simulations again
but this time including a variation in $n_{\rm pairs}, \epsilon_{\rm w}$ and $p$ in the range representative of those in
Table \ref{table:lcfits}; i.e. $5 \le n_{\rm pairs} \le 35$, $-6 \le \log{\epsilon_{\rm w}}\le -4.5$ and $2 < p \le 3$. 
We see that the trend seen in previous simulations with a limited
parameter ranges starts to vanish. 
In fact  using $\sigma_{\log{B_{\rm NS}}}=0.5$  instead of $\sigma_{\log{B_{\rm NS}}}=0.2$ obtained from best fits to the selected LGRBs,
 gives a significantly larger scatter in the 500 simulations than 
 shown in Figure  \ref{fig:amati-multiple-simulations}  and Figure \ref{fig:amati-everything-simulations}.  
Using only bright LGRBs  with high photon
peak energy (thus eliminating high-$z$ and faint LGRBs), the phenomenological Yonetoku and Amati laws
re-appear in our model. Observations would  select bright and  high $E_{\rm peak}$ LGRBs likely
throwing out high-$z$ and faint ones thus reproducing the phenomenological laws.

 \subsection{Discussion and predictions}
\label{sec:grb-discussion}

\subsubsection{The SN/GRB connection}
\label{sec:QN-Type-Ic-BL}

As discussed in \S ~ \ref{sec:intro-SN-association},  all  SNe associated with LGRBs are classified as 
broad-line Type Ic with photospheric velocities exceeding $\sim 10,000$-$20,000$ km s$^{-1}$ 
 (e.g. \citealt{modjaz_2016}). These 
   are reminiscent of hypernovae with kinetic energy  of order $10^{52}$ ergs (\citealt{iwamoto_1998}).
  It is also important to note that not all Type Ic-BL SNe are accompanied by a LGRB
which may be attributed to viewing effect (a review of GRB-SNe can be found in \citet{hjorth_2012}).
 These seem to explode with less energy, showing a lower luminosity
  and mass ejected when compared to those associated with LGRBs (e.g. \citealt{nomoto_2006}).

 In our model, a Type Ic-BL SN can occur simultaneously with a LGRB in the blow-out
stage  if the filamentary PWN-SN shell is on average
dense enough (i.e. there are enough filaments with $n_{\rm F}> n_{\rm w, RS}\sim 2.2\times 10^7$ cm$^{-3}$; see Eq. (\ref{eq:chunk-RS}))
 for the RS shock into the chunk to take place. 
The RS will convert  the kinetic energy of the chunks hitting dense filaments to internal
energy in a chunk's crossing time $A_{\rm c, T}^{1/2}/c$ (e.g. \citealt{sari_1995}) which is a fraction of  a second.
The result is an optically thick chunk fireball (with a very large $(e^+,e^-)$ density) expanding at $\sim c/\sqrt{3}$ inside
 the optically thin wall. 
This late time (i.e. at time $t_{\rm QN}$ after the SN), instantaneous, energy injection as thermal energy 
 into the PWN-SN shell should yield a 
  luminous SN  (\citealt{leahy_2008,ouyed_2012}) with properties reminiscent of a Type Ic-BL SN.

   We offer the following scenario for SN-LGRBs which will be explored elsewhere:

    \begin{enumerate}
   
 \item  A ``normal" 
   SN Ic has formed from the collapse of a massive star stripped 
   of its hydrogen and helium (e.g. \citealt{filippenko_1997,heger_2003});
   
   \item It is followed by the SN interaction with a pulsar wind 
   with $E_{\rm SpD}>E_{\rm SN}$ creating a turbulent PWN-SN shell;

 \item  A QN follows the SN after time $t_{\rm Plat.} < t_{\rm QN}=t_{\rm SpD} < t_{\rm QN, RS}$ when the PWN-SN
  shell is already blown-out by the PWN.   Setting $t_{\rm QN}=t_{\rm SpD}\le t_{\rm QN, RS}$ with  $t_{\rm QN, RS}$ given by Eq. (\ref{eq:tQN-RS})
  means $B_{\rm NS}\ge 10^{13}$ G when $P=2$ ms. In other words, LGRBs associated
  with Type Ic-BL SNe occur when $10^{13}\ {\rm G} \le B_{\rm NS}< 2.4\times 10^{13}$ G (see Table \ref{table:GRB-regimes})
  with the upper $B_{\rm NS}$ value corresponding to $t_{\rm QN}=t_{\rm SLSN}$ above
  which the SN ejecta is optically thick (see Eq. (\ref{eq:Brange}));

  \item The QN chunks interacting with the densest 
   filaments (i.e. those with $n_{\rm F} > n_{\rm w, RS}$) shed their kinetic energy via the RS. 
  Even if we assume that only $\zeta_{\rm QN}=10$\% of the chunks' kinetic energy is converted by the
  RS into accelerating the entire SN ejecta, this gives an ejecta's  
 velocity of 
     $v_{\rm SN}\sim 1.4\times 10^4\ {\rm km\ s}^{\-1}\times \left(\zeta_{\rm QN, -1} E_{\rm QN, 53}M_{\rm SN, 34}^{-1}\right)^{1/2}$;
   
    \item Seeing a LGRB along the observer's line-of-sight (i.e. a primary chunk  colliding with 
    a filament with $n_{\rm F} <  n_{\rm w, RS}$) means the chunk's Lorentz factor would have decreased
     before reaching the subsequent filaments. For example, a decrease of the LGRB Lorentz factor
     from $10^{3.5}$ to $10^3$ would increase the critical density for the RS trigger 
    by a factor of 10 (i.e. $n_{\rm w, RS}=2.2\times 10^8$ cm$^{-3}$; see Eq. (\ref{eq:chunk-RS})). This means that the LGRB-generating chunk
     will less likely be subject to the RS but  will instead yield an afterglow;
    
      \item On the other hand, a Type Ic-BL SN with no LGRB association would result if the filaments along the observer's line-of-sight have  
    $n_{\rm F}>  n_{\rm F,  RS}$;
   
\item For a range $10^{13}\ {\rm G} \le B_{\rm NS}< 2.4\times 10^{13}$ G and using the lognormal distribution
in $B_{\rm NS}$ peaking at $10^{12.5}$ G with variance $\sigma_{\log{B_{\rm NS}}}=0.3$ gives
only a few percent of all LGRBs predicted to be associated with Type Ic-BL SNe
 (see Table \ref{table:GRB-regimes} and \S ~ \ref{sec:LGRB-rate}).

 \end{enumerate}

\subsubsection{``SN-less" LGRBs}
\label{sec:SN-less-LGRBs}
 
 Some LGRBs, in particular those found in metal-rich environments with little star formation (e.g. \citealt{tanga_2018}),
  show no underlying Type Ic-BL SNe.
 In our model, every LGRB is associated with a ``faded" SN; i.e. the original type-Ic SN
 which is to be differentiated from the Type Ic-BL SN occurring at $t_{\rm QN}$ after the SN (see Table \ref{table:GRB-regimes}). However,
 the time delay of years to decades between the SN and the QN means that the underlying Type Ic SN
  is  too faint to detect. We predict that  eventually extremely faint  type-Ic SNe will be associated
  with nearby LGRBs.
  In our model  the formation of an LGRB should be independent of metallicity. As long
 as a SN leaves behind a NS powerful enough to blow-out the 
 SN ejecta and massive enough to undergo a QN event, a LGRB should result.
 
 \subsubsection{``Smooth" LGRBs}
\label{sec:smooth-LGRBs}

For $E_{\rm SN}\sim E_{\rm SpD}$, or equivalently when $P_{\rm NS}\sim P_{\rm NS, cr.}$ (see Eqs. (\ref{eq:tPlat}) and (\ref{eq:PNScr})),
 then $t_{\rm QN}=t_{\rm SpD}\sim t_{\rm Plat.}$, as in the $P=4$ ms and
  $E_{\rm SN}\sim 10^{51}$ erg case presented in \S ~ \ref{sec:grbs-non-filamentary}. This means 
  that the QN occurs at the interface between the pre-blow-out
 and blow-out stages in the blow-out regime (see Table \ref{table:GRB-regimes}). In this case, $R_{\rm F, out}\sim R_{\rm w}(t_{\rm Plat.})$
  and the PWN-SN is  less turbulent than in the fully blow-out stage. These LGRBs should yield relatively smoother 
  light-curves according to our model with a wide variation in duration with overlap with SGRB duration (i.e. $< 0.1$ s)
  as clearly shown in panel A in Figure \ref{fig:amati-single-theory1} .   These short duration LGRBs
   are more likely be associated with a Type Ic-BL SN since the PWN-SN ejecta is relatively dense as discussed in \S ~ \ref{sec:QN-Type-Ic-BL}. 
    These should  be easily
   distinguishable from SN-less SGRBs which are associated with mergers (see \S ~ \ref{sec:SGRBs}).
   We speculate that  SGRB 051221A and SGRB 070724A may be two candidates of short duration LGRBs as discussed here
   since their  duration and hardness hint at a massive star origin (see \citealt{bromberg_2013}).

\subsubsection{LGRB rate in our model}
\label{sec:LGRB-rate}

The range in NS magnetic field applicable to LGRBs in our model is given in Eq. (\ref{eq:Brange}).
Furthermore,  our best fits to light-curves is consistent with  a lognormal distribution with
 standard deviation of $\sigma_{\log{B_{\rm NS}}}\sim 0.2$. 
Making use of the normal distribution of birth periods of NSs with mean
of $\mu_{\rm P_{\rm NS}}=300$ ms and standard deviation of $\sigma_{\rm P_{\rm NS}}=150$ ms
 (e.g. \citealt{faucher_2006}), 
we can estimate the rate of LGRBs as

\begin{align}
\label{eq:GRB-rate}
r_{\rm LGRB} &= \frac{ \int_{12.2}^{13.4} e^{-\frac{(\log{B_{\rm NS}}-\mu_{\log{B_{\rm NS}}})^2}{2\sigma_{\log{B_{\rm NS}}}^2} } d\log B_{\rm NS}}{\int_{11}^{15} e^{-\frac{(\log{B_{\rm NS}}-\mu_{\log{B_{\rm NS}}})^2}{2\sigma_{\log{B_{\rm NS}}}^2} }d\log B_{\rm NS}} \times\\\nonumber
&\times \frac{ \int_{1.5}^{4} e^{-\frac{(P_{\rm NS}-\mu_{\rm P_{\rm NS}})^2}{2\sigma_{\rm P_{\rm NS}}^2}} dP_{\rm NS}}{\int_{1.5}^{\infty} e^{-\frac{(P_{\rm NS}-\mu_{\rm P_{\rm NS}})^2}{2\sigma_{\rm P_{\rm NS}}^2}} dP_{\rm NS}}
\times   r_{\rm CCSNE}\times r_{\rm M_{\rm NS, c.}}\ ,
\end{align}
where $r_{\rm CCSNE}\sim 1/100$ per year per galaxy is the core-collapse SN rate  (e.g. \citealt{cappellaro_2015}) and $r_{\rm M_{\rm NS, cr.}}$ the percentage
of CCSNE giving birth to NSs with mass exceeding $M_{\rm NS, c.}$. 
 The lower value of 1.5 ms in the period distribution takes into consideration the 
 constraints of r-mode instability on rapidly rotating accreting NSs (\citealt{andersson_1999,andersson_2000}).

We get $r_{\rm LGRB}\sim 0.5\times 10^{-2}\times 10^{-2}\times 10^{-2} \sim 5\times 10^{-7}$ yr$^{-1}$ galaxy$^{-1}$
(or close to one LGRB per million years per Galaxy) if roughly 1 in 100 CCSNE yield NSs massive enough to explode as QNe.
 Of these, less than $\sim 5$\% yield SN-LGRBs\footnote{The probability of a NS to be born with a magnetic field in the range $B_1 \le B_{\rm NS} \le B_2$ is
\begin{align}
\label{eq:erfBNS}
&\frac{1}{\sqrt{2\pi}\sigma_{\rm \log{B_{\rm NS}}}} \int_{\log{B_1}}^{\log{B_2}} \exp{\left( - \frac{(\log{B_{\rm NS}}-\mu_{\rm \log{B_{\rm NS}}})^2}{2 \sigma_{\rm \log{B_{\rm NS}}}^2} \right)} d\log{B_{\rm NS}} = \\\nonumber
& \frac{1}{2}\left(    \erf{ \left( \frac{\log{B_2}-\mu_{\log{\rm B_{\rm NS}}}}{\sqrt{2}\sigma_{\log{\rm B_{\rm NS}}}}\right)}  -  \erf{ \left(\frac{\log{B_1}-\mu_{\log{\rm B_{\rm NS}}}}{\sqrt{2}\sigma_{\log{\rm B_{\rm NS}}}}\right)}  \right)\ .
\end{align}} (i.e. LGRB associated with Type Ic-BL SNe; see \S ~ \ref{sec:QN-Type-Ic-BL} and Table \ref{table:GRB-regimes} for the
corresponding  $B_{\rm NS}$ range).

 \subsubsection{The Blackbody component}
 \label{sec:BBs}

   In the early stages of the chunk's evolution, before colliding with the wall, 
   the primary chunk is thermalized up to the transparency radius given by Eq. (\ref{eq:RT}).  The time evolution of the 
   properties of this BB precursor is given in Eq. (\ref{eq:BBprime}). The 
     maximum observed BB photon peak energy, when $\theta_{\rm P}=0$, is $E_{\rm BB, max.} = 2 \Gamma_{\rm QN} \times (3 k_{\rm B}T_{\rm c, T}^\prime)$. Or,

  \begin{equation}
   \label{eq:BBobs}
E_{\rm BB, max.}  \simeq
       4.3\ {\rm keV} \times \Gamma_{\rm QN, 3.5} m_{\rm c, 22.5}^{-0.194} \kappa_{\rm c, -1}^{-3.3/6.7} \ .
       \end{equation}

For $t^\prime < t_{\rm T}^\prime$,  the chunks would hit the wall before they become optically thin; 
particularly in the turbulent PWN-SN scenario where inner filaments form well within the PWN-SN wall.
 This suggest that the early light-curves of LGRBs should show spikes with 
  a hybrid spectrum which would consist of  a BB component (from the chunk proper) and Synchrotron emission
  (from the FS).

   \subsubsection{Predictions}
   \label{sec:GRB-predictions}
   
   \begin{itemize}
   
\item {\bf Super LGRB}: Because it is due to a single  chunk (the primary), the observed prompt LGRB luminosity in our model 
can be extreme even for an isotropic engine. Eq. (\ref{eq:LGRB-range}) gives
a maximum value of

  \begin{equation}
  \label{eq:LGRB-max}
 L_{\rm GRB, max.} \sim  3.7\times 10^{58}\ {\rm erg\ s}^{-1}\ .
\end{equation}

 Eq. (\ref{eq:LGRB-max}) also implies that the  observed isotropic energy of the chunk  
 far exceeds the QN total isotropic energy of $10^{53}$ ergs.  
 However we should keep in mind that:
 
  (i) The combination of parameters yielding very short delays   (i.e. $P_{\rm NS}< 4$ ms and $B_{\rm NS}> 10^{13}$ G) 
    and the requirement of a massive NS mass is rare. Assuming a log-normal distribution in $B_{\rm NS}$ with a mean of $10^{12.5}$ G  and
 standard deviation $\sim 0.2$, and defining a super LGRB as $L_{\rm GRB} > 10^{56}$ erg s$^{-1}$ (i.e. $B_{\rm NS} > 10^{13}$ G),
we estimate a fraction of super LGRB to be  $\sim 10^{-4}$ of LGRBs according to our model.;

(ii)  For  $t_{\rm Syn.}^\prime< t_{\Gamma}^\prime$, in the 
  efficient and fast cooling Synchrotron regime  (the case in our model), we see all of the kinetic energy of
  a given chunk harnessed during sweeping. 
   If $t_{\rm Syn.}^\prime\ge t_{\Gamma}^\prime$, some of the sweeping energy is stored in the chunk instead of being radiated 
  and we see only a fraction ($L_{\rm c, p}\times (t_{\Gamma}^\prime/t_{\rm Syn.}^\prime)$) of the sweeping luminosity;
  thus reducing the upper limit given above;

 \item {\bf Super Flare}:  In principle a Super flare with $L_{\rm Flare} > L_{\rm GRB}$ (recall that the flare is from the secondary and the prompt is from the primary)  is possible if the secondary, viewed at an  angle which is close
to that of the primary, collides with a filament (or a region
of the same filament) which is much denser than the one the primary crosses (see discussion in \S ~ \ref{subsec:flare-II}).
 A second possibility is to interpret these as a consequence of the chunk's mass distribution, where
 the secondary chunk is much more massive than the primary. In both cases the primary emission
would be interpreted as a precursor and the secondary emission as the prompt. About 10\% of LGRBs have precursors
 (e.g. \citealt{lazzati_2005,burlon_2008,troja_2010}) 
which we interpret as emission from the primary chunk while the prompt is from the secondary chunk (i.e.
a mis-identified super Flare);

\item {\bf The parent core-collapse SN}: We assumed that the SN is a type-Ic  (based
on observations). In principle,
our model should work regardless of the type of core-collapse SN as long as a NS forms with properties prone to a QN.
  At this point, it is not clear why nature would favor some type-Ics as QN progenitors.

\item {\bf The pre-GRB SN}: Our fits to the 48 selected LGRBs yields a mean time delay $t_{\rm QN}$
of about 35 years with about 10\% with $3\ {\rm yrs} < t_{\rm QN} < 20$ yrs (see panel A  in  Figure \ref{fig:distributions-t90}).
 Assuming these numbers apply to the thousands of known GRBs (which remains to be confirmed
 by fits), then about 1 in $\sim 10^4$ SNe observed with future large surveys should 
 have a LGRB in the following few decades. Conversely, archival data could reveal 
 past SNe  at the location of known LGRBs, a few years or few decades prior to the LGRBs.

\end{itemize}

%
%
\section{Fast Radio Bursts in our model (a proof-of-principle study)}
\label{sec:frbs}

 In this section we suggest  a mechanism for FRB
emission based on coherent synchrotron emission (CSE) and postulate
 that some FRBs (including repeating ones) can occur  in the wake of core-collapse SNe.
  When a QN occurs in a non-turbulent, weakly magnetized PWN-SN we appeal to CSE in the
chunk's FS to generate the FRB  (a mechanism
for repetition is presented in Appendix \ref{appendix:RFRBs}).  As shown 
in \S~ \ref{sec:FRB-rate}, the FRBs we discuss here  have rates 
closer to that of LGRBs than to the much higher rate of observed FRBs. 
I.e. the emission mechanism for FRBs we suggest here cannot be the only one and
other mechanisms are needed to compliment it (which may be related to FRBs from QNe in other environments).

\subsection{The Weibel instability and the coherent synchrotron emission (CSE)}

 Let us define $\sigma_{\rm w}= B_{\rm w}^2/(4\pi n_{\rm w}m_{\rm H}c^2)$
 as the magnetization of the upstream region (i.e. the wall in this case). The Weibel instability (hereafter WI; \citealt{weibel_1959,fried_1959,yoon_1987,medvedev_1999}; see also \citealt{achterberg_2007,lemoine_2010} and references therein) may develop
 on timescales faster than the shock crossing time  if $\sigma_{\rm w}\le \zeta_{\rm WI}/\Gamma_{\rm FS}^2$ where $\zeta_{\rm WI}\sim 10^{-2}$
 is the fraction of incoming energy transferred into electromagnetic fluctuations (e.g.  \citealt{kato_2007,spitkovsky_2008,nishikawa_2009}). Making use of Eq. (\ref{eq:Bwall}),  the WI would occur  when

  \begin{equation}
 \label{eq:Weibel-condition-SN}
 \epsilon_{\rm w} <  \epsilon_{\rm w, WI}=3.6\times 10^{-6}\times \frac{\zeta_{\rm WI, -2}}{\Gamma_{\rm FS, 3.5}^2V_{\rm w, 8.7}^2} \ .
 \end{equation}
 The upper limit is effectively set by the chunk's Lorentz factor (and thus controlled by the
 QN ejecta) since $V_{\rm w}$ varies very little in our model (see Eq. (\ref{eq:Vwall})).

In the blow-out regime, FRBs are only possible before turbulence saturation when $\epsilon_{\rm w}$ reaches
its maximum value. This is expected to occur early in the evolution of the PWN-SN ejecta,
on timescales $<< t_{\rm Plat.}$  (e.g. \citealt{blondin_2017} and references therein),  much  before
the QN occurs.
 
   We assume that the condition  given in Eq. (\ref{eq:Weibel-condition-SN}) is satisfied in the non-blow-out regime
   (i.e. when $E_{\rm SN}> E_{\rm SpD}$)  
   with a non-turbulent or  weakly turbulent PWN-SN shell where turbulence saturation is unlikely to happen. 
   In this regime the 
 PWN cannot overpower the SN ejecta (see Table \ref{table:GRB-regimes}). It stalls and becomes frozen to the SN expansion, never reaching the edge of the plateau.  Thus in this
 regime $t_{\rm Plat.}$ is meaningless and  the QN  can occur any time in the range $t_{\rm SLSN} < t_{\rm QN}= t_{\rm SpD} < t_{\rm SN, RS}$.
     For $P_{\rm NS}=2$ ms and $B_{\rm NS}=10^{12.5}$ G (i.e.  $t_{\rm QN}\sim 25.9$ years), corresponding to
 $n_{\rm w}\sim 10^5$ cm$^{-3}$ and $R_{\rm w}\sim 5\times 10^{17}$ cm, we find FRBs properties that are similar 
 to observed values.

\subsection{Bunching length}

Once the WI sets in it induces coherent structures that allows for electron bunching to occur. In particular, 
in the magnetized chunk frame,  if the wavelength of the
 synchrotron radiation, $\lambda_{\rm Sync.}^\prime$,  exceeds the length of the bunch $l_{\rm b}^\prime$
 then the bunch can  radiate coherently (see Appendix \ref{appendix:CSE});  the primed quantities refer to the shock frame. 

The  magnetic field in the in the forward-shocked wall material saturates when $\omega_{\rm B}^\prime/\omega_{\rm p}^\prime= \gamma_{\rm e}^{1/2}$
where $\omega_{\rm B}^\prime$ and $\omega_{\rm p}^\prime$ are the electron angular cyclotron frequency  and
the plasma frequency, respectively (e.g. \citealt{medvedev_1999}; see also recent Particle-in-cell (PIC) simulations by \citealt{kato_2007,spitkovsky_2008,nishikawa_2009}). In other words, if we associate electron bunching length in the shock
with the correlation length of the magnetic field (i.e. effectively a coherence length $l_{\rm b}^\prime\sim c\gamma_{\rm e}^{1/2}/\omega_{\rm p}^\prime$). The bunching length
is then

\begin{equation}
l_{\rm b}^\prime \sim 1.4\times 10^3\ {\rm cm} \times n_{\rm pairs, 1}^{-1} n_{\rm w, 5}^{-1/2}\ ,
\end{equation}
with  $c/\omega_{\rm p}^\prime=5.31\times 10^5\ {\rm cm}/{n_{\rm e}^\prime}^{1/2}$, $n_{\rm e}^\prime = (7\Gamma_{\rm FS} 2 n_{\rm pairs})\times n_{\rm w}$ and $n_{\rm w}$ given by
Eq. (\ref{eq:nwall}); the shocked gas' adiabatic index is taken as  4/3. We set $\gamma_{\rm e}= (\Gamma_{\rm FS}\times m_{\rm p}/m_{\rm e})/2n_{\rm pairs}$ and recall that the no-pairs case is recovered mathematically by setting $2n_{\rm pairs}=1$.

\subsection{CSE frequency}
\label{sec:FRB-frequency}

The characteristic CSE frequency  in the observer's frame  ($\nu_{\rm CSE}=D(\Gamma_{\rm FS}, \theta_{\rm c})\nu_{\rm CSE}^\prime$,
 with $\nu_{\rm CSE}^\prime \simeq c/l_{\rm b}^\prime$ as given in Appendix \ref{appendix:CSE}), is
      
  \begin{align}
  \label{eq:CSE-frequency}
  \nu_{\rm CSE}&\sim 60\ {\rm GHz}\times \frac{\Gamma_{\rm FS}}{l_{\rm b}^\prime  f(\theta_{\rm c})} \\\nonumber
     &\sim \frac{140\ {\rm GHz}}{ f(\theta_{\rm P})} \times \Gamma_{\rm FS, 3.5} n_{\rm pairs, 1} n_{\rm w, 5}^{1/2}\ ,
  \end{align}
   with an average value $\bar{ \nu}_{\rm CSE}^{\rm obs.} \sim 7.8\ {\rm GHz}$ 
   for $\bar{\theta}_{\rm P}=(4/3)/N_{\rm c}^{1/2}$ (i.e. $f(\bar{\theta}_{\rm P})=17.9$). The above is always larger then the 
   plasma frequency ($\nu_{\rm p, med.}\simeq 9\times 10^{3}\ {\rm Hz}\times n_{\rm med.}^{1/2}$; e.g. \citealt{lang_1999}) in the unshocked medium
   ahead of the CSE radiation (i.e. the wall) which is always in the MHz range.

\subsection{Luminosity}
\label{sec:FRB-luminosity}
 
 The power per bunch, $L_{\rm b}^\prime$, in the shock frame is given by Eq. (\ref{eq:CSE-L}) in Appendix \ref{appendix:CSE}.  
 To derive it  we first need to estimate the relevant factors:
 
 (i) We first estimate the ratio between the bunching length and the electron's Larmor radius  to be
 \begin{equation}
 \frac{l_{\rm b}^\prime}{r_{\rm L, e}^\prime}\sim 0.14 \zeta_{WI, -2}^{1/2}\ .
 \end{equation}
 To calculate the Larmor radius $r_{\rm L, e}^\prime$ we use the WI saturated wall's magnetic field  ${B_{\rm WI}^\prime}^2/8\pi =\epsilon_{w, WI}\times \Gamma_{\rm FS} n_{\rm w}^\prime m_{\rm p}c^2$.
  This is another key difference between the LGRB case where the wall's magnetic field
 is simply shock amplified versus the CSE case where the magnetic field is  larger since it reaches equipartition values  (e.g. \citealt{medvedev_1999}).
   The ratio above
 is independent of $n_{\rm pairs}$ since $\gamma_{\rm e}\propto n_{\rm pairs}^{-1}$ and
 $n_{\rm e}^\prime = 2n_{\rm pairs} n_{\rm w}^\prime \propto n_{\rm pairs}$;
 
 (ii)  The ratio $N_{\rm b}/l_{\rm b}^\prime$  can
 be calculated by noting that  $N_{\rm b}= A_{\rm b}^\prime l_{\rm b}^\prime n_{\rm e}^\prime$
 where $ A_{\rm b}^\prime= \pi {l_{\rm b}^\prime}^2$ is the bunch's cross-section.  
 Using Eq. (\ref{eq:CSE-L}) together with the $l_{\rm b}^\prime/r_{\rm L, e}^\prime$ ratio 
we get $L_{\rm b}^\prime \simeq 9.7\times 10^{11}\ {\rm erg\ s}^{-1}\times\zeta_{WI, -2}^{1/3}\times \Gamma_{\rm FS, 3.5}^2 n_{\rm pairs, 1}^{-2}$.
With  Doppler boosting the observed luminosity per bunch, $L_{\rm b}=D(\Gamma_{\rm FS},\theta_{\rm c}^4)L_{\rm b}^\prime$, is

\begin{equation}
 L_{\rm b} \simeq \frac{1.6\times 10^{27}\ {\rm erg\ s}^{-1}}{f(\theta_{\rm P})^4}\times\zeta_{WI, -2}^{1/3}\times \Gamma_{\rm FS, 3.5}^6 n_{\rm pairs, 1}^{-2}\ .
\end{equation}
 The observed luminosity per bunch is independent of the wall's density and thus of the time
delay between the QN and SN;

(iii) We estimate the number of bunches per chunk  as $N_{\rm b, T} \sim A_{\rm c, T}/ A_{\rm b}^\prime $ 
(for the chunk's cross-sectional area $A_{\rm c, T}$, 
see Eq. (\ref{eq:AcT}))
or

\begin{align}
N_{\rm b, T} &\simeq 4.3\times 10^{15}\times   \left( m_{\rm c, 22.5}\kappa_{\rm c, -1}\right) n_{\rm pairs, 1}^2 n_{\rm w, 6}\ ;
\end{align}

(iv) Finally we arrive at the CSE luminosity $L_{\rm FRB}=N_{\rm b, T}\times  L_{\rm b}$ of 

\begin{align}
\label{eq:LFRB}
 L_{\rm CSE} &\simeq  \frac{6.6\times 10^{41}\ {\rm erg\ s}^{-1}}{f(\theta_{\rm c})^4}\times \zeta_{WI, -2}^{1/3}\times\\\nonumber
&\times \left(m_{\rm c, 22.5}\kappa_{\rm c, -1} \Gamma_{\rm FS, 3.5}^{6}\right)\times n_{\rm w, 5}  \ ,
\end{align}
which is independent of $n_{\rm pairs}$.   Ignoring pair production in
 Eq. (\ref{eq:CSE-frequency}) (by setting $2 n_{\rm pairs}=1$) gives $\nu_{\rm CSE}$ 
 of the order of a few GHz which agrees better with observed FRB frequencies.

\subsection{Duration and total isotropic energy}
\label{sec:FRB-duration-energy}

The CSE duration is  the time it takes the chunk to cross the unperturbed (i.e. turbulently ``quiet") wall ($\Delta R_{\rm w}/D(\Gamma_{\rm FS},\theta_{\rm P})\Gamma_{\rm FS}$
with $\Delta R_{\rm w}=R_{\rm w}/12$. This is equivalent to 
 Eq. (\ref{eq:tGRB}) which we reproduce here for CSE emission to get

\begin{equation}
 \Delta t_{\rm CSE} \sim \frac{1}{60}\ {\rm s}\times f(\theta)\times  
  \frac{R_{\rm w, 17}}{\Gamma_{\rm FS, 3.5}^{2}}\ .
   \end{equation}
   The above is an upper limit since the wall's thickness may  be $< R_{\rm w}/12$.

The implied isotropic (effectively an upper limit) CSE energy 
$E_{\rm CSE}=   L_{\rm CSE}\times \Delta t_{\rm CSE}$  is

\begin{align}
\label{eq:E-CSE-SN}
E_{\rm CSE} &\simeq   \frac{1.1\times 10^{40}\ {\rm erg\ s}^{-1}}{f(\theta_{\rm c})^3}\times \zeta_{WI, -2}^{1/3}\times\\\nonumber
&\times \left(m_{\rm c, 22.5}\kappa_{\rm c, -1} \Gamma_{\rm FS, 3.5}^{4}\right)\times \left(n_{\rm w, 5}R_{\rm w, 17}\right)  \ .
\end{align}

\subsection{Source dispersion and rotation measures}

 The Dispersion Measure (DM) and Rotation Measure (RM) associated with the
 SN ejecta ahead of the CSE photons  we calculate using $DM = \int n_{\rm e}(l) dl/(1+z)$  and $RM =0.81 \int n_{\rm e}(l) \cdot B_{\parallel}(l) dl/(1+z)^2$
where $n_{\rm e}(l)$ is the electron density (in cm$^{-3}$),  $l$ is the distance (parsecs) and,  $B_{\parallel}(l)$ is the line-of-sight 
magnetic field strength (in $\mu$G).  Effectively we have $(1+z)DM_{\rm source} \simeq n_{\rm Plat.} (R_{\rm Plat.}-R_{\rm w})\simeq n_{\rm Plat.} R_{\rm Plat.}$.  Using Eqs. (\ref{eq:RPlat}) and (\ref{eq:nPlat}) we get

\begin{equation}
(1+z) DM_{\rm source}  \sim 810\ {\rm pc\ cm}^{-3}\times  n_{\rm w, 5} R_{\rm Plat., 17}\ ,
\end{equation}
where we used $n_{\rm Plat.}=n_{\rm w}/4$.

Given the low magnetic field strength  we expect $RM_{\rm source}\sim 0$.

\subsection{Rate}
\label{sec:FRB-rate}

Given the narrow period range in NSs exploding as QNe in our model, 
the division between the blow-out regime (i.e. GRBs when  $E_{\rm SN}<E_{\rm SpD}$) and the non-blow-out regime (i.e.
 CCSN-FRBs with $E_{\rm SN}>E_{\rm SpD}$) depends mostly on the distribution of $E_{\rm SN}$.
 For $P_{\rm NS}=2$ ms, for example, the non-blow-out regime occurs when $E_{\rm SN}>E_{\rm SpD}\sim 4.8\times 10^{51}$ erg$\ \times P_{\rm NS, -2.7}^{-2}$ (see Eq. (\ref{eq:tPlat})). In this case,   
  for the FRB rate, $r_{\rm FRB}$, to exceed the LGRB rate, $r_{\rm LGRB}$ (given in \S ~ \ref{sec:LGRB-rate}), we require a distribution
in $E_{\rm SN}$ with a peak at or above $\sim 4.8\times 10^{51}$ erg.  One may argue that:
(i) the division between the blow-out
and non-blow-out regimes is not precisely defined; (ii) the range in $E_{\rm SN}$ is not well known  from observations;
 (iii) the value of $E_{\rm SpD}$ depends on an uncertain moment of inertia of the NS, $I_{\rm NS}$.  Together these effects
  leaves room for a scenario where  
 $E_{\rm SN}>10^{51}$ erg may be the dominant regime. Nevertheless, the estimated
 rate for CCSNe-FRBs in our model is close to that of LGRBs.

  The properties of CSE are consistent with FRBs. However, the corresponding rate estimated above
 is too small to accommodate the rate of the general FRB population which is  $\sim 10^3$-$10^4$
   detectable over the whole sky every day (e.g. \citealt{champion_2016,ravi_2019};
  see also Table 3 in  \citealt{petroff_2019} and Table 1 in \citet{cordes_2019}). This suggests an  FRB rate
  which is about $10^3$ times greater than the GRB rate or a sizeable fraction of  the overall CCSN  rate.
  We will explore alternative models for FRBs, involving QNe occurring in different environments, elsewhere.

\subsection{Discussion and predictions}
\label{sec:frbs-predictions}

\begin{itemize}

\item {\bf The Type Ic-BL SN/FRB connection}: For $t_{\rm QN}< t_{\rm QN, RS}$, the PWN-SN shell
is dense enough for the RS into the chunk to take effect (see Eq. (\ref{eq:tQN-RS})).  Thus the
association of  some FRBs with Type Ic-BL SNe is a possibility in our model. However,
 if  Fermi acceleration of particles
to UHECRs during the FRB (see \S ~ \ref{sec:UHECRs}) is  efficient and acts faster
 than the chunk's crossing time then the chance for a Type Ic-BL SN is reduced;

\item {\bf Orphan afterglows and CCSN-FRBs}: 
FRBs may be accompanied by afterglows  in our model if after the FRB (and the 
UHECR) phase the chunks still have enough kinetic energy to yield an afterglow (via the chunk's FS Synchrotron emission)
during the interaction with the ambient medium. 
    CSE ceases when the chunks exit 
  the FRB site (i.e. the SN ejecta) since the density drops by a few orders of magnitude
  effectively shutting-off the WI; i.e. $\sigma_{\rm amb.} >> \sigma_{\rm w, WI}$ (see \S ~ \ref{sec:frbs}). 
  We speculate that Orphan Afterglows seen by GRB detectors and assumed to be associated with GRBs may instead be associated with FRBs.

Since GRB detectors' solid angle (e.g. Swift/BAT has a 2 sr field of view; \citealt{barber_2006}) exceed those
of FRB detectors (e.g. the Parkes 64-m telescope at 1.4 GHz has a primary beam of $\sim 10$ arcminutes.), then 
about $\sim 2\pi/4\pi\sim 1/6$ of FRBs should be associated with orphan afterglows
(i.e. no GRB association). However, if there is efficient acceleration of wall's particles to UHEs during the FRB (see \S ~ \ref{sec:UHECRs}), 
 then the chunk's Lorentz factor after exiting the wall may be reduced as to yield effectively no orphan afterglows;

\item  {\bf  The phenomenological Yonetoku and Amati laws}: It is  interesting that the expression for the FRB luminosity in our model is,
  except for the factor $\zeta_{WI}^{1/3}$, the exact same expression as the LGRB luminosity given
  in Eq. (\ref{eq:LGRB}). 
  Furthermore, the dependency of the FRB peak frequency on $n_{\rm w}^{1/2}$ while
   $L_{\rm FRB}\propto n_{\rm w}$  means that FRBs may obey the   phenomenological
   Yonetoku and Amati laws (see discussion in \S ~ \ref{sec:yonetoku-amati-theory} and \S ~ \ref{sec:yonetoku-amati-simulations});

  \item {\bf Super FRBs}:  Similarly to super-LGRBs described in \S ~ \ref{sec:GRB-predictions}, 
  super FRBs with luminosity exceeding $10^{45}$ erg s$^{-1}$ when  $B_{\rm NS}> 10^{13}$ G
   are possible in  our model. The rate of Super FRBs is
  $\sim 10^{-4}$ of that of FRBs related to core-collapse SNe;

\item {\bf FRB Flares and super Flares}:  Similarly to flares in LGRBs described in \S ~ \ref{subsec:flare-I} and \S ~ \ref{subsec:flare-II},
 flares can occur in FRBs.   It is possible that  the double-peaked FRBs (e.g. FRB 121002 ; \citealt{champion_2016}) are a manifestation of
  FRB flares as described here.  FRB flares  can be brighter than 
  that from the primary at $\theta_{\rm P}$.   Since FRBs occur in non-filamentary PWN-SN a super FRB flare can only be caused
by a chunk's mass distribution. In analogy with the rate of super Flares in
LGRBs (see \S~ \ref{sec:GRB-predictions}), it is reasonable to assume that the rate of super Flares is about 10\% of FRB flares.
 These means that roughly 10\% of our FRBs should show a ``precursor";

\item  {\bf Repeating mechanism  (see  Appendix \ref{appendix:RFRBs})}:
 We speculate that a  plasma shell surrounding the SN, and thus the QN, site
 can act as a refractor  bending off-line-of-sight FRBs towards the observer.  
 Clustered events occur when multiple beams are bent towards the observer  by  inhomogeneities
 in the shell.

 \end{itemize}

 \section{Other astrophysical implications of our model} 
 \label{sec:other-implications}
 
 Here we discuss some general ideas that
 may have implications to high-energy astrophysics. We suggest
  other  predictions to add to those listed previously in the
 LGRB and FRB parts of this paper.

\subsection{Post-LGRB/FRB QN chunks}

A typical QN chunk exits the LGRB/FRB site (i.e. the SN ejecta) with a Lorentz factor $\Gamma_{\rm c}\sim 10^3$
 but is subject to deceleration in the ambient medium (we consider this here to be
  ISM with typical density of 1 cm$^{-3}$ and magnetic field $B_{\rm ISM}=10^{-5}$ G).
   The post-LGRB/FRB fragments would slow down, for example, from $\Gamma_{\rm c}$   to $\Gamma_{\rm c}/10\sim 10^2$ for 
 $t^\prime\sim100 t_{\Gamma}^\prime\sim  254\ {\rm yrs}/(n_{\rm ISM}\Gamma_{\rm c, 3}^2\kappa_{\rm c, -1})$  
and reach $\Gamma_{\rm c}/100\sim10$ after $t^\prime \sim 10^4 t_{\Gamma}^\prime\sim   2.5\times 10^{4}\ {\rm yrs}/n_{\rm amb.}\Gamma_{\rm c, 3}^2 \kappa_{\rm c, -1}$.
In the NS frame they would have travelled on average a maximum  distance of 
  $\sim 78\ {\rm pc}$ and $7.8\ {\rm kpc}$, respectively. In addition, the Synchrotron cooling timescale  
  $t_{\rm Syn.}^\prime= 4.1\times 10^7\ {\rm s}\times n_{\rm pairs, 1}/(\Gamma_{\rm c, 3}^3 B_{\rm ISM, -5}^2)$ is
  of the same order as the dynamical timescale $t_{\Gamma}^\prime= 9.9\times 10^6\ {\rm s}/(n_{\rm ISM} \Gamma_{\rm out, 3}^2\kappa_{\rm c, -1})$.
  Thus applying the radiative cooling solution, 
  the fragment would radiate at a rate of 

\begin{align}
L_{\rm c,  ISM}(t) &\simeq \frac{1.7\times 10^{42}\ {\rm erg\ s}^{-1}}{f(\theta_{\rm c})^4}\times \\\nonumber
 &\times m_{\rm c,22.5}  \kappa_{\rm c, -1} n_{\rm ISM, 0}\times  \Gamma_{\rm c, 2}(t)^6 \ .
\end{align}

As the  chunk slows down, the solid angle of the beam ($1/\Gamma_{\rm c}^2$) increases thus
increasing the probability of detection but at the expense of a decreasing luminosity.
Thus unless the fragment is travelling directly towards the observer, these
``wandering chunks" may not be easily detectable. 
  However, 
 if ever detected, to the observer a ``wandering chunk" would appear as a continuous source of Synchrotron emission with a peak at

\begin{equation}
\nu_{\rm c,  ISM} \sim \frac{1.4\times 10^{14}\ {\rm Hz}}{f(\theta_{\rm c})}\times \Gamma_{\rm c, 2}^3 B_{\rm ISM, -5} n_{\rm pairs, 1}^{-2}\ ,
\end{equation}
which is in the Infrared band.

It is possible that a very long exposure to ISM 
could erode and deform  these chunks to smaller objects
acquiring unusual shapes. We have
in mind the asteroid  1I/2017 U1 ($^{`}$Oumuamua; \citealt{meech_2017}) 
with its extremely elongated shape. In a Hubble time of order $10^{10}$ 
QN chunks would have formed in a galaxy like ours. An important fraction of them become wanderers, many of these in intergalactic space, once they leave the FRB/LGRB sites.

 \subsection{Ultra-High Energy Cosmic Rays}
 \label{sec:UHECRs}

 Ultra-High Energy Cosmic Rays (UHECRs; \citealt{auger_1935,linsley_1963}) have puzzled 
 physicists and astrophysicists since their discovery. Despite decades of  
 observations and modelling (see e.g. \citealt{kotera_2011,abbasi_2012} and \citealt{aloisio_2018}  for recent reviews)
  the underlying source remains uncertain.

 In our model, the onset of the WI may create conditions for Fermi acceleration (\citealt{fermi_1949,peacock_1981,vietri_1995})  of the ions
 in the wall and boosting them by a Lorentz factor of $\sim 2\Gamma_{\rm FS}^2$ (e.g. \citealt{gallant_1999,achterberg_2001};
see also \citealt{bykov_2012} and references therein). Hadronic losses are negligible in our model since the hadronic-hadronic mean-free path
 is $\lambda_{\rm HH}= 1/n_{\rm w}\sigma_{\rm HH}\sim 10^{22}\ {\rm cm}/n_{\rm w, 5} >> R_{\rm w}$;
 here  $\sigma_{\rm HH}$ of the order of milli-barns is the hadronic-hadronic cross-section (e.g. \citealt{letaw_1983}).
 During an FRB, with the Fermi mechanism  in action, accelerated particles   can reach  energies of

 \begin{equation}
 E_{\rm UHECR}\sim 3.2\times 10^{17}\ {\rm eV}\times  \Gamma_{\rm FS, 3.5}^2 A_{\rm UHECR, 16} \ ,
 \end{equation}
 where $A_{\rm UHECR}\sim 16$ is the atomic weight of Oxygen which is representative of
  SN-Ic ejecta. For $A=56$  $E_{\rm UHECRs}\sim 1.1\times 10^{18}\ {\rm eV}$
 and exceed this value if we take into account r-process elements in the SN ejecta. 
 
 The connection between FRBs and UHECRs proposed here warrants more detailed studies that 
 we leave for the future. For now we note that:
 
 \begin{itemize}
 
 \item   The measured composition of UHECRs may be representative of that of a 
  Type Ic SN ejecta including  the heavier  r-process  elements. 
  It will be interesting to search for these two compositions in  Auger data  (e.g. \citealt{aab_2014});

\item   A rate of  one QN per million years per galaxy means an available power of $\sim 5\times 10^{46}\ {\rm erg}$ yr$^{-1}$ per galaxy which 
 amounts to  $\sim 10^{45}\ {\rm erg}$ yr$^{-1}$ Mpc$^{-3}$ using the estimate of  galaxy number density of $0.01$ Mpc$^{-3}$ (e.g. \citealt{conselice_2005}).  
Assuming 50\% of QNe occur in the non-blow-out regime,
this is more than enough  power
  to account for  UHECRs beyond the knee  (\citealt{waxman_1999,berezinsky_2006,berezinsky_2008,murase_2009});

\item UHECRs  would not be associated with  LGRB according to our model since 
  conditions in the PWN-SN shell are not favorable for the WI to set in (i.e. $\epsilon_{\rm w} > \epsilon_{\rm w, WI}$ 
  as given in Eq. (\ref{eq:Weibel-condition-SN})). UHECRs 
  would instead be  associated with FRBs.  Nevertheless, 
   the deflection of UHECRs by the Inter-Galactic and Galactic 
magnetic field (e.g. \citealt{batista_2017} and reference therein) may wash out the
    direct spatial correlation between FRBs and UHECRs suggested in our model.

\end{itemize}

 \subsection{Magnetars in our model}
    
For $t_{\rm QN}<  t_{\rm SLSN}=1.8$ years, the SN is optically thick 
 and the QN chunks' kinetic
energy is deposited as thermal energy yielding a SLSN (see \S ~ \ref{sec:QN-Type-Ic-BL}; see also \citealt{leahy_2008,ouyed_2009a}). 
 For $P_{\rm NS}\sim 4$ ms  with $t_{\rm QN}=t_{\rm SpD}$ for example, the SN ejecta
 is optically thick when  (using Eq. (\ref{eq:appendix-tauSN}))
 
  \begin{equation}
   B_{\rm NS} > 2.4\times 10^{13}\ {\rm G}\times P_{\rm NS, -2.4}{E_{\rm SN, 51}}^{1/4} M_{\rm SN, 34}^{-1/2}\ .
   \end{equation}
  This has the intriguing consequence in our model that NSs with magnetar magnetic field strength (\citealt{duncan_1992,thompson_1993}) cannot yield FRBs/LGRBs.
  Instead, these yield SLSNe. 
 More specifically:
 
 \begin{itemize}

    \item Magnetars are  not engines, but instead a side effect (i.e. the highly magnetized
     QS which is the QN compact remnant) of the LGRB/FRB proper.  The QN compact remnant
     is born with $10^{14}$-$10^{15}$ G magnetic fields since  
     such strong fields are readily achievable during the hadronic-to-quark-matter
     phase transition  (\citealt{iwazaki_2005,dvornikov_2016a,dvornikov_2016b});
     
\item The NS period (inherited by the newly born QS at $t_{\rm QN}$) is $P_{\rm QS}\sim P_{\rm NS}\times (1+t_{\rm QN}/t_{\rm SpD})^{2/3}$ 
  which for a NS birth period of 4 ms, as an example,  gives a QS birth period of $P_{\rm QS}\sim 6.4$ ms. The corresponding
 spin-down power is $L_{\rm SpD, QS}\sim 6.2\times 10^{43}\ {\rm erg\ s}^{-1}\times P_{\rm QS, -2.2}^{-4}B_{\rm QS, 14}^2$
  and a characteristic spin-down timescale $L_{\rm SpD, QS}\sim 2.6\times 10^4\ {\rm s}\times P_{\rm QS, -2.2}^{2}B_{\rm QS, 14}^{-2}$. 
     The GRB-QS connection can in principle be tested by searching for this post-GRB spin-down power signatures and searching for 
   the  corresponding QS wind nebulae (QWNe). However, such a signal may be too weak to detect and furthermore 
   QSs according to the QN model do not pulse in radio since they are born as aligned rotators (\citealt{ouyed_2004,ouyed_2006});

\item Our best fits to GRB light-curves and spectra
suggest time delays of years to decades between the SN and the QN (whose compact remnant, the QS, is
a ``magnetar").   This may be one explanation for  the discrepancy found
when comparing spin-down age and SN's remnant age in magnetars reported in the literature;  these studies    
 assume that the magnetar is formed concurrently with the SN (see discussion in \citealt{dass_2012} on this topic and references therein).
 As reported in \citet{leahy_2007,leahy_2009}, 
 a time delay of tens of years between the SN and the magnetar formation,
 in agreement with our findings here for LGRBs/FRBs, removes the age 
 discrepancy.

\end{itemize}

 \subsection{SGRBs in our model: A QN in a binary NS merger}
\label{sec:SGRBs}


With an isotropic equivalent energy in  the $10^{49}$-$10^{51}$ erg range,
SGRBs are  less energetic than their long duration counterpart which can exceed
$\sim 10^{54}$ erg. SGRBs have less luminous afterglows than LGRBs.
 
 SGRBs are not associated with star forming regions and are not accompanied with 
 core-collapse SNe. Their spatial distribution is different from that of LGRBs, preferring
 instead outskirts of galaxies with some SGRBs occurring in 
 elliptical galaxies  (see \citealt{nakar_2007,berger_2014,davanzo_2015} for a review). 
  This points to a binary-merger origin for SGRBs (\citealt{blinnikov_1984,paczynski_1986,eichler_1989}).
  The gravitational wave event GW170817 (\citealt{abbott_2017a,abbott_2017b}) gave support to the 
 binary-merger origin of SGRBs but the engine behind this SGRB is still being analyzed and studied.

Many groups have simulated NS binary mergers, obtaining different outcomes for the remnant compact object
(e.g. \citealt{ruffert_1996,ruffert_1999,shibata_2000,rosswog_2003,shibata_2005,rezzolla_2010,bauswein_2010,sekiguchi_2011,hotokezaka_2011,baiotti_2013,hotokezaka_2013}). Some studies find that 
a long-lived  (stable to gravitational collapse) rapidly
rotating NS is one possible outcome  of NS mergers (e.g. \citealt{zhang_2011b,giacomazzo_2013,gao_2016,ciolfi_2017,piro_2017,ai_2018,fujibayashi_2018,piro_2018});
 such a fate depends on the poorly known Equation-of-State of neutron matter (\citet{hebeler_2013}).
 
 SGRBs as we investigate here, may be associated with NS mergers
if the  long-lived rapidly rotating  NS is born with mass above $M_{\rm NS, cr.}$ so it can undergo a QN event
after $t_{\rm SpD}$.  The QN relativistic ejecta 
  would collide with the merger's  sub-relativistic ejecta. The latter  
  in this case plays the role of the SN ejecta (hereafter labeled with ``SN") which interacts with the
  ambient medium and with the PWN. We take the merger ejecta to have 
   a typical mass $M_{\rm ``SN"} \sim 10^{-3}M_{\odot}\simeq 10^{30.3}$ gm and to expand 
  at a typical speed of $v_{\rm ``SN"}\sim 0.3c$; i.e. $E_{\rm ``SN"}\sim  10^{50}$ erg.


    If we take the merger's ejecta to be in free expansion, we can adopt the solutions given in \S ~ \ref{sec:QN-SN}. 
  For a NS born with  a magnetic field  $B_{\rm NS, 14}$, in units of $10^{14}$ G,  and 
   a period $P_{\rm NS, -2.7}$, in units of 2 ms, the characteristic timescales (see \S ~ \ref{sec:QN-SN} and
   \S ~ \ref{sec:grbs-non-filamentary}) applied to the merger case are 
\begin{align}
t_{\rm SpD}&\sim 9.5\ {\rm days}\times P_{\rm NS, -2.7}^2 B_{\rm NS, 14}^{-2}\\\nonumber
t_{\rm Plat.} &\sim (0.02 E_{\rm SN, 50}  P_{\rm NS, -2.7}^2)\times t_{\rm SpD}\\\nonumber
t_{\rm QN, RS}&\sim 3.6\ {\rm days}\times \left(E_{\rm ``SN", 50}^{-1/2} M_{\rm ``SN", 30.3}^{5/6}\right)\times\\\nonumber
 &\times  \left( m_{\rm c, 22.5}^{1/6} \kappa_{\rm c, -1}^{1/2}\Gamma_{\rm c, 3.5}^{2/3}\right)\\\nonumber
t_{\rm SLSN}&\sim 0.4\ {\rm days}\times E_{\rm ``SN", 50}^{-1/2}M_{\rm ``SN", 30.3}\ . 
\end{align}
The above implies that the QN occurs ($t_{\rm QN}=t_{\rm SpD}$) days following the merger 
while the  ejecta is blown out much earlier by the PWN at  $t_{\rm Plat.}\simeq 0.02 t_{\rm SpD} \sim 4.6$ hours after
 the merger. The ``SN" reverse shock can be ignored because the ratio between the pressure in the PWN
  and behind the RS (e.g.  Eq. (9) in \citealt{blondin_2001};
  see also \citealt{vanderswaluw_2001}) is, in the merger case, $P_{\rm PWN}/P_{\rm RS} >> 1$.
 The RS into the chunks, as they plow through the merger's
 ejecta,  is triggered if the QN occurs on timescales less
  than $t_{\rm QN, RS}$ (see \S ~ \ref{sec:QN-Type-Ic-BL}).

 The light ejecta mass  and the large QN energy implies
  that the merger ejecta will expand at a speed ($\sim c/\sqrt{3}$) when it is 
  blown-out. The size and  baryon density  are then found using $R_{\rm F, out}\sim (c/\sqrt{3})\times t_{\rm SpD}$ and $n_{\rm ``SN"}\simeq (M_{\rm ``SN"}/m_{\rm H})/ (4\pi R_{\rm F, out}^3/3)$ which gives
 \begin{align}
 R_{\rm F, out}&\sim 1.4\times 10^{16}\ {\rm cm}\times P_{\rm NS, -2.7}^2 B_{\rm NS, 14}^{-2}\\\nonumber
 n_{\rm ``SN"}&\sim 3.4\times 10^4\ {\rm cm}^{-3}\times M_{\rm ``SN", 30.3}  P_{\rm NS, -2.7}^{-6} B_{\rm NS, 14}^{6}\ .
 \end{align}
 
 Compared to the moment of blow-out (i.e. when $R_{\rm w}(t_{\rm Plat.})=R_{\rm Plat.}(t_{\rm Plat.})$), the size of 
 the ejecta when the QN occurs is 
 
 \begin{equation}
 \frac{R_{\rm F, out}}{R_{\rm Plat.}(t_{\rm Plat.})}\sim 14\times E_{\rm ``SN", 50}^{-1/2}M_{\rm ``SN", 30.3}^{1/2}\ .
 \end{equation}
  with $R_{\rm Plat.}(t)\propto t$  given in Eq. (\ref{eq:RPlat}).

 The resulting luminosity (Eq. (\ref{eq:LGRB})), duration ($ \Delta t_{\rm SGRB}=R_{\rm F, out}/D(\Gamma_{\rm QN},\theta_{\rm P})\Gamma_{\rm QN}c)$, isotropic energy ($E_{\rm SGRB} =L_{\rm SGRB}  \Delta t_{\rm SGRB}$) and photon peak energy (Eq. (\ref{eq:Epeak}))  are

 \begin{align}
  L_{\rm SGRB} & \simeq \left( \frac{1.7\times 10^{55}\ {\rm erg\ s}^{-1}}{ f(\theta_{\rm p})^4}\right) \times \\\nonumber  
     &\times (m_{\rm c, 22.5}\kappa_{\rm c, -1} \Gamma_{\rm QN, 3.5}^6)\times \left( n_{\rm ``SN", 4}\right)\\\nonumber  
    \Delta t_{\rm SGRB}&\simeq \left( 0.02\ {\rm s}\times f(\theta_{\rm P})\right)\times (\Gamma_{\rm QN, 3.5}^{-2})\times \left( R_{\rm F, out, 16}\right)\\\nonumber
       E_{\rm SGRB}     &\simeq \left( \frac{3.4\times 10^{53}\ {\rm erg}}{ f(\theta_{\rm p})^3}\right)\times \left(m_{\rm c, 22.5}\kappa_{\rm c, -1} \Gamma_{\rm QN, 3.5}^4\right) \times \\\nonumber  
 & \times \left( n_{\rm ``SN", 4} R_{\rm F, out, 16}\right)\\\nonumber  
E_{\rm \gamma, p, SGRB}&\simeq \left(\frac{12\ {\rm MeV}}{f(\theta_{\rm P})}\right)\times \left(\Gamma_{\rm QN, 3.5}^4\right)\times \left(\frac{g(p)/g(2.4)}{n_{\rm pairs, 1}}\right)^2 \times \\\nonumber 
   &\times \left(n_{\rm ``SN", 4}^{1/2} \epsilon_{\rm w, -6}^{1/2}V_{\rm w, 9.7}\right) \ .
 \end{align}

 Figure \ref{figure:lcSGRB} is our model's fit to light-curves of 4  selected SGRBs   with the ``best-fit" parameters listed in Table \ref{table:SGRBs}.
 The fits are constrained by matching the model's $E_{\rm peak}$ to their  measured   $E_{\rm peak}$ (see Table 3 in \citealt{davanzo_2014}). Fitting the afterglow requires ambient density which is  higher than expected 
 for the ambient medium surrounding late evolution of NS binary merger. More detailed fits to SGRB
 light-curves (including flares) and  spectra will be presented elsewhere.

\begin{table}
\caption{Preliminary fit parameters$^\mathbf{a}$  from the QN model for 4 SGRB sources (see \S ~ \ref{sec:SGRBs}).}
\label{table:SGRBs}
\centering
\begin{tabular}{|c|c|c|c|c|c|c|c|c|c|}\hline
  \#&
  Source&
  $\theta_{\rm P} [\textrm{rad}]$&
  $\log{n_\textrm{amb.}}[\textrm{cm}^{-3}]$&
  $\log{B_\textrm{amb.}}[\textrm{G}]$&
  $n_\textrm{pairs}$&
  $\log{\epsilon_{\rm w}}$&
  $p$&
  $F_0^\mathbf{b}$&
  $F_1^\mathbf{c}$\\  
 \hline
1&051221A&5.00E-4&0.00&-8.00&50.00&-6.00&2.40&0.02&9.71\\
2&070714B&6.00E-4&1.50&-8.00&25.00&-6.00&2.60&0.11&6.73\\
3&100816A&6.00E-4&0.20&-8.00&60.00&-6.00&2.50&0.08&7.17\\
4&101219A&9.50E-4&3.50&-5.00&25.00&-6.00&2.40&0.05&1.79\\\hline
\end{tabular}\\
$^\mathbf{a}$ Other parameters are set to their fiducial values (see \S ~ \ref{sec:SGRBs}).\\
$^\mathbf{b}$ $F_0$ = location of the first filament (i.e. $R_{\rm F, in}/R_{\rm w}(t_{\rm Plat.})$).\\
$^\mathbf{c}$  $F_1$ = location of the last filament  (i.e. $R_{\rm F, out}/R_{\rm w}(t_{\rm Plat.})$).\\
\end{table}

  \subsubsection{Discussion and predictions}
\label{sec:SGRBs-discussion}

  \begin{itemize}

  \item {\bf Light-curve variability}: The PWN-``SN" is expected to be turbulent in the blow-out regime, producing irregularly spaced filaments and chaotic-looking LGRB light-curves;
  
    \item {\bf Flares} are an outcome of our model (because of the secondary chunks; see \S ~ \ref{subsec:flare-I} and \S ~ \ref{subsec:flare-II}) and should thus be
  seen in SGRBs from binary NS mergers;
  
  \item {\bf Long duration LGRBs vs SGRBs}:   A SGRB is thus a shortened version of a long duration ($>> 1$ s) LGRB according to our model since the
QN occurs days after the merger (i.e. in a more compact ejecta because of the high $B_{\rm NS}$) instead of years after the SN in the case of LGRBs. 
 Consistent with our model, \citet[and references therein]{ghirlanda_2011} find that  the luminosity and the spectral properties of SGRBs resemble the first few seconds of LGRBs;

 \item {\bf The two-component relationships}: The phenomenological Yonetoku and Amati laws should still apply to SGRBs
  because we still have a range in $B_{\rm NS}$ and $\theta_{\rm P}$;

\item {\bf The QN-induced SN (QN-``SN")}: When $t_{\rm QN}< t_{\rm QN, RS}$,  following the analysis in \S ~ \ref{sec:QN-Type-Ic-BL}, 
a re-brightening of the merger's ejecta may occur yielding a QN-``SN". 
 If  $\zeta_{\rm QN}=10$\% of the chunks' kinetic energy is converted by the
  RS into accelerating the entire merger's ejecta, the resulting ejecta's  
  maximum velocity is  $\sim 0.75c$. 
 However, because the ejecta is optically thin (i.e. $t_{\rm QN}> t_{\rm SLSN}$) we expect a percentage of QN-``SN" 
 energy to be radiated on timescales of $\Delta t_{\rm QN-``SN"}\simeq R_{\rm F, out}/c\sim 5.4\ {\rm days}\times P_{\rm NS, -2.7}^2 B_{\rm NS, 14}^{-2}$
 with a luminosity $L_{\rm QN-``SN"}\simeq \zeta_{\rm QN} E_{\rm QN}/\Delta t_{\rm QN-``SN"}\sim 2.1\times 10^{46}\ {\rm erg\ s}^{-1} \zeta_{\rm QN, -1} E_{\rm QN, 53}P_{\rm NS, -2.7}^{-2} B_{\rm NS, 14}^{2}$. Thus  ``SNe" associated with SGRBs
   will have  rapidly decaying light-curves and spectra with  
 extremely broad lines.  

There are two scenarios:
 
\begin{itemize}

 \item When the primary chunk has no RS,
  the outcome is a SGRB associated with a QN-``SN". 
  
   \item When the primary chunk has RS, no SGRB results.
 Here, the QN-``SN" should appear as an isolated  transient in no star-forming environments. In general, the spatial distribution
 of isolated QN-``SNe" should  follow that of NS binary mergers.

 \item Our preliminary  fits to 4 SGRB light-curves yield parameters in favor of a no
 QN-``SN" scenario.

\end{itemize}

   \item {\bf Short duration LGRBs vs SGRBs}: Both SGRBs and short duration  (i.e.  $< 1$ s)  LGRBs (see \S ~ \ref{sec:smooth-LGRBs})  can have $t_{\rm QN}<t_{\rm QN,RS}$ but SGRBs have very low mass, 
so the re-brightening will not produce a Type Ic-BL SN (see \S ~ \ref{sec:QN-Type-Ic-BL});

 \item {\bf FRBs and NS binary mergers}:  The interaction of QN chunks with a weakly magnetized merger ejecta  
 should yield an FRB in a manner similar to that described for a non-blow-out SN ejecta (see \S ~ \ref{sec:frbs}).
 Refraction (i.e. repeating FRBs) are less likely in NS binary mergers;

\item {\bf The quark star radio signal}: 
The  QN compact remnant (i.e. the QS), born at time $t_{\rm SpD}$ after the
merger, is rapidly rotating. The corresponding spin-down power should yield a radio signal 
 similar to that predicted in the NS case (e.g. \citet{nakar_2011}).
 The lack or presence of such a signal in SGRBs (\citealt{metzger_2014,horesh_2016})
 may constrain our model.  However, our model for SGRBs is preliminary and key
 parameters remain to be explored before it can be refined;

 \item {\bf The GW170817 SGRB}: We interpret that the SGRB associated with the GW170817 event is a no-QN event which may have
resulted in a BH  formation following the merger.

\end{itemize}

Our model for SGRBs is preliminary and has not been fully explored. However, a QN following a binary
 merger provides a framework to unify SGRBs and LGRBs.

\section{General discussion and model's limitations}
\label{sec:general-discussion}

\subsection{A unifying model: FRBs, XRFs, XRR-GRBs, GRBs and SLSNe}
\label{sec:unification}

 X-ray Flashes (XRFs) and X-ray rich GRBs (XRR-GRBs) are in many ways very similar to GRBs  except that the flux comes mostly 
 in the 2-30 keV band  (e.g. \citealt{heise_2001,barraud_2003}; see also  \citealt{ripa_2016}
 and references therein).  There is evidence that the properties of GRBS, XRR-GRBs and XRFs form a
continuum (e.g. Figure 2 in \citealt{sakamoto_2005})  and that they have similar duration and sky distributions supporting the suggestion that they
are the same phenomenon.

 Figure \ref{fig:unification} is a schematic unification of  FRBs, XRFs, XRR-GRBs, LGRBs
  and SLSNe  in our model.
  XRFs, XRR-GRBs and LGRBs can be unified and explained  as the same phenomenon
     (i.e. emission induced by the interaction of the QN chunks with the PWN-SN shell)
     in the blow-out stage of the blow-out regime where $\epsilon_{\rm w}>\epsilon_{\rm w, WI}$ (see Table \ref{table:GRB-regimes}).
       One evolves from a LGRB to an XRR-GRB to finally an XRF
       in a continuous transition by increasing $t_{\rm QN}$ (i.e. decreasing $B_{\rm NS}$). The higher the 
        $t_{\rm QN}$ (the lower $B_{\rm NS}$) the more  extended (and  less dense) the PWN-SN shell is when it is hit by the QN chunks.  
        The viewing angle $\theta_{\rm P}$ will create overlap in  properties of these phenomena. 
     Within the LGRBs category, there is a subset of LGRBs associated with Type Ic-BL SNe  occurring when $B_{\rm NS}> 10^{13}$ G
     (see \S ~ \ref{sec:QN-Type-Ic-BL}) 
      and the ``SN-less" LGRB for lower NS magnetic filed (see also Table \ref{table:GRB-regimes}). 
       At the opposite end of this classification,
       next to SN-LGRBs, are SLSNe which occur in a young, very dense and optically thick, PWN-SN shell.
    SLSNe also occur in the non-blow-out regime for $t_{\rm QN}< t_{\rm SLSN}$ while
    for longer time delays an FRB results.

Other noteworthy points:

 \begin{itemize}
 
 \item An increase in $t_{\rm QN}$ 
  is  associated with an increase in $\epsilon_{\rm w}$ and a decrease in $n_{\rm w}$. This means that  as one evolves from  LGRBs to XRFs,   the
  photon peak energy on average decreases from MeVs to  keVs. 
  The lower limit (i.e. cut-off) in photon peak energy allowed in our model in the blow-out regime (i.e. in the turbulent PWN-SN case
  when $\epsilon_{\rm w} \ge \epsilon_{\rm w, WI}$)
 can be obtained from  Eq. (\ref{eq:Epeak})
 by setting $\theta_{\rm P}=2/N_{\rm c}^{1/2}$ (i.e. the maximum
 viewing angle allowed for the primary chunk with a corresponding $f(\theta_{\rm P})=41$), $n_{\rm w}=12.6$ cm$^{-3}$ (i.e.
  the minimum wall density; see Eq. (\ref{eq:nwrange})) and $\epsilon_{\rm w}=\epsilon_{\rm w, WI}$
 (as given by Eq. (\ref{eq:Weibel-condition-SN})).  Eq. (\ref{eq:Epeak}) then yields:
 
  \begin{equation}
 E_{\rm \gamma, p} \sim 2\ {\rm keV}\times \zeta_{\rm WI, -2}^{1/2}\Gamma_{\rm QN, 3.5}^3 \times \left(\frac{g(p)/g(2.4)}{n_{\rm pairs, 1} }\right)^2
     \ .
 \end{equation}
 
 As depicted in  Figure \ref{fig:unification}, this cut-off
  means that in our model no bursts can occur at wavelength between the radio (i.e. the FRBs
  in the non-blow-out regimes) and X-ray bands;
  
  \item XRFs, which are connected with longer time delay  than LGRBs, are associated with
  more extended and more filamentary PWN-SN resulting in more variable light-curves. In the classification
  suggested in Figure \ref{fig:unification}, XRR-GRBs lie between XRFs and LGRBs and 
   should show intermediate properties (variability, frequencies etc...);
   
   \item Since flares are echoes of the prompt emission (induced by the secondary chunks), all of the
 points listed above should in principle apply to the flaring phases in XRFs/XRR-GRBs/LGRBs;

 \item  The classification we suggest here assumes a narrow distribution of $\epsilon_{\rm w}$ from
 filament to filament in a given PWN-SN ejecta. Relaxing this assumption would allow the co-existence of 
 $\epsilon_{\rm w}< \epsilon_{\rm w, WI}$  and $\epsilon_{\rm w}\ge \epsilon_{\rm w, WI}$ filaments in the
 same PWN-SN ejecta.  This suggests the intriguing possibility of the co-existence of FRBs concurrently with XRFs/XRR-GRBs/LGRBs 
 (specifically, the occurrence of FRB
pulses related to filaments with $\epsilon_{\rm w}<\epsilon_{\rm w, WI}$ within light-curves of XRFs/XRR-GRBs/GRBs).
This requires a wide distribution of $\epsilon_{\rm w}$ for a given PWN-SN ejecta.

 \end{itemize}

 \subsection{FRBs/XRFs/XRR-GRBs/GRBs as probes of turbulence in PWN-SN ejecta}
\label{sec:unification}

 The classification suggested above, if verified, implies that the variability in the prompt emission   (as well
as in flares when present) of XRFs/XRR-GRBs/LGRBs may be a probe of the  filamentary  structure
 of PWN-SN ejecta years to decades after the SN. They could be used to help understand turbulence 
 in PWN-SN interaction. At the other end, FRBs probe a relatively non-turbulent PWN-SN ejecta. 
 
 The spectrum of $\Delta R_{\rm F}$ and $n_{\rm F}$ derived from fits to light-curves
 are probably related to turbulence.  
 The bottom panel in Figure \ref{fig:distribution-nwdRw} shows the distribution of the relative  column
 density $\sum{n_{\rm F}\Delta R_{\rm F}}$
 (normalized to ($n_{\rm Plat.} R_{\rm w}(t_{\rm Plat.})/12$)) when adding up all of the filaments along the line-of-sight, 
 one for each of the 48 selected LGRBs.   The resulting distribution seems to agree with  the distribution of
  column densities resulting from the 3-Dimensional simulations of the PWN-SN shell (see Figure 8 in  \citealt{blondin_2017}).

 \subsection{Neutrino and Gravitational Wave signals in QNe}
 
According to our preliminary calculations (\citealt{keranen_2005,ouyeda_2018a,ouyeda_2018b}), 
a QN is associated with a neutrino burst that is distinct from
that of a SN. The total energy release is similar between QN and SN but the neutrino
energy is higher in a QN with a neutrino-sphere temperature of
$\sim 20$ MeV for a proto-QS compared to a $\sim 5$ MeV for the case of a
proto-NS. Thus during the first few milliseconds that follow a QN, the neutrino signal  is harder and brighter
(by a factor of about $(20/5)^4\sim 200$) than the SN one. This translates to a peak
detector count for QNe that is about a thousand times higher than the peak
detector count for SNe (see \citet{ouyeda_2018,ouyeda_2019}).
 Neutrino observatories, such as Super-Kamiokande-III (\citealt{ikeda_2007}) should  in principle
distinguish between the SN and  QN neutrino signals.
 Detailed numerical simulations are required to confirm the properties of the QN neutrino signal (see details in  \citet{ouyeda_2018}).

Another property of a QN is its gravitational wave (GW) signal 
which is  distinguishable from that of the preceding SN (see Appendix in \citealt{staff_2012}).
 The GW wave signal is expected to be stronger in a QN because it is a
more compact explosion than a SN.  If the hadronic-to-quark-matter conversion
front leading to the QN is asymmetric, it should emit a gravitational wave signal 
with an integrated luminosity of the order of $10^{46}$-$10^{48}$ ergs or $\sim 0.01$\% of
binding energy of a neutron star (\citealt{staff_2012}). This signal should be detectable by the Advanced LIGO (\citealt{ligo_2015}) if 
 bursts (i.e. FRBs/XRFs/XRR-GRBs/LGRBs) occur  a few kpcs away.
 These are preliminary results and a 
 more detailed analysis of the GW signal from a QN  requires advanced numerical simulations to track the
evolution of the asymmetric burning front.

  The neutrino and GW signals occurring years to tens of years following the SN should be
 common to  XRFs/XRR-GRBs/LGRBs/FRBs/SLSNe according to our model.

 \subsection{Model's limitations}
 \label{sec:limitations}
  
  While our model captures key features of LGRBs and FRBs, it has 
 some  simplifying assumptions (organized by topics below) that require scrutiny before firm conclusions can be reached. For example:

\begin{itemize}

\item When fitting the LGRB light-curves:

\begin{enumerate}

\item  We have kept most parameters fixed when fitting the light-curves and spectra,
varying mainly $t_{\rm QN}$ (which translates to a variation in $B_{\rm NS}$
for a fixed $P_{\rm NS}$) and the viewing angle $\theta_{\rm P}$.  Nevertheless, our fits to data 
suggest that it may be consistent with the QN being 
a universal explosion (i.e. $E_{\rm QN}$ and $M_{\rm QN}$ the same from source to source). 
It is not unrealistic to assume that the quark deconfinement density $\rho_{\rm NS, cr.}$ (a property
of Quantum-Chromo-Dynamics; e.g. \citealt{weber_2005}), is universal which in principle can translate to  
  a universal NS mass, $M_{\rm NS, cr.}$. This implies that  the number of neutrons to convert to
quarks is fixed during a QN and thus the energy released, $E_{\rm QN}$. On the other hand, $M_{\rm QN}$ (and thus 
the ejecta's Lorentz factor $\Gamma_{\rm QN}=E_{\rm QN}/M_{\rm QN}c^2$)
may be less straightforward since it involves complex ejection mechanisms (see discussion in  \citealt{ouyed_leahy_2009}). 
 Nevertheless, a narrow distribution in $\Gamma_{\rm QN}$ peaking at $10^{3.5}$, as suggested from our
 fits to 48 LGRB light-curves and spectra, supports the idea
of a narrow distribution in $M_{\rm QN}$ and thus a plausible universal QN explosion.   At this point,
 we can only state that our assumption of fixed $E_{\rm QN}$ and $M_{\rm QN}$ together with the resulting successful fits to many
 LGRBs based on fiducial values of our parameters may be considered a self-consistency check;

\item We assumed $\zeta_{\rm p}=1$ or that 
  all of the swept-up proton energy in the FS is transferred to, and radiated by, leptons
(electrons and pairs).  This gave extreme LGRB brightness (see Eq. (\ref{eq:LGRB-max})) which can be
relaxed by considering $\zeta_{\rm p}<1$. However, in this regime part of the
sweeping power is transferred into the chunk's internal energy requiring a treatment 
  beyond the scope of this paper;

 \item  We assumed that the secondary chunks ($\sim$ 6 per primary chunk
 in a spherical geometry) are collapsed into,
  and represented by, one effective secondary chunk at $\theta_{\rm S}$. Taking into
  account individual secondary chunks will allow for complex effects not accounted
  for in our model in its current version. For example, 
    separate secondary chunks would contribute at different
   times and different luminosities to flares. The resulting flare will therefore be a sum of all these ``mirrored" light-curves at longer times and lower fluxes
   than in the single secondary assumption.  We should note that multiple secondaries will allow for repetitive X-ray flares
   as seen in some GRBs;

 \item Assuming that the chunks have reached their maximum size prior to their
  interaction with the wall is another simplification. In the turbulent PWN-SN ejecta presented in \S ~ \ref{sec:grbs-filamentary}, the wall is torn into
 filaments with the innermost filaments at radius $R_{\rm F}<< R_{\rm w}$ which means that
 chunks start interacting with the PWN-SN before they reach their maximum size (see \S ~ \ref{subsec:optically-thin-regime}).
 
 \begin{itemize}
  \item When fitting the early light-curves, the chunks are assumed bigger than they should be and this is  compensated by artificially decreasing the density $n_{\rm F}$. In reality
   the true filament density is higher than the fit density.    This can be seen in panel G in Figure \ref{figure:fits-histograms} showing the distribution of filaments
  densities where  a low density peak detaches itself (i.e. shifted to the left)  from the main peak at $\sim 10^3$ cm$^{-3}$.
  
    This led us to introduce the fitting parameter $\alpha_{\rm F}$ (see Eq. (\ref{eq:filament-density}) in Appendix \ref{section:denstiy_algorithm}).
    We find  $1<\alpha_{\rm F}<3$ from the fits to the 48 selected LGRBs  when scaling the filament density in the pre-peak luminosity phase.

 \item  Past the peak luminosity,  scaling the filament density using 
    $\alpha_{\rm F}=1$ gives natural fits  to the light-curves which agrees very well with our model
   for the constant chunk area with $n_{\rm F}\propto L_{\rm F}$ as given in
   Eq. (\ref{eq:Lcsw}).  This suggests that once a chunk enters the densest filament (i.e.
    with $n_{\rm F}\sim n_{\rm w}$) it expands to its transparency radius and continues without expanding 
    thereafter.

  \item  Using $A_{\rm c}(t^\prime)$ (see Eq. (\ref{eq:Ac})) instead of the maximum area $A_{\rm c, T}$, we argue, could remove the need for the $\alpha_{\rm F}$ parameter. However, 
    a time-dependent chunk area requires re-integrating Eqs. (\ref{eq:Lcsw})
   and (\ref{eq:Gammac}) to derive $L_{\rm c, sw.}(t^\prime)$ and $\Gamma_{\rm c}(t^\prime)$, accordingly. 
   This  treatment  is left for another paper.

   \end{itemize}

\item  Fitting each LGRB is a lengthy process and we have not fully explored the degeneracy in parameters
 for any single LGRB. We fit 48 LGRBs and assume that their parameters distributions are somewhat representative of the whole  population of LGRBs.

\end{enumerate}

\item When fitting the spectrum:

\begin{enumerate}

\item We find it necessary to include pair creation in order to 
simultaneously fit the spectrum and the light-curve of a given LGRB in our model. 
The pair-production mechanism remains to be better understood; 

 \item Assuming that the chunks have reached their maximum size prior to their
  interaction with the wall ignores the fact that  the chunks are still thermal (i.e. emit as BBs) when they start
  colliding with the innermost filaments. This  may modify the early spectra (see \S ~ \ref{sec:BBs}) since 
  a hybrid (BB and a Band) spectrum is the more likely outcome;

\item In the LGRB case,  we assumed an electron energy distribution with a power-law index of $p\sim 2.4$.
However, since Fermi acceleration may not take place in LGRBs, because  the WI is suppressed, 
the electrons may acquire a different distribution in energy.  It remains to be shown that the convolution of distributions
other  than the one we adopt here for LGRBs  could yield 
 the Band function (see \S ~ \ref{subsec:spectrum-multiple-filaments}); 
 
 \item Extremely steep high-energy spectral indices (i.e. $\beta> 3.0$) measured in
some GRBs would require $p>5$ (i.e. $(p+1)/2> 3$; see e.g.
 GRB 080319B listed as \#31 in Table \ref{table:lcfits} in \S ~ \ref{subsec:spectrum-multiple-filaments}).
  This also suggests an electron energy distribution other than the power-law one adopted here.

\end{enumerate}

\item To connect CSE emission to FRBs in the context of CCSNe:

\begin{enumerate}

\item We assumed the bunching length to be set by coherence scales linked to 
 Weibel saturated magnetic field in the shocked chunk frame. A proper treatment of the development
 of the instability with proper analysis of the coherence scale formation and the 
  magnetic field amplification would require
  PIC simulations before particles bunching can be firmly established in our model;

\item  Ignoring pair creation altogether 
    for the CSE  gives a frequency $\nu_{\rm CSE}$ of the order of a few GHz in  better agreement with
    data;  the luminosity  is independent of $n_{\rm pairs}$
    (see Eq. (\ref{eq:LFRB})).    This difference
between LGRBs and FRBs in our model, in addition to understanding the pair production and suppression 
mechanism in itself, is unclear at the moment;

\item To make  FRBs  repeat we appeal to a refracting ionized plasma (e.g. an HII region) surrounding the
SN site (see Appendix \ref{appendix:RFRBs}).  The existence of an appropriate screen remains unclear.

\end{enumerate}

\item Other assumptions include:

\begin{enumerate}

\item Our evaluation of the SN optical depth (see Appendix \ref{appendix:tauSN}), may be an oversimplification (see for example \citealt{bietenholz_2017})
 which may affect, and shift, our estimates  of the range in $t_{\rm QN}$ applicable to LGRBs and FRBs. However,
  this will not change our overall findings and conclusions;

\item We have argued that a type-Ic-BL SN results when the QN chunks collide with a turbulent PWN-SN
shell with filaments dense enough to trigger the RS into the chunks (see \S ~ \ref{sec:QN-Type-Ic-BL}). 
 We assumed that  the kinetic energy of the chunks which interacted with dense filaments
 is converted into kinetic energy of the surrounding PWN-SN shell material
 yielding a type-Ic-BL SN.  However, we lack a complete physical picture of how the process
  occurs;

\item We assumed that the QN ejecta fragments into chunks with single mass $m_{\rm c}$.
A more realistic scenario would consist, for example, of a mass distribution close to a log-normal distribution
with peak at $m_{\rm c}$, as
expected in debris from explosions (e.g. the Weibull distribution; \citealt{weibull_1993,brown_1995} ; see also \citealt{aastroem_2004}); 
 The main effects are: (i) The mass of primary and secondary chunks can be significantly larger or smaller than the fiducial $10^{22.5}$ gm value.
For example, if the primary chunk  is much less massive than the secondary, the prompt emission may be interpreted as a precursor as discussed  in \S~ \ref{sec:GRB-predictions};  (ii) The angular separation between chunks $\theta_{\rm sep.}$ would take on values
that are different from the one shown in Figure \ref{fig:stencil}.

\end{enumerate}
    

\end{itemize}

\section{Conclusion}
\label{sec:conclusion}

Assuming a QN event occurring years to decades following the core-collapse 
 of a massive star (e.g. a Type Ic SN as assumed in this work), we built a model capable of explaining many
 of the key characteristics of LGRBs and  a subset of FRBs (i.e. those
 related to CCSNe). 
  The time delay between the QN and the SN is the key parameter in our model
  since it defines the level of turbulence (thus the number of filaments) and the induced magnetization
  of the PWN-SN when it is sprayed by the millions of relativistic QN chunks.
   A shocked QN chunk emitting synchrotron radiation as it passes through successive filaments  can explain the light-curves   of many observed LGRBs including the flares (induced by secondary chunks) and the afterglow (from the
   interaction of the chunk with the medium surrounding the SN). 
  We successfully fit the light-curves in the XRT-band (including the afterglow and the flares when present) simultaneously with the spectrum
  for each of the 48 LGRBs we selected.
    Specifically, the time-averaged fast cooling
   synchrotron spectra from the interaction of the chunk with successive filaments yields
   a Band-like spectrum which for a given burst can be fit simultaneously with the resulting light-curve.
   
    In our model, the Yonetoku and Amati laws are not fundamental but are instead phenomenological 
    because the LGRB properties (i.e. $L_{\rm iso, peak}, E_{\rm peak}$ and $E_{\rm iso}$) depend on multiple physical parameters, 
which each have a limited range of scatter.

     The FRBs described here result from the interaction of QN chunks with a non-turbulent or a weakly-turbulent (and thus
    weakly magnetized) PWN-SN shell with conditions prone to the development of the Weibel instability 
    in the shocked chunk frame. The coherence length associated with the Weibel
    amplified magnetic field in the shocked chunk frame leads to electron and pair bunching triggering
    coherence synchrotron emission, in contrast to the LGRB case.
     The resulting frequency, luminosity and timescale  are consistent with those
     of observed FRBs.

Besides the limitations listed in \S ~ \ref{sec:limitations}, our model relies  on the feasibility of a delayed explosive transition of
    a massive NS to a QS years to decades following the SN explosion
    of a massive star. While such a transition is already  hinted at by 
    analytical (e.g. \citealt{keranen_2005,vogt_2004,ouyed_leahy_2009}) and by one-dimensional numerical simulations (\citealt{niebergal_2010,ouyeda_2018a,ouyeda_2018b}; see also \citealt{ouyeda_2019}), 
     detailed multi-dimensional simulations  are required to prove or disprove our working hypothesis (\citealt{niebergal_2011,ouyeda_2018}).
    Furthermore,   a full treatment of the interaction between the relativistic QN ejecta and the turbulent and non-turbulent PWN-SN shell
    would require  detailed hydrodynamical simulations beyond the scope of this paper.
Despite these limitations, our model seems successful at capturing key properties of
 LGRBs and FRBs and at unifying them with other related phenomena such as XRFs, XRR-GRBs and
 SLSNe.

  If our model is the correct representation of these phenomena, it can be used to  probe 
   the structure and physics of
 collisionless relativistic shocks and of the Weibel instability  
 and related coherence lengths. The connection between FRBs and UHECRs as we suggest here (see \S ~ \ref{sec:UHECRs})
  means that FRBs can also  be a vehicle to understanding Fermi acceleration.
  
  Our model and findings suggest that : (i) a catastrophic event (i.e. the QN)
      is behind LGRBs (including SGRBs) and some FRBs;
      (ii)  NSs born  with periods in the range $1.5\ {\rm ms} < P_{\rm NS} \le P_{\rm NS, cr.}$ and with mass
       $M_{\rm NS}\ge M_{\rm NS, cr.}$ can explode as QNe releasing $\sim 10^{53}$ ergs
       in kinetic energy; (iii)  spontaneous strange-quark nucleation can occur during
       quark deconfinement (e.g. \citealt{bombaci_2009}) induced by spin-down in massive NSs.
  
   Confirming the QN as the engine driving LGRBs and FRBs means that other implications
 of QNe to Cosmology (e.g. to re-ionization (\citealt{ouyed_2009c}) and to type Ia SNe 
calibration (\citealt{ouyed_2014,ouyed_2015d})), to  binary evolution  (\citealt{ouyed_2016,ouyed_2018a}) 
and to AXPs/SGRs (e.g. \citealt{ouyed_2007a,ouyed_2007b,ouyed_2018b})  warrant further studies.


\section{Acknowledgements}

This research is supported by operating grants from the National Science and Engineering Research Council of Canada (NSERC).
This work made use of data supplied by the UK Swift Science Data Centre at the University of Leicester.

\begin{appendix}

\section{Reference frames}
\label{appendix:frames}

Here we list the three reference frames
 involved: (i) the chunk's  (i.e. co-moving) frame where the quantities are primed  (the subscript ``c"  stands for ``chunk"  and is used to denote chunk's parameters); (ii) the exploding  NS frame (also the GRB cosmological rest frame) where quantities are unprimed; (iii) the observer's frame denoted by the
 superscript ``obs." in which quantities are angle  dependent. The Doppler factor is $D_{\rm c}(\Gamma_{\rm c}(t^\prime),\theta_{\rm c})= 1/(\Gamma_{\rm c}(t^\prime)(1-\beta_{\rm c}(t^\prime)\cos{\theta_{\rm c}}))$  where $\Gamma_{\rm c}(t^\prime)$ is the chunk's Lorentz factor,  $\beta_{\rm c}(t^\prime)=v_{\rm c}(t^\prime)/c$ with $v_{\rm c}(t^\prime)$ the chunk's speed 
  and $\theta_{\rm c}$ the chunk's viewing angle. The chunk's initial Lorentz factor is $\Gamma_{\rm c}(0)= \Gamma_{\rm QN}$.
 The transformation
 from the local NS frame to the chunk's frame is given by  $dt = \Gamma_{\rm c}(t^\prime)  dt^\prime$ while 
 the transformation from the chunk's frame to the observer's frame (where the emitted light is being observed) are
 $dt^{\rm obs.}= (1+z) dt^\prime/D_{\rm c}(\Gamma_{\rm c}(t^\prime),\theta_{\rm c})$, $\nu^{\rm obs.} = D_{\rm c}(\Gamma_{\rm c}(t^\prime),\theta_{\rm c})\nu^\prime/(1+z)$ where $z$ is the source's redshift. The NS frame and the observer frame share the same spatial co-ordinates, except for the $(1+z)$ factor, but not time which is subject
 to the additional Doppler factor.

\section{Some properties of the neutron-rich ultra-relativistic QN ejecta}
\label{appendix:qnejecta}
\setcounter{equation}{0}

\subsection{Chunks angular distribution and statistics}
\label{appendix:fireworks}

The QN causes the outermost NS crust to be ejected, which then breaks into 
$N_{\rm c} =10^6$ chunks, each with $\Gamma_{\rm c}(0)=\Gamma_{\rm QN} = 10^{3.5}$  and equally spaced in solid angle centered on the explosion site. I.e. many bright small pieces ejected radially outward from the explosion center.

For a distribution of $N_{\rm c}$ chunks that is uniform in solid angle we have $dN_{\rm c}/d\Omega =const.= N_{\rm c}/4\pi$ with
$d\Omega = 2\pi \sin{\theta} d\theta$ so that

\begin{equation}
\label{eq:Ntheta}
\frac{dN_{\rm c}}{d\theta} = \frac{N_{\rm c}}{2} \sin{\theta}\simeq \frac{N_{\rm c}}{2} \theta\ ,
\end{equation}
where the last expression applies the small angle approximation; i.e. $N_{\rm c}(\theta)\propto \theta^2$.
We note the following characteristics of the distribution:

(i) The solid angle covered by each chunk is given by $\pi \theta_{\rm c}^2 = 4\pi/N_{\rm c}$ .
This gives an  angular separation between chunks of $\theta_{\rm sep.}=2\theta_{\rm c} = 4/N_{\rm c}^{1/2} = 4\times 10^{-3}/N_{\rm c, 6}^{1/2}$.
  For our fiducial value of $\Gamma_{\rm c}=10^{3.5}$, the angular separation between chunks 
is about $12.6/\Gamma_{\rm c}$ (with $\Gamma_{\rm c}=\Gamma_{\rm QN}$ initially). Each chunk emits radiation into a narrow beam with half angle $\simeq 1/\Gamma_{\rm c}$
(i.e. a beam fullwidth of $2/\Gamma_{\rm c}= 6.3\times 10^{-4}/\Gamma_{\rm c, 3.5}$) which is about 1/6 of the angular spacing between chunks.
 This means emission pattern on the sky is $\sim 10^6$ narrow radial beams, one for each chunk, spaced over the whole sky;

(ii) The chunk aligned most closely toward to the observer is call the primary chunk. 
 The observed mean angle for the primary chunk is 
 $\bar{\theta}_{\rm P}= \int_{0}^{\theta_{\rm c}}2\pi \theta^2 d\theta/\int_{0}^ {\theta_{\rm c}}2\pi \theta d\theta = (2/3)\theta_{\rm c} =(4/3)/N_{\rm c}^{1/2}=1.3\times 10^{-3} N_{\rm c, 6}^{-1/2}$ which is close to $\sim 4/\Gamma_{\rm c}$;

(iii) There are 6 peripheral chunks we refer to as secondary chunks. We define the viewing angle of the secondary chunks as $\theta_{\rm S}$ with
$\theta_{\rm S}(\theta_{\rm P})= 2\theta_{\rm c}-\theta_{\rm P}$ which has a
 mean value of  $\bar{\theta}_{\rm S}= \int_{\theta_{\rm c}}^{2\theta_{\rm c}}2\pi \theta^2 d\theta/\int_{\theta_{\rm c}}^ {2\theta_{\rm c}}2\pi \theta d\theta\simeq  (14/9) \theta_{\rm c} =(28/9)/N_{\rm c}^{1/2}=3.1\times 10^{-3} N_{\rm c, 6}^{-1/2}$ which is close to $\sim 10/\Gamma_{\rm c}\sim 3.1\times 10^{-3}/\Gamma_{\rm c, 3.5}$;

(iv) We refer to emission  with $\theta_{\rm P}> 2/\Gamma_{\rm c}$  as ``off-axis" The ``off-axis'' solid angle is about 36 times the on-axis solid angle. 
I.e. there will be one bright burst for every $\sim 36$ faint bursts. 
 For an observer, this means most bursts will be ``off-axis".

 \subsection{The QN ejecta as an r-process site}
 \label{appendix:rprocess}

The extremely neutron-rich, relativistically expanding, QN ejecta 
is converted to unstable r-process material in a few milliseconds  (\citealt{jaikumar_2007,kostka_2014b,kostka_2014c,kostka_2014}).
     Figures 5 and 6 in \citet[see also \citealt{jaikumar_2007}]{kostka_2014c} show examples of the final composition of the expanding QN ejecta.
     The different Lorentz factor, $\Gamma_{\rm QN}$, of the QN ejecta, correspond to  different expansion timescales
       thus the differences in the final abundances.   For 
     $M_{\rm QN}< 10^{-4}M_{\odot}$ (i.e. $\Gamma_{\rm QN}> 10^{3}$),  
      the abundances  are dominated by elements with atomic weight $A<100$.    We adopt a chunk's opacity of $\kappa_{\rm c}=0.1$ cm$^2$ gm$^{-1}$ in this work since Lanthanides 
        (which would otherwise yield a much higher opacity; see \citealt{kasen_2013,tanaka_2013}) are not present in large quantities in the QN r-process yield for low QN mass ejecta.

The  $\beta$-decay luminosity in the chunk's frame can be defined by the following equations  (e.g. \citealt{korobkin_2012})

\begin{align}
  L_{\rm c, \beta}^\prime (t^\prime)    &=  2\times 10^{18}\ {\rm erg\ g}^{-1}\ {\rm s}^{-1}\times \epsilon_{\rm th., 0.5}\times m_{\rm c}\times rp(t^\prime )\\
 rp(t^\prime ) &=  \left(\frac{1}{2}- \frac{1}{\pi} \arctan{\left(\frac{t^\prime -t_{\rm F}^\prime }{0.11}\right)}\right)^{1.3}  \ ,
 \end{align}
  where   $t_{\rm F}^\prime$ which of the order of a second is the freeze-out timescale  and $\epsilon_{\rm th., 0.5} = \epsilon_{\rm th.}/0.5$  
  the percentage of $\beta$-decay energy which thermalizes in the chunk in units of 0.5.
  
When  $t^\prime$ exceeds  a few times the freeze-out time (the case in our model) the $\beta$-decay contribution can be expressed as  (e.g.  \citealt{li_1998,metzger_2010b})

\begin{equation}
\label{eq:L-beta}
L_{\rm c, \beta}^\prime(t^\prime) \sim 
  9\times 10^{38}\ {\rm erg\ g}^{-1}\ {\rm s}^{-1}\times m_{\rm c, 22.5}\times  \left(\frac{t^\prime}{1\ {\rm s}}\right)^{-1.3}\ .
 \end{equation}

\subsection{The chunk's sweeping luminosity}
\label{appendix:sweeping}

 The evolution in the chunk's rest frame
  of the sweeping luminosity $L_{\rm c, sw.}^\prime(t^\prime)$  and of the Lorentz factor $\Gamma_{\rm c}(t^\prime)$ are given by the two following fundamental equations

 \begin{align}
 \label{eq:Lcsw}
 L_{\rm c, sw.}^\prime(t^\prime) &= E_{\rm p}(t^\prime) \times  A_{\rm c}(t^\prime) \times  (\Gamma_{\rm c}(t^\prime) n_{\rm amb.}) \times \beta_{\rm c} c\\
  \label{eq:Gammac}
 -\frac{d\Gamma_{\rm c}(t^\prime)}{\Gamma(t^\prime)^2} &= \frac{4 A_{\rm c}(t^\prime)\times \rho_{\rm amb.}\times (\beta_{\rm c} c \times (\Gamma_{\rm c}(t^\prime) dt^\prime))}{3 m_{\rm c}} \ ,
 \end{align}
 where hereafter $\beta_{\rm c}=v_{\rm c}/c=1$. The above is for 
  a chunk of mass $m_{\rm c}$ and area $A_{\rm c}(t^\prime)$ sweeping
 protons and electrons in an ambient medium of baryon number density $n_{\rm amb.}$. These
 equations assume  is  the radiative case and $\Gamma_{\rm c}>>1$ (e.g. \citealt{peer_2012} and references therein).

  The unprimed quantities are in the NS's frame,  with $dt=\Gamma(t^\prime) dt^\prime$.
The adiabatic index of the swept-up material is taken to be 4/3 and 
  $E_{\rm p}(t^\prime) = \zeta_{\rm p}\Gamma_{\rm c}(t^\prime) m_{\rm p}c^2$ is the 
  the fraction of proton energy transferred to electron-positron pairs. We take $\zeta_{\rm p}=1$
  for simplicity, effectively assuming  efficient thermalization  of dissipated kinetic
     energy in the shocks. This means that  electrons (and positrons) are accelerated to a Lorentz factor
      of $\gamma_{\rm e}\sim (1/2 n_{\rm pairs})\times \Gamma_{\rm c}(t^\prime) m_{\rm p}/m_{\rm e}$. 
      Here $n_{\rm pairs}$ is the number of paris created per proton  (see \S ~ \ref{subsec:spectrum-single-filament}).
  This is representative of the radiative case 
  (where most of the swept energy is promptly radiated)  and is associated 
  in our model with  the regime where the chunk is optically thin
    (see \S ~ \ref{subsec:optically-thin-regime}).  These simplifying assumptions allow us to provide analytical solutions in our model.
    In particular, for a constant sweeping area $ A_{\rm c}(t^\prime)=A_{\rm c, T}$ (see Eq. (\ref{eq:AcT})), Eq. (\ref{eq:Gammac}) becomes
    
    \begin{equation}
     \label{eq:Gammac2}
    d\left( \frac{\Gamma_{\rm c}(0)^2}{\Gamma_{\rm c}(t^\prime)^2} \right) = d \left( \frac{t^\prime}{t_{\Gamma}^\prime}\right)\ ,
    \end{equation}
    with $t_{\Gamma}^\prime$ and the solution, $\Gamma_{\rm c}(t^\prime)$, to equation above given
    in Eqs. (\ref{eq:tGamma}) and (\ref{eq:GammacT}), respectively. 
    

 \subsection{QN chunks inside a PWN}
 \label{appendix:PWN}
  
  Since the QN involves the explosion of a NS, it is natural to consider
 the evolution of the QN ejecta inside a PW bubble. Before the chunks collide with the
  PWN-SN shell, the density is such that  the sweeping luminosity is 
    dwarfed by heating from $\beta$-decay; i.e. $L_{\rm c, sw.}(t) << L_{\beta}(t)$. 
   The time evolution of the chunk's cross-section area $A_{\rm c}(t)$  and temperature $T_{\rm c}(t)$ during the time that the chunk 
   is optically thick
    (i.e. when $A_{\rm c}(t^\prime) < A_{\rm c, T}$ where $A_{\rm c, T}$ is the area of the chunk when it becomes optically thin;
    see Eq. (\ref{eq:AcT})) is found from 
  
  \begin{align}
 A_{\rm c}(t^\prime) &= \pi (R_{\rm c, 0}^\prime+c_{\rm s, c}^\prime t^\prime)^2 \\
4 A_{\rm c}(t^\prime)\sigma_{\rm SB} {T^\prime}^4  &=  L_{\beta}^\prime(t^\prime)  + L_{\rm c, sw.}^\prime(t^\prime)\sim L_{\rm c, \beta}^\prime(t^\prime)  \ ,
 \end{align}
 where $c_{\rm s, c}^\prime=\sqrt{\gamma_{\rm ad., c} \frac{k_{\rm B}T^\prime}{\mu_{\rm c} m_{\rm H}}}$ is the sound speed in the chunk and $R_{\rm c, 0}^\prime$ is the chunk's  initial radius.
 The constants are the
 Boltzmann constant $k_{\rm B}$, the hydrogen mass $m_{\rm H}$ and, the Stefan-Boltzmann constant $\sigma_{\rm SB}$.
The chunk's adiabatic index is taken as $\gamma_{\rm ad., c}=5/3$, and  the chunk's
mean molecular weight (for heavy composition) is  $\mu_{\rm c}\simeq2$.

  Analytical solutions can then be found for timescales relevant to the QN-SN interaction,  $t^\prime >> R_{\rm c, 0}^\prime/c_{\rm s, c}^\prime$ and  $t^\prime>t_{\rm A}^\prime$. In this case, the $\beta$-decay contribution
  is given by Eq. (\ref{eq:L-beta}) which  allows us to  solve for:

\begin{align}
\label{eq:Tc}
 k_{\rm B}T_{\rm c}^\prime(t^\prime) &\sim 
   0.6\ {\rm keV}\ m_{\rm c, 22.5}^{1/5}\times  (t^\prime)^{-3.3/5}\\
   \label{eq:Ac}
 A_{\rm c}^\prime(t^\prime) &\sim
   1.5\times 10^{15}\ {\rm cm}^2\ m_{\rm c, 22.5}^{1/5} \times (t^\prime)^{6.7/5}\ .
 \end{align}

   Setting $A_{\rm c}(t_{\rm T}^\prime)= A_{\rm c, T}$  gives  a critical time  which defines the end
 of the chunk's optically thick expansion and the start of the optically thin regime, where the
 chunk stops expanding. This time is

 \begin{equation}
 t_{\rm T}^\prime \simeq 4.3\times 10^4\ {\rm s}\times m_{\rm c, 22.5}^{4/6.7} \kappa_{c, -1}^{5/6.7}\ .
 \end{equation}
 The above is an upper limit  on the time it would take the expanding chunk  to reach transparency, because it only takes into account
 heating from $\beta$-decay. Including heating from sweeping  yields higher
 temperatures which makes the chunk expand faster and yields a smaller transparency time.
 
In the NS frame the transparency time is
  $t_{\rm T} = \Gamma_{\rm QN} t_{\rm T}^\prime$  corresponding to a distance from the NS of  
  \begin{equation} 
  \label{eq:RT}
  R_{\rm T} = c t_{\rm T}  < 4.1 \times 10^{18}\ {\rm cm} \times m_{\rm c, 22.5}^{4/6.7}\kappa_{\rm c, -1}^{5/6.7}\Gamma_{\rm QN, 3.5}\ .
  \end{equation}

   \section{The SN ejecta's optical depth}
   \label{appendix:tauSN}
   \setcounter{equation}{0}

   The optical depth of the SN ejecta is $\tau_{\rm SN}=\int_{R_{\rm w}}^{R_{\rm SN}} \sigma_{\rm Th.} n_{\rm e, SN} dr
   = \tau_{\rm SN, inner} + \tau_{\rm SN, outer}$, where $\sigma_{\rm Th.}$ is the Thomson optical depth
    and $n_{\rm e, SN} =\rho_{\rm SN}/m_{\rm H}$, with
   
   \begin{align}
   \label{eq:tausn1}
   \tau_{\rm SN, inner} &= \int_{R_{\rm w}}^{R_{\rm Plat.}} \sigma_{\rm Th.} n_{\rm e, Plat.} dr \\\nonumber
    &=  
   \sigma_{\rm Th.}n_{\rm e, Plat.}  R_{\rm Plat.} \left(1-\frac{R_{\rm w}}{R_{\rm Plat.}}\right)
   \end{align}
   and 
   
   \begin{align}
   \label{eq:tausn2}
   \tau_{\rm SN, outer}  &=\int_{R_{\rm Plat.}}^{R_{\rm SN}} \sigma_{\rm Th.} n_{\rm e, SN} dr \\\nonumber
   &= \int_{R_{\rm Plat.}}^{R_{\rm SN}} \sigma_{\rm Th.} n_{\rm e, Plat.} \left(\frac{R_{\rm Plat.}}{r}\right)^{-n}dr \\\nonumber
   &\simeq   \sigma_{\rm Th.} n_{\rm e, Plat.}R_{\rm Plat.}\times \frac{1}{8}\left(1- \left( \frac{R_{\rm SN}}{R_{\rm Plat.}}\right)^{-8} \right)\ .
   \end{align}
  
  where  $n_{\rm e, Plat.}=\rho_{\rm e, Plat.}/m_{\rm H} = At^{-3}/m_{\rm H}$
   and $R_{\rm Plat.}=v_{\rm t}t$. 
  Adding  Eqs. (\ref{eq:tausn1}) and (\ref{eq:tausn2}) for $R_{\rm SN}>> R_{\rm Plat.}$
  yields 
  
  \begin{equation}
   \tau_{\rm SN} \sim    (3.15\times 10^{15} t^{-2})\times {E_{\rm SN, 51}}^{-1}M_{\rm SN, 34}^2\ .
   \end{equation}

  The conditions $\tau_{\rm SN} <1$ yields
  
  \begin{equation}
  \label{eq:appendix-tauSN}
   t_{\rm QN} > t_{\rm SLSN}=1.8\ {\rm years}\times {E_{\rm SN, 51}}^{-1/2}M_{\rm SN, 34} \ .
   \end{equation}

\section{The thick wall case}  
\label{appendix:thick-wall}
\setcounter{equation}{0}

  For the thick wall case, the forward shock (FS) Lorentz factor varies in time as $\Gamma_{\rm FS}(t^\prime) \simeq \Gamma_{\rm c}(t^\prime)$ with $\Gamma_{\rm c}(0)=\Gamma_{\rm QN}$. Thus the main differences between the thin wall and thick wall cases are: 
 (i) integrating time variable 
 quantities that depend on the decreasing Lorentz factor $\Gamma_{\rm c}(t^\prime)$); (ii) time-averaging of
 quantities such as the time-dependent photon peak energy and luminosity; (iii)  setting 
  the typical GRB duration, in the NS frame, to be $3 t_{\Gamma}$.
 
 The  peak frequency  of a single chunk is averaged over time, weighted by photon number:
  
  \begin{equation}
\bar{E}_{\rm \gamma, p}= \frac{\int_0^{t_{\rm w}} E_{\rm \gamma, p}(t) N(t)dt}{\int_0^{t_{\rm w}} N(t)dt}\ ,
\end{equation}
where the photon rate is $N(t)= L(t)/h\nu\propto \Gamma_{\rm FS}^6/\Gamma_{\rm FS}\propto \Gamma_{\rm FS}^5(t)$
so that 
\begin{equation}
\bar{E}_{\rm \gamma, p}= \frac{\int_0^{t_{\rm w}} E_{\rm \gamma, p}(t) \Gamma(t)^5dt}{\int_0^{t_{\rm w}^{\rm obs.}} \Gamma(t)^5dt}\ .
\end{equation}
  The peak luminosity occurs at $\Gamma_{\rm c}(0)=\Gamma_{\rm QN}$, i.e.
  
  \begin{equation}
  L_{\rm GRB, p} = D(\Gamma_{\rm QN},\theta_{\rm P})^4 {L^\prime}_{\rm c, sw.}(0)\ ,
   \end{equation}
  
 where the chunk's sweeping luminosity is $L^\prime_{\rm c, sw.}(t^\prime)= C_1^{\prime\prime}\times \Gamma_{\rm c}(t^\prime)^2 n_{\rm w}\kappa_{\rm c}$ ($C_1^{\prime\prime}$  is a constant; see Eq. (\ref{eq:LcswT})) with
 $\Gamma_{\rm c}(t^\prime)=\Gamma_{\rm c}(0)/ (1+t^\prime/t_{\Gamma}^\prime)^{1/2}$ (see Eq. (\ref{eq:Gammac})).
 
 The isotropic energy $E_{\rm GRB}$ from a single chunk is:
  
\begin{equation}
\label{eq:wall-energy}
E_{\rm GRB}=  \int_0^{t_{\rm w}^\prime} D(\Gamma_{\rm c}(t^\prime),\theta_{c})^3 L_{\rm c, sw.}^\prime(t^\prime) dt^\prime
 = C_1^{\prime\prime}\times 2^3 n_{\rm w}\kappa_{\rm c} \int_{0}^{t_{\rm w}^\prime}   \frac{\Gamma_{\rm c}(t^\prime)^5}{(1+(\Gamma_{\rm c}(t^\prime) \theta_{\rm c})^2)^3} dt^\prime \ ,
\end{equation}
    
 where we made use of $D(\Gamma_{\rm c}(t^\prime),\theta_{\rm c}) = 2\Gamma_{\rm c}(t^\prime)/(1+(\Gamma_{\rm c}(t^\prime) \theta_{\rm c})^2)$.
 One can show that

\begin{equation}
\frac{dt^\prime}{t_{\Gamma}^\prime} = - 2 \left( \frac{\Gamma_{\rm c}(0)}{\Gamma_{\rm c}} \right)^2 \frac{d\Gamma_{\rm c}}{\Gamma_{\rm c}}\ ,
\end{equation}

so the integral above becomes

\begin{equation}
E_{\rm GRB} = - C_1^{\prime\prime}\times 2^4\Gamma_{\rm c}(0)^2 n_{\rm w}\kappa_{\rm c} t_{\Gamma}^\prime \int_{\Gamma_{\rm c}(0)}^{\Gamma_{\rm c, F}}   \frac{\Gamma_{\rm c}^2}{(1+(\Gamma_{\rm c} \theta_{\rm c})^2)^3} d\Gamma_{\rm c}\ ,
\end{equation}
where $\Gamma_{\rm c, F}$ is the chunk's Lorentz factor at the exit of the filament.

Since $t_{\Gamma}^\prime = C_2^{\prime\prime}\times  (n_{\rm w}\Gamma_{\rm c}(0)^2\kappa_{\rm c})^{-1}$ (see Eq. (\ref{eq:Gammac})) we can rewrite the above as

\begin{equation}
\label{eq:amati-integral}
E_{\rm GRB} = - C_1^{\prime\prime\prime}\times 2^4\times\int_{\Gamma_{\rm c}(0)}^{\Gamma_{\rm c, F}}   \frac{\Gamma_{\rm c}^2}{(1+(\Gamma_{\rm c} \theta_{\rm c})^2)^3} d\Gamma_{\rm c}\ ,
\end{equation}
with $C_1^{\prime\prime\prime}=C_1^{\prime\prime} C_2^{\prime\prime}$.

\section{Light-curve and spectrum simulation algorithms}
\label{appendix:LC-algorithms}
\setcounter{equation}{0}

As described in section \S ~ \ref{sec:grbs-filamentary}, the variability of each LGRB light-curve is determined by the spatial (location/thickness) and density distributions of the filaments.  In order to successfully model a specific LGRB light-curve, we must therefore determine these distributions which will be dependent on our fitting parameters. 

\subsection{Filament location and thickness generation}\label{section:filament_algorithm}

The algorithm for finding the location and thickness of each filament (using observed data points) during a simulation is given below.  It  gives a good approximation to the observed light-curves:

\begin{enumerate}

\item Create a first ``filament", $F_0$, to represent the outer edge of the pulsar wind bubble inside of which the density is set to 0.  Set its position, $d_0$ to 0.

\item For each subsequent data point, $i$, in the observed set (i.e. light-curve) do the following:

\begin{enumerate}[a.]

\item Transform the point time, $t_i\obs$, to the rest frame time, $t_i^\prime$, of the chunk:

\begin{equation}
\label{eq:appendix-Fi}
t^\prime_i = \frac{(t_i\obs - t_{i-1}\obs) D(\Gamma_{i,0},\theta_{\rm c})}{(1+z)} + t_{i-1}^\prime
\end{equation}

where $t_{i-1}^{\rm obs}$ and $t^\prime_{i-1}$ is the observed and chunk frame time, respectively, when the chunk entered the previous filament ($F_{i-1}$); here $\Gamma_{i,0}$ is the value of the Lorentz factor when the chunk exits the previous filament.

\item Calculate the distance, in the NS frame, the chunk traveled in this time:

\begin{equation} \label{dns}
\Delta d_i = 2 c \Gamma_{i,0} t_{\Gamma_i}^\prime \left[\sqrt{1+\frac{t_i^\prime}{t_{\Gamma_i}^\prime}} - \sqrt{1+\frac{t_{i-1}^\prime}{t_{\Gamma, i}^\prime}} \right]
\end{equation}

Where $t_{\Gamma_i}^\prime$ is given by Eq. (\ref{eq:tGamma}). Set the end position of our filament, $F_i$, to:

\begin{equation}
d_{i,1} = d_{i,0}  + \Delta d_i\ .
\end{equation}

\item If $\Delta d_i$ is  greater than $\Delta R_{\rm F, max.} =R_{\rm w}(t_{\rm Plat.})/3$ (the maximum filament's thickness), create a ``gap" filament, $F_g$ with $n_g = 0$ and adjust $d_{i,1}$ of $F_i$ accordingly. {\it In this way, no filament can exceed a width of $R_{\rm w}/12$}. Effectively, the
filament's thickness is $\Delta R_{\rm F_i} = {\rm Min}[\Delta d_i,\Delta R_{\rm F, max.}]$.

\item Create a new filament, $F_{i+1}$, with its start position at $d_{i+1,0}= d_{i,1}= d_{i,0} + \Delta d_i$. 
\end{enumerate}

\item Create a final ``filament"  with infinite thickness, $F_\textrm{amb.}$, to represent the ambient medium (e.g. ISM or the low density SN ejecta
overlaying the wall) with density $n_\textrm{amb.}$ and magnetic field $B_\textrm{amb.}$.

\end{enumerate}

\subsection{Filament density generation}\label{section:denstiy_algorithm}

 Instead of fitting a large number of individual filament densities (about a hundred per LGRB),
  we chose to fit the peak luminosity (assigned a density $n_{\rm w}$) of the selected LGRB then scale all other filament
  densities $n_{\rm F}$ using the following power-law 

\begin{equation}
 \label{eq:filament-density}
n_{\rm F}=
\begin{cases}
n_{\rm w}\times \left(\frac{L_{\rm F}}{L_{\rm GRB, p}}\right)^{\alpha_{\rm F}}\, &   \text{for\ } t^{\rm obs.} <  t^{\rm obs.}_{\rm p}\\
n_{\rm w}\times \left(\frac{L_{\rm F}}{L_{\rm GRB, p}}\right)\, &   \text{for\ } t^{\rm obs.} > t^{\rm obs.}_{\rm p}\ .
\end{cases}
 \end{equation}
 Here, $t^{\rm obs.}_{\rm p}$ is the location of the peak luminosity, $L_{\rm GRB, p}$, in the light-curve. The parameter 
  $\alpha_{\rm F}>0$ is a constant for each burst and is a consequence of
  our assumption of constant chunk area (given by Eq. (\ref{eq:AcT})) which is invalid in the pre-peak luminosity phase during the interaction with the filaments (see discussion
   in \S ~ \ref{sec:limitations}). For $ t^{\rm obs.} > t^{\rm obs.}_{\rm p}$,
    $\alpha_{\rm F}=1$ gives good fits to light-curves and agrees well with the maximum  chunk's area regime 
   where $n_{\rm F}\propto L_{\rm GRB}$ (see Eq. (\ref{eq:LcswT})).

\subsection{Light-curve generation} \label{section:lightcurve_algorithm}

The observed light-curve for the GRB is calculated by the following algorithm:

\begin{enumerate}
\item Generate a list of time points in the rest frame of the chunk. The simulation uses 500 evenly spaced time intervals in log scale between $-3 \leq \log(t^\prime) \leq 10$. In order to assure adequate sampling of each filament, we also calculate emission for 100 equally spaced time intervals in log scale between $\log(t^\prime_0) \leq \log(t^\prime) \leq \log(t^\prime_1)$ where $t^\prime_0$ and $t^\prime_1$ are the time the chunk enters and exits the filament, respectively. Each filament is resolved into 100 time-steps in order to capture the slowing down of the chunk within a filament.

\item For the primary chunk in the simulation, step through the time points and calculate : The  corresponding observed time, and the observed luminosity.  The observed time for the {\it i}th time point is calculated as:

\begin{equation}
\label{eq:appendix-Fi-inverse}
t\obs_i = \frac{(1+z)(t^\prime_i - t^\prime_{i-1})}{D(\Gamma_{\rm c},\theta_{\rm c})} + t\obs_{i-1} \ .
\end{equation}

Eq. (\ref{eq:appendix-Fi-inverse}) above is the inverse of Eq. (\ref{eq:appendix-Fi}). However,
 the observed time in Eq. (\ref{eq:appendix-Fi-inverse}) refers to time in the observer's frame based on our model
 while $t^{obs}$ in Eq. (\ref{eq:appendix-Fi}) means the actual observed time (i.e. for each data point) for the light-curve
 being fit.

The luminosity in the XRT band ($0.3 \leq E_{\gamma}\obs \leq 10$ keV) is calculated as:

\begin{equation}
\begin{split}
L\obs(E_{\gamma, 0}\obs, E_{\gamma, 1}\obs) &=  (1+z) D(\Gamma_{\rm c},\theta_{\rm c})^4 L^\prime(E_{\gamma, 0}^\prime, E_{\gamma, 1}^\prime) \\
L^\prime(E_{\gamma, 0}^\prime, E_{\gamma, 1}^\prime) &=  C_{\rm XRT} \int_{E_{\gamma, 0}^\prime}^{E_{\gamma, 1}^\prime} {L^\prime(E_{\gamma}^\prime) dE_{\gamma}}^\prime\ .
\end{split}
\end{equation}

With

\begin{equation} \label{equation:Erest}
E_{\gamma}^\prime = \frac{E_{\gamma}}{D(\Gamma_{\rm c},\theta_{\rm c})}=\frac{(1+z) E_{\gamma}\obs}{D(\Gamma_{\rm c},\theta_{\rm c})}
\end{equation} 

and an XRT bolometric correction (the BAT emission in the fit light-curve was converted to XRT band in \citet{evans_2010})

\begin{equation}
\label{eq:cXRT}
C_{\rm XRT} = \frac{L^\prime_\textrm{c., sw.}}{L^\prime(E_{\gamma, 0}^\prime, E_{\gamma, 1}^\prime)}\ ,
\end{equation}

is a constant with the chunk's sweeping luminosity $L^\prime_\textrm{c, sw.}$ given by Eq. (\ref{eq:LcswT}) 
 being the bolometric luminosity. The fast and slow cooling regimes are defined each by their  luminosity density given in the following equations

\begin{equation} \label{equation:LE}
    L^\prime(E_{\gamma}^\prime) =
\begin{cases}
    L^\prime_\textrm{slow}(E_{\gamma}^\prime),& \text{if } E_{\gamma, p}^\prime \leq E_{\gamma, c}^\prime \\
    L^\prime_\textrm{fast}(E_{\gamma}^\prime),& \text{if } E_{\gamma, p}^\prime > E_{\gamma, c}^\prime    
\end{cases}
\end{equation}

\begin{equation}\label{equation:Lslow}
    L^\prime_\textrm{slow}(E_{\gamma}^\prime) = C_{\rm XRT}\ {\rm erg\ s^{-1}\ keV^{-1}}\times 
\begin{cases}
    (E_{\gamma}^\prime/E_{\gamma, p}^\prime)^{1/3},& \text{if } E_{\gamma}^\prime < E_{\gamma, p}^\prime \\
    (E_{\gamma}^\prime/E_{\gamma, p}^\prime)^{-(p-1)/2},& \text{if } E_{\gamma, p}^\prime \leq E_{\gamma}^\prime \leq E_{\gamma, c}^\prime \\
    (E_{\gamma, c}^\prime/E_{\gamma, p}^\prime)^{-(p-1)/2} (E_{\gamma}^\prime/E_{\gamma, c}^\prime)^{-p/2},& \text{if } E_{\gamma}^\prime > E_{\gamma, c}^\prime
    
\end{cases}
\end{equation}

\begin{equation}\label{equation:Lfast}
    L^\prime_\textrm{fast}(E_{\gamma}^\prime) = C_{\rm XRT}\ {\rm erg\ s^{-1}\ keV^{-1}}\times
\begin{cases}
    (E_{\gamma}^\prime/E_{\gamma, c}^\prime)^{1/3},& \text{if } E_{\gamma}^\prime < E_{\gamma, c}^\prime \\
    (E_{\gamma}^\prime/E_{\gamma, c}^\prime)^{-1/2},& \text{if } E_{\gamma, c}^\prime \leq E_{\gamma}^\prime \leq E_{\gamma, p}^\prime \\
    (E_{\gamma, p}^\prime/E_{\gamma, c}^\prime)^{-1/2} (E_{\gamma}^\prime/E_{\gamma, p}^\prime)^{-p/2},& \text{if } E_{\gamma}^\prime > E_{\gamma, p}^\prime
    
\end{cases}
\end{equation}

In the above $E_{\gamma, c}^\prime = E_{\gamma, c}/D(\Gamma_{\rm c},\theta_{\rm c})$ and $E_{\gamma, p}^\prime = E_{\gamma, p}/D(\Gamma_{\rm c},\theta_{\rm c})$  with
$E_{\gamma, c}$ and $E_{\gamma, p}$  given by Eqs. (\ref{eq:Ecool}) and (\ref{eq:Epeak}), respectively.

\item Create observed time bins between $t\obs$ = 0 and the last BAT  time point in the observed data.  The width of each bin during the prompt is set to 64 ms, and 100 s during the afterglow.

\item For each time bin ($t\obs_{\rm bin}$) created in step 3, add the calculated observed flux ($F\obs = D(\Gamma_{\rm c},\theta_{\rm c})^4L_{\rm c, sw.}^\prime/(4\pi d_{\rm L}^2)$) for each chunk.  If the chunk does not have a calculated flux for the $t\obs_{\rm bin}$, use linear interpolation in time to find it.

\end{enumerate}

\subsection{Spectrum}\label{section:spectrum_algorithm}

The final spectrum is created between energies of 0.2 and $10^6$ keV (in the observer's frame) by taking an average of spectra sampled at each observed time point generated for the light-curve (see step 1 in \S ~ \ref{section:lightcurve_algorithm}; i.e. using the same sample points as in
light-curve generation).  The algorithm for creating a single spectrum at $t\obs$ is the following
(using the primary chunk):

\begin{enumerate}

\item For the primary chunk in our simulation, generate an observed spectrum at $t\obs$:

\item For each energy in our observed spectrum, $0.2 \leq E_{\gamma}\obs \leq 10^6$ keV, calculate the observed flux density:

\begin{equation}
F\obs(E_{\gamma}\obs)=D(\Gamma_{\rm c},\theta_{\rm c})^3 (1+z) \frac{L^\prime(E_{\gamma}^\prime)}{4 \pi d_{\rm L}^2}
\end{equation} 

which is the observed flux at $E_{\gamma}^\prime$ given by Eq. (\ref{equation:Erest}) and $L^\prime(E_{\gamma}^\prime)$ by Eq. (\ref{equation:LE}).

\item Multiply the observed flux by the observed frequency, $\nu_{\gamma}\obs = E_{\gamma}\obs/h$, to get $\nu_{\gamma}\obs F\obs(\nu_{\gamma}\obs)$.

\end{enumerate}

\section{Coherent synchrotron emission (CSE)}
\label{appendix:CSE}
\setcounter{equation}{0}

 A relativistic electron beam moving in a circular orbit in free space can radiate coherently if the wavelength of the
 synchrotron radiation, $\lambda_{\rm Sync.}^\prime$,  exceeds the length of the bunch $l_{\rm b}^\prime$;  here  the primed quantities refer to the
 shock frame.  One can picture each electron emitting an electromagnetic wave with just a small phase difference with respect  to the other emitting electrons
 in the beam.
 If $N_{\rm b}$ is the number of electrons in a bunch then it can be shown that the 
  intensity  of the CSE scales as $N_{\rm b}^2$ instead of $N_{\rm b}$ as in 
   the incoherent case (\citealt{schiff_1946,schwinger_1949,nodvick_1954}; see also \citealt{goldreich_1971}). 
 For a bunch where the longitudinal density function is Gaussian with r.m.s $l_{\rm b}^\prime$,
 the spectral distribution is (e.g. \citealt{novokhatski_2012})
 
 \begin{equation}
 I_{\rm b}(\omega^\prime) = I_{\rm s}(\omega^\prime)N_{\rm b}\left(1+ N_{\rm b}\exp{\left(-\left(\omega^\prime\frac{l_{\rm b}^\prime}{c}\right)^2\right)} \right)\ ,
 \end{equation}
 
 where $I_{\rm s}(\omega^\prime)$ is the single particle spectrum and $\omega^\prime$ the angular frequency.
 The equation above implies that 
CSE  dominates when 
\begin{equation}
\label{eq:CSE-condition}
N_{\rm b}\exp{\left(- \left( \omega^\prime\frac{l_{\rm b}^\prime}{c}\right)^2\right)}>1 \ ,
\end{equation}
which translates to  $\omega^\prime < (c/l_{\rm b}^\prime)\times \sqrt{\ln{N_{\rm b}}}$. 

 \subsection{CSE characteristic frequency}

 Since $\sqrt{\ln{N_{\rm b}}}$
 is of the order of a few for a very wide range of $N_{\rm b}$ (e.g. $4.3 < \sqrt{\ln{N_{\rm b}}} < 8.5$
 for $10^{8} < N_{\rm b} < 10^{30}$) hereafter we write 
 the peak  CSE frequency as
 
 \begin{equation}
 \label{eq:CSE-nu}
 \nu_{\rm CSE}^\prime =  \frac{c}{l_{\rm b}^\prime}\times \frac{\sqrt{\ln{N_{\rm b}}}}{2\pi}\sim \frac{c}{l_{\rm b}^\prime} \ .
 \end{equation} 
This shows that $ \nu_{\rm CSE}^\prime$ is set by the length of the electrons bunch in the shock frame.
 
  \subsection{CSE power}

The total coherent power per bunch  is (\citealt{schwinger_1949}; see also Eq. (16) in \citealt{novokhatski_2012})

\begin{equation}
 \label{eq:CSE-L}
L_{\rm b}^\prime \simeq 5.4\times 10^{-23}\ {\rm erg\ s}^{-1}\ \left(\frac{N_{\rm b}}{l_{\rm b}^\prime}\right)^2 \left( \frac{l_{\rm b}^\prime}{r_{\rm L, e}^\prime}\right)^{2/3}\ ,
\end{equation}
where $r_{\rm L, e}^\prime = c/\omega_{\rm e}^\prime$ is the electron's Larmor radius 
with $\omega_{\rm e}^\prime=e B^\prime/\gamma_{\rm e} m_{\rm e}c$ and $\gamma_{\rm e}$ the electron's
 thermal Lorentz factor; $e$ and $m_{\rm e}$ are the electron's charge and mass, respectively.   

 \section{Repeating FRBs}
\label{appendix:RFRBs}
\setcounter{equation}{0}

We argue that these are ``twinkling FRBs" due to refractive ionized plasma (e.g. HII regions) in the
vicinity of the SN explosion.  
   Let us assume that the QN is surrounded by a thin shell of ionized plasma (e.g. an HII region) at
  radius $R_{\rm sh.}$. The shell has a refraction index $n_{\rm sh.}=\sqrt{1-\nu_{\rm sh., p}/\nu}$ with 
  $\nu_{\rm sh., p}$ the shell's plasma frequency. We assume that each emission beam (with initial
  beam-width $\sim 1/\Gamma_{\rm QN}$) from the $N_{\rm c}$ QN
  chunks is bent by an angle $\Delta \theta_{\rm sh.}\le \theta_{\rm max.}$ in  a random direction. Here
  $\theta_{\rm max.}$ is the maximum bending angle (see Figure \ref{fig:RFRBs}). Thus the
  probability of seeing any one beam (i.e. a beam scattered towards the observer) is
  
  \begin{equation}
  P_{1}=\frac{\pi(1/\Gamma_{\rm QN}^2)}{\pi \theta_{\rm max.}^2}\ .
  \end{equation}
  
  As can be seen from Figure \ref{fig:RFRBs}, beams in the $\theta>\theta_{\rm max.}$ quadrant cannot be seen by the 
  observer. For $\theta < \theta_{\rm max.}$ the number of beams that are scattered randomly is
  
  \begin{equation}
  N_{\rm scat.}= \frac{\pi \theta_{\rm max.}^2}{4\pi} N_{\rm c}\ .
  \end{equation}
  
  The above implies that the total probability of seeing one beam  scattered towards the observer is
  
  \begin{equation}
  P_{\rm T, 1}= P_{1}\times N_{\rm scat.}= \frac{N_{\rm c}}{4 \Gamma_{\rm QN}^2}\ .
  \end{equation}

  In our model thus, a repeating FRB occurs only if $P_{\rm T, 1}\ge 1$ or when
  
  \begin{equation}
  \label{eq:RFRBs-condition}
  N_{\rm c}> 4 \Gamma_{\rm QN}^2 = 4\times 10^7 \times \Gamma_{\rm QN, 3.5}^2\ .
  \end{equation}
  
  The implication of Eq. (\ref{eq:RFRBs-condition}) above is that a typical chunk in a QN where a
   a repeating FRB occurs has a mass $m_{\rm c, RFRB} < m_{\rm c, FRB}/40$; for our fiducial
  values $N_{\rm c}=10^6$. 
  Since the FRB luminosity is linearly proportional to the chunk's mass (see Eq. \ref{eq:LFRB}),
   this means that the luminosity of an RFRB is such that $L_{\rm RFRB}< L_{\rm FRB}/40$.

  The maximum duration of the entire repeating episode is the time delay between $\theta=0$ and $\theta_{\rm max.}$, or
  
  \begin{equation}
  \Delta t_{\rm max}= \frac{R_{\rm sh.}(1-\cos{\theta_{\rm max.}})}{c}\sim 32.6\ {\rm years}\times R_{\rm sh., 1}(1-\cos{\theta_{\rm max.}})\ .
  \end{equation}
 with $R_{\rm sh., 1}= R_{\rm sh.}/10\ {\rm pc}$. We can estimate a minimum repeating timescale by using the 
 typical separation in angle between chunks of $\theta_{\rm sep.}=(4/N_{\rm c}^{1/2})$ which gives
  
   \begin{equation}
  \Delta t_{\rm min}= \frac{R_{\rm sh.}(1-\cos{\theta_{\rm sep.}})}{c}\simeq \frac{R_{\rm sh.}\theta_{\rm sep.}^2}{2c}\sim 0.23\ {\rm hours}\times \frac{R_{\rm sh., 1}}{N_{\rm c, 7}}\ .
  \end{equation}
  
  Assuming the refraction process is Poissonian in nature (i.e. when $P_{\rm T, 1}>1$) we can estimate the probability of detecting $k$ bursts during $ \Delta t_{\rm max}$
  as $P(k)= \frac{(P_{\rm T, 1})^k e^{-P_{\rm T, 1}}}{k!}$ which has a peak at $k\sim 5$ if $P_{\rm T, 1}\sim 1$.
  To explain clustered events (like observed repeating FRBs) we must appeal to coherence inhomogeneities in the refracting shell 
   capable of refracting adjacent beams towards the observer. Defining $\theta_{\rm coh.}$ as the angular scale 
   of the coherence scale, to get $\sim 10$ FRBs within a time interval of a  few times $\Delta t_{\rm min}$,
   the coherence angular scale must be a few times $\theta_{\rm sep.}$. The corresponding coherence scale is $R_{\rm sh.}\theta_{\rm coh.}= R_{\rm sh.}\times 6/\Gamma_{\rm QN}\sim
   4000$ AU.

\end{appendix}



\newpage

%
%
\setcounter{figure}{0}

\begin{figure*}[t!]
\centering
\includegraphics[scale=0.15]{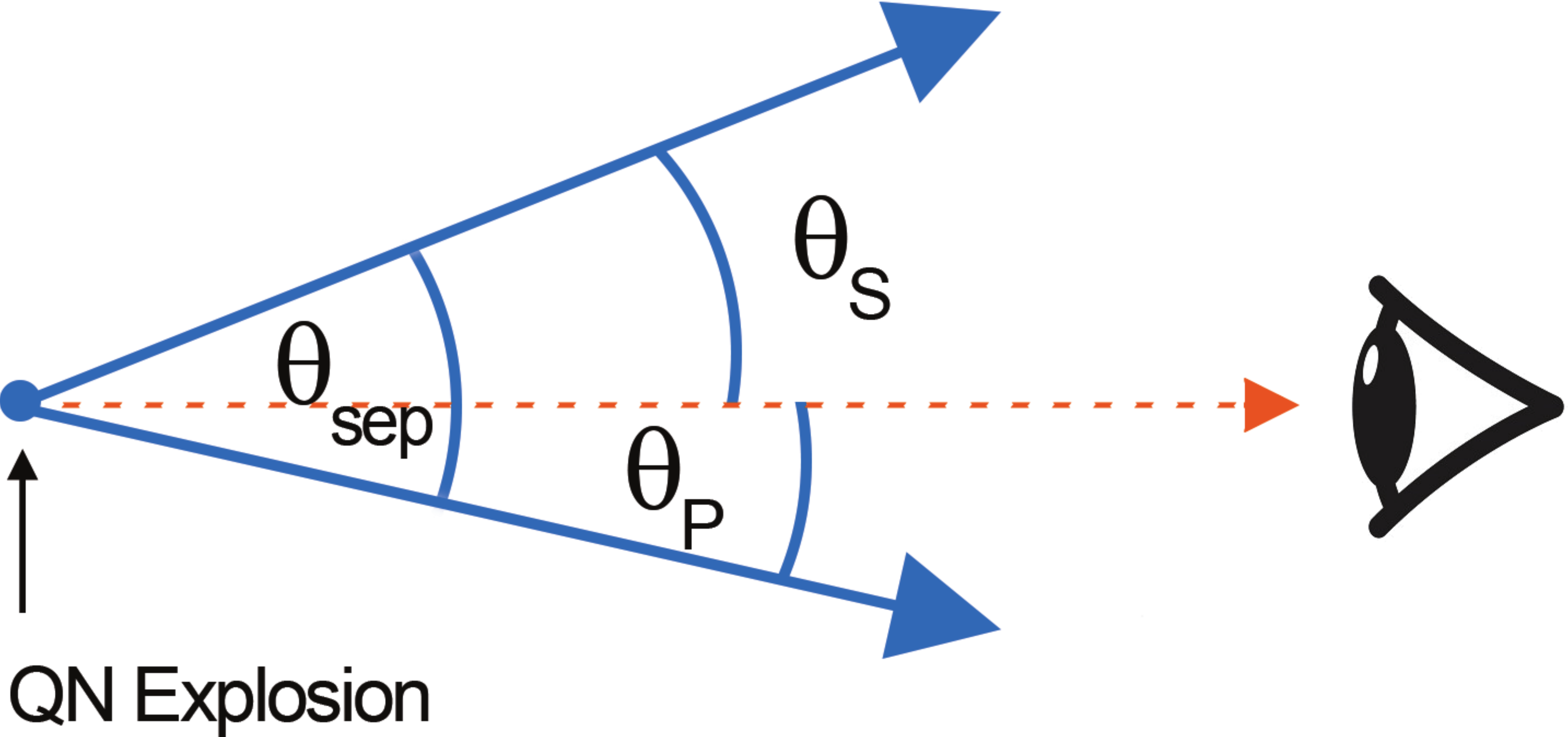}
\caption{{\bf Viewing geometry}: Illustration of the velocity vectors of two chunks, the primary chunk at $\theta_{\rm P}$ and the 
secondary chunk at $\theta_{\rm S}=\theta_{\rm sep.}-\theta_{\rm P}$.}
\label{fig:stencil}
\end{figure*}

\newpage
 
 \begin{figure*}[t!]
\centering
\includegraphics[scale=0.4]{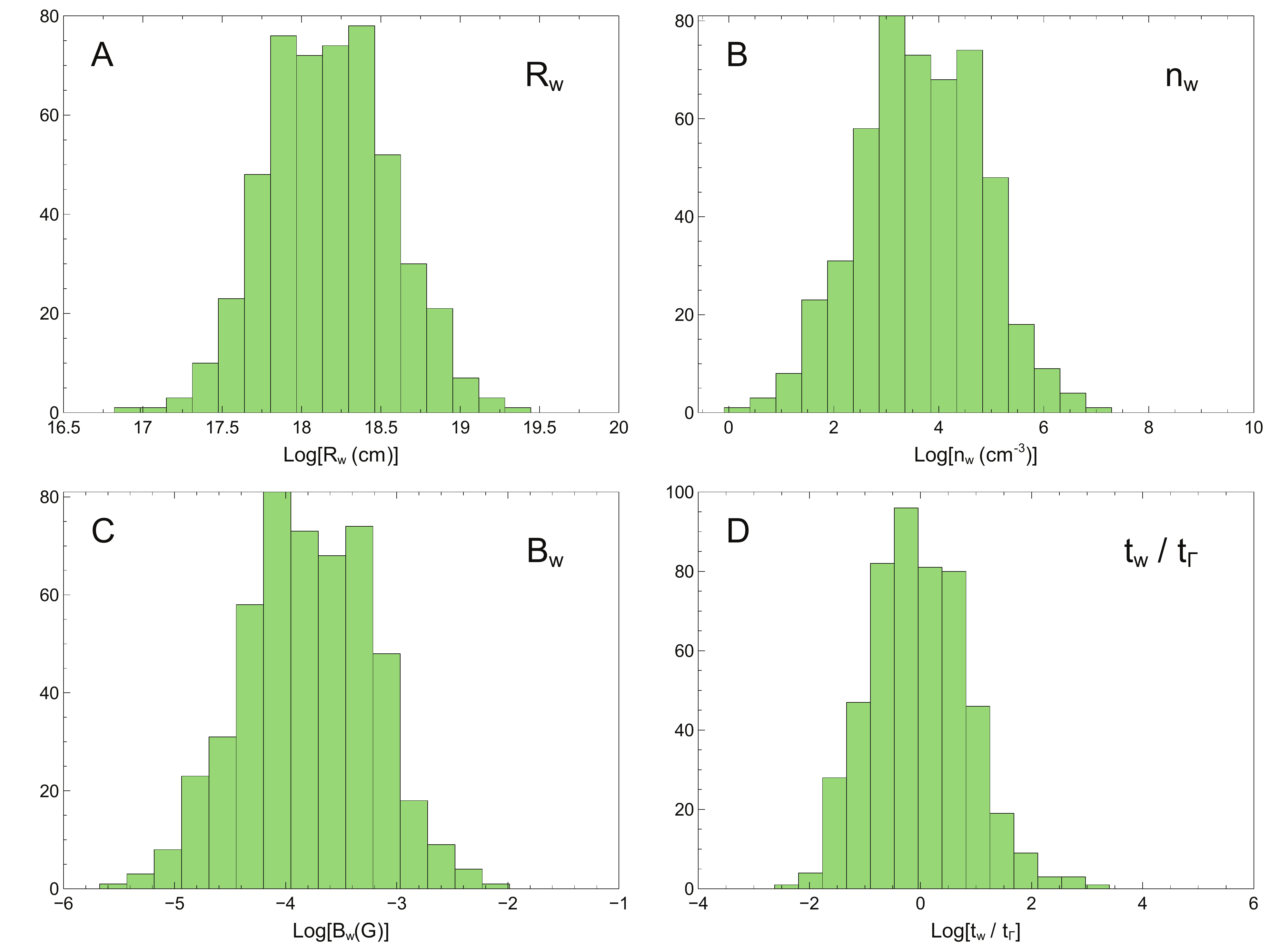}
\caption{{\bf PWN-SN shell (``the wall") properties}: Distributions of the wall's parameters (radius $R_{\rm w}$, density $n_{\rm w}$, magnetic field
$B_{\rm w}$ and thickness parameter $t_{\rm w}/t_{\Gamma}$) for a distribution of $t_{\rm QN}$ (i.e. $B_{\rm NS}$) in our model. We use Rice's rule for binning}
 \label{fig:Wall-distributions}
\end{figure*}

\newpage
\begin{figure*}[t!]
\centering
\includegraphics[scale=0.4]{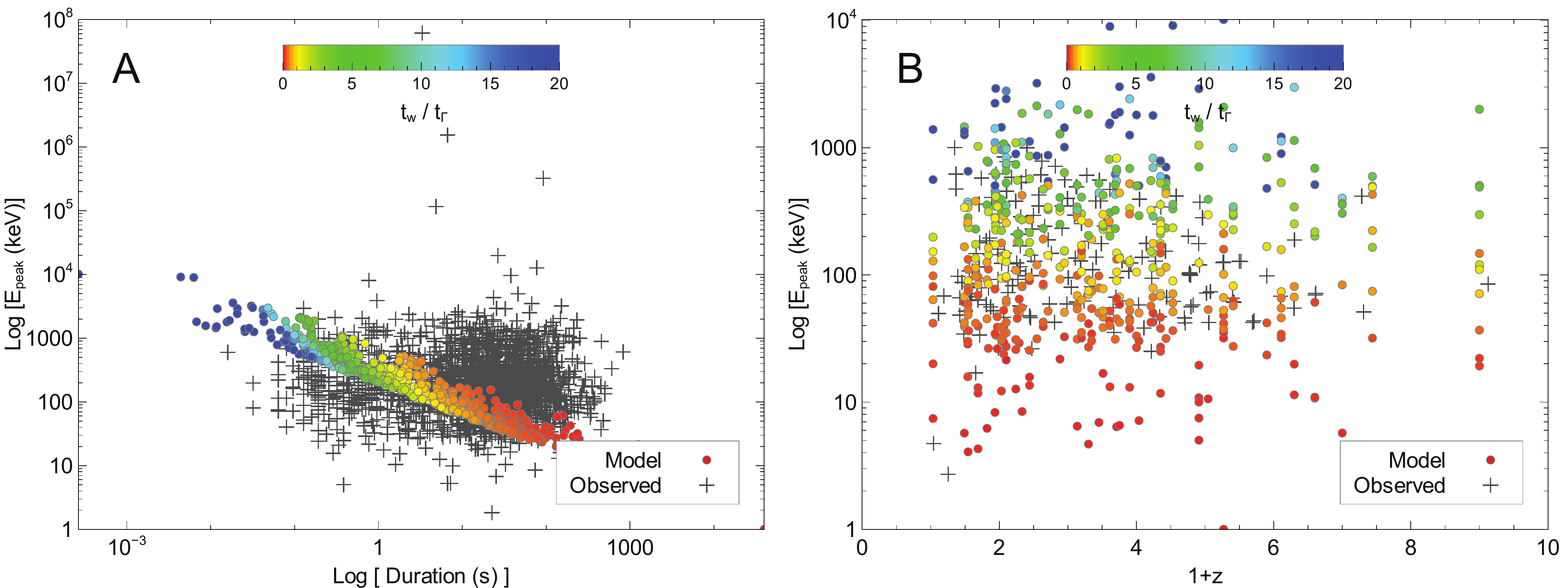}
\caption{{\bf Comparison of our analytical model (\S ~ \ref{sec:yonetoku-amati-theory}) to LGRB data}: Shown are the single 
thin (i.e. $t_{\rm w}\le 3t_{\Gamma}$) and thick (i.e. $t_{\rm w} > 3t_{\Gamma}$) wall  runs compared to observed properties of LGRBs (pluses; from \citealt{ghirlanda_2009}).  The color palette shows the range of the wall thickness parameter $t_{\rm w}/t_{\Gamma}$. 
 There are 500 analytical runs (one per dot) using  fiducial parameters 
given in Table \ref{table:parameters}.  Each run is obtained by varying $t_{\rm QN}$ (i.e. $B_{\rm NS}$) and the viewing angle $\theta_{\rm P}$
  with ($0 < \theta_{\rm P} < 2\times 10^{-3}/N_{\rm c, 6}^{1/2}$). 
  The left panel shows the duration 
 compared to the observed $t_{90}$ (data from \url{https://swift.gsfc.nasa.gov/archive/grb_table}) while the
  right panel shows the photon peak energy versus the source redshift. The redshift was obtained by cross-referencing the
 LGRBs from \citet{ghirlanda_2009} with data in \url{https://swift.gsfc.nasa.gov/archive/grb_table}. 
 The redshift $z$ for each run is obtained by randomly selecting a GRB
 from a global list of 350 GRBs (those with known redshifts) from \url{https://swift.gsfc.nasa.gov/archive/grb_table}.\\}
 \label{fig:amati-single-theory1}
\end{figure*}

\begin{figure*}[t!]
\centering
\includegraphics[scale=0.4]{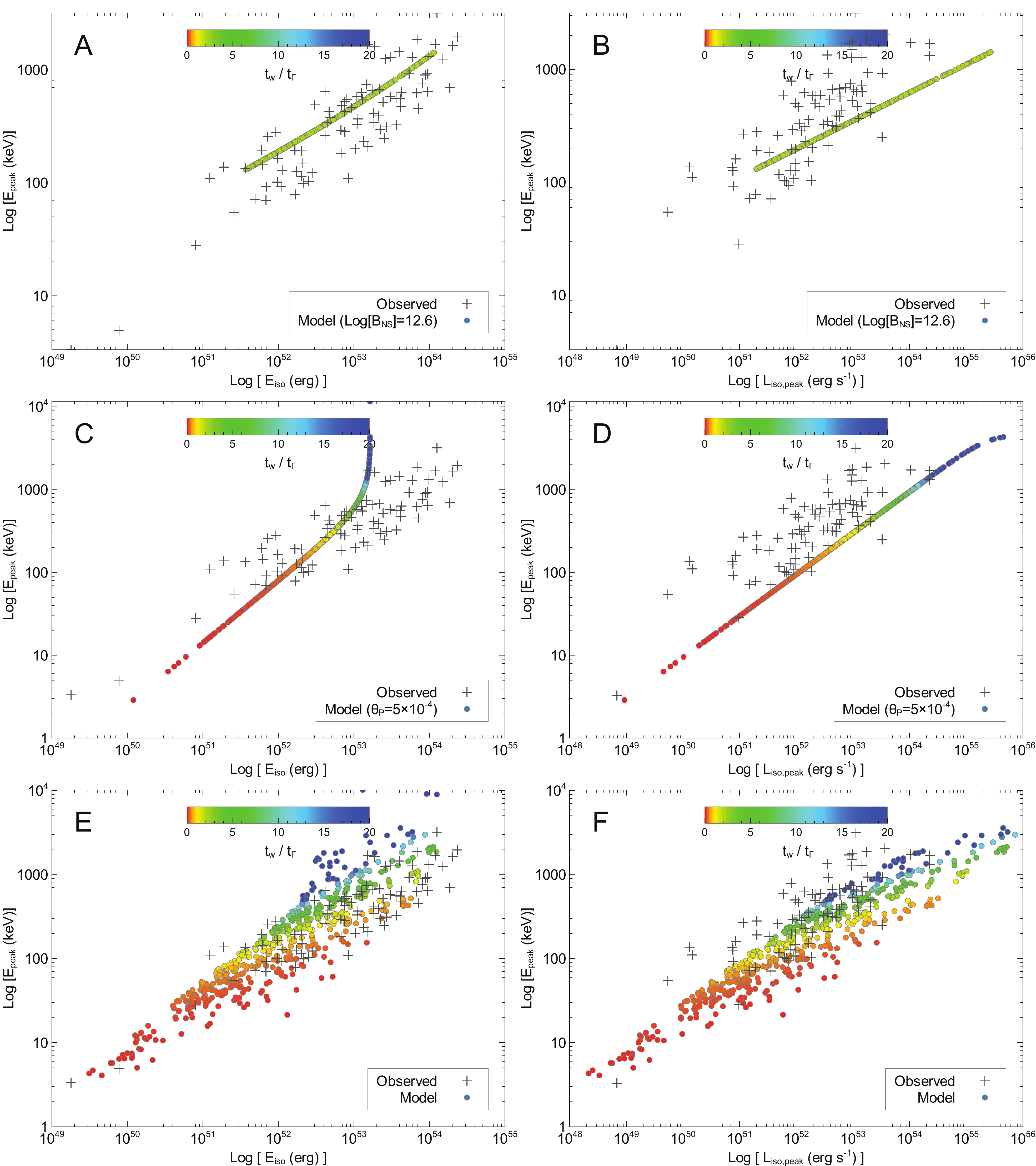}
\caption{{\bf Amati plot (left panels) and Yonetoku plot (right panels) for the single wall analytical model (\S ~ \ref{sec:yonetoku-amati-theory})}: 500 runs (the dots; the pluses are data from \citealt{ghirlanda_2009}) of the analytical model (including deceleration of chunk for large values of $t_{\rm w}/t_{\Gamma}$).  For each run the primary chunk passes through a single wall. 
Each dot is generated by varying $t_{\rm QN}$ (i.e. $B_{\rm NS}$) and $\theta_{\rm P}$ with ranges similar to those used in Figure \ref{fig:amati-single-theory1}. 
Other parameters are set to their fiducial values (see Table \ref{table:parameters}). 
The redshift $z$ for each simulation point is obtained by randomly selecting a GRB
 from a global list of 350 GRBs (those with known redshifts) from \url{https://swift.gsfc.nasa.gov/archive/grb_table}.
  {\bf Top panels}: Effects of varying the viewing angle $\theta_{\rm P}$ for a fixed $B_{\rm NS}$. {\bf Middle panels}: Effects of varying $B_{\rm NS}$ for a fixed viewing angle $\theta_{\rm P}$. {\bf Bottom panels}: Effects of varying both the viewing angle $\theta_{\rm P}$ and the NS magnetic field $B_{\rm NS}$.\\
}
\label{fig:amati-single-theory2}
\end{figure*}

\begin{figure*}[t!]
\centering
\includegraphics[scale=0.5]{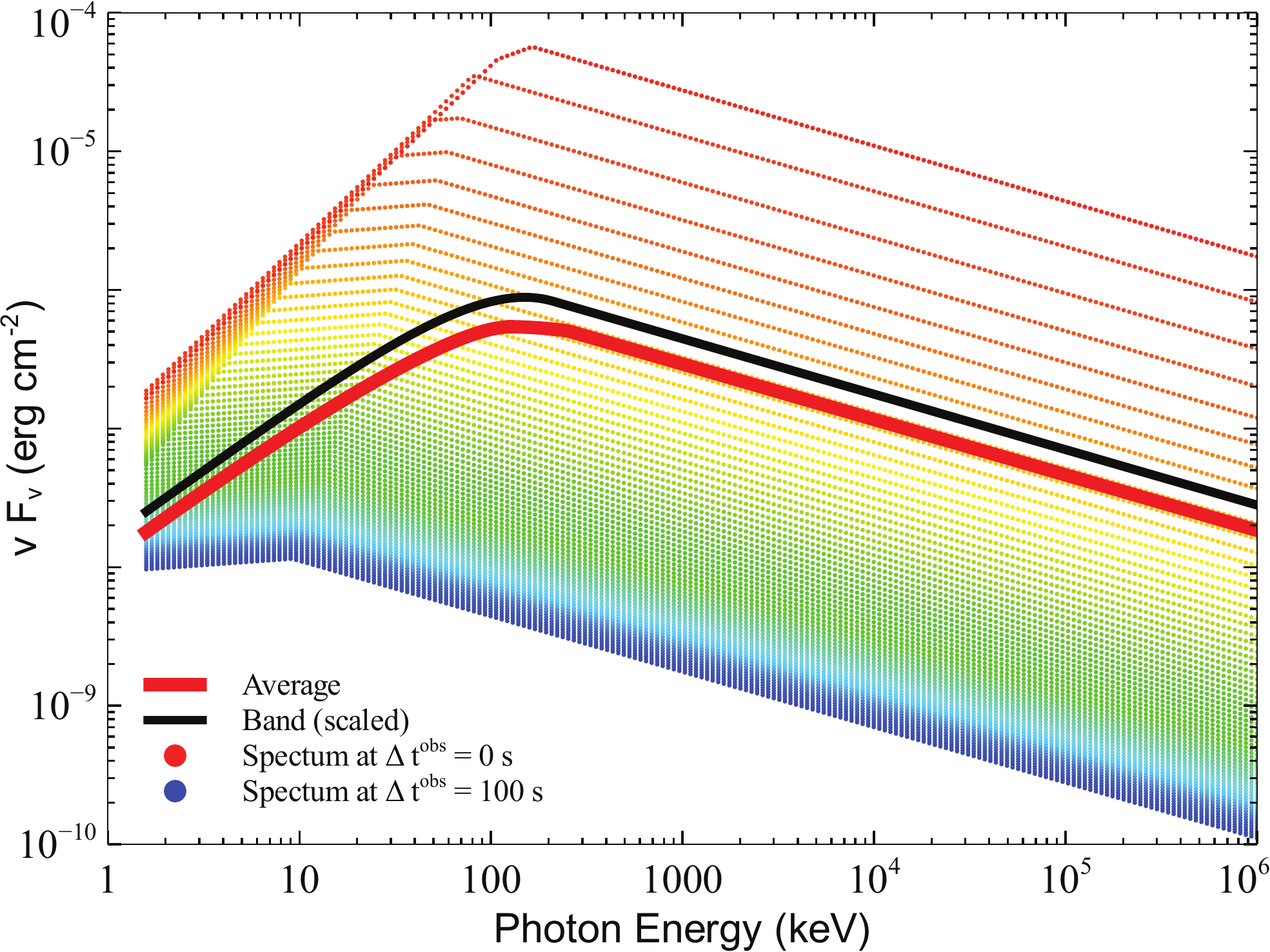}
\includegraphics[scale=0.5]{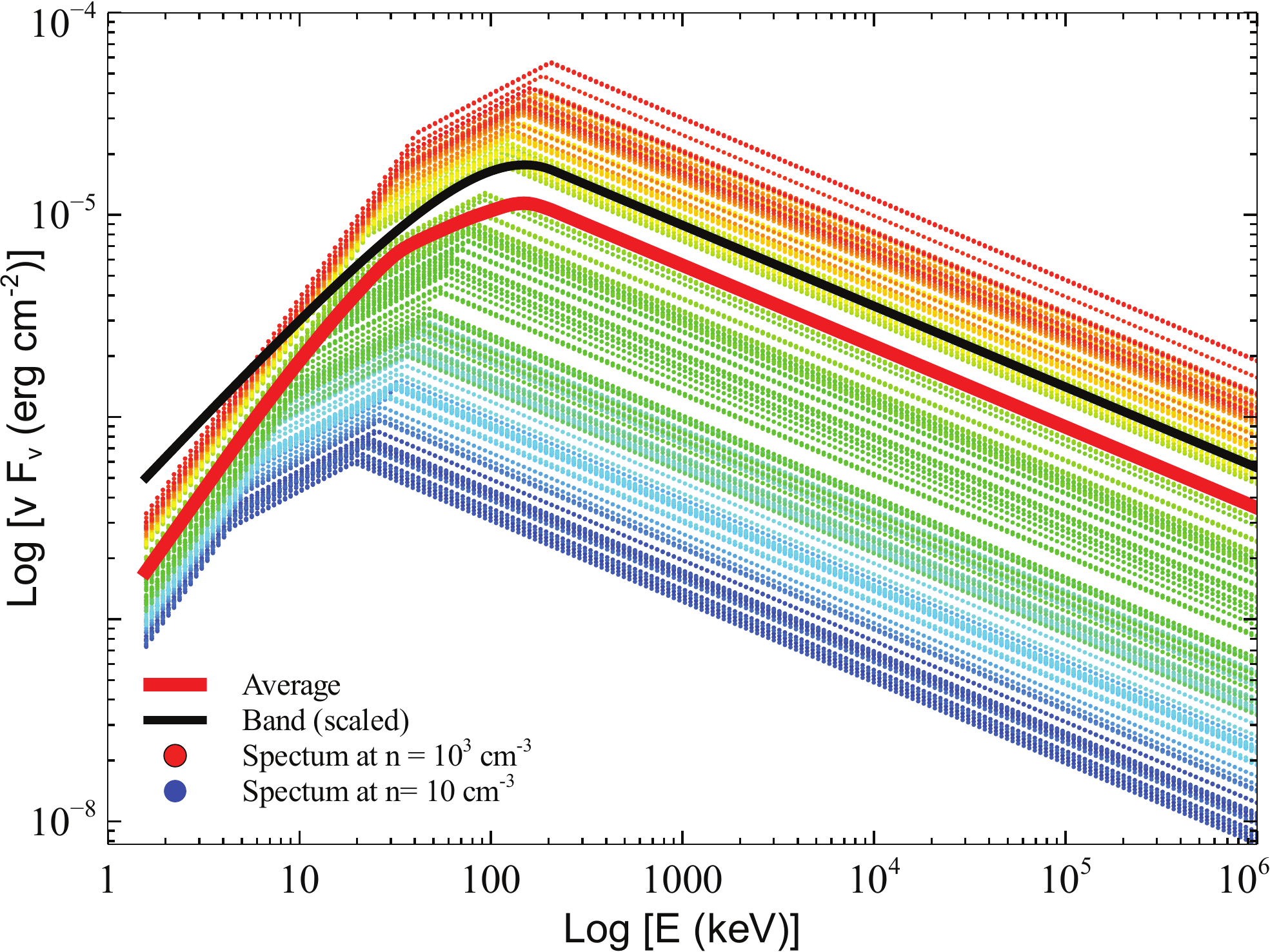}
\caption{{\bf The Band-like spectrum in our model}: {\bf Top panel}: A single chunk going through one high density wall ($n_{\rm w} = 10^5$ cm$^{-3}$, $R_{\rm w} = 10^{18}$ cm
and $\Delta R_{\rm w} = 10^{17}$ cm).  Different colors correspond to increasing time. The thick red curve is the time-averaged model
spectrum. The thick black curve is a generic Band function.
{\bf Bottom panel}: A single chunk going through multiple thin filaments ($\sim$ 120) with density randomly drawn between $n_{\rm F} = 10$ cm$^{-3}$ and $n_{\rm F} = 1000$ cm$^{-3}$.  The individual spectra are taken at the beginning of each filament.\\}
 \label{fig:spectrum-band}
\end{figure*}


\begin{figure}[ht!]
\centering
\includegraphics[scale=0.65]{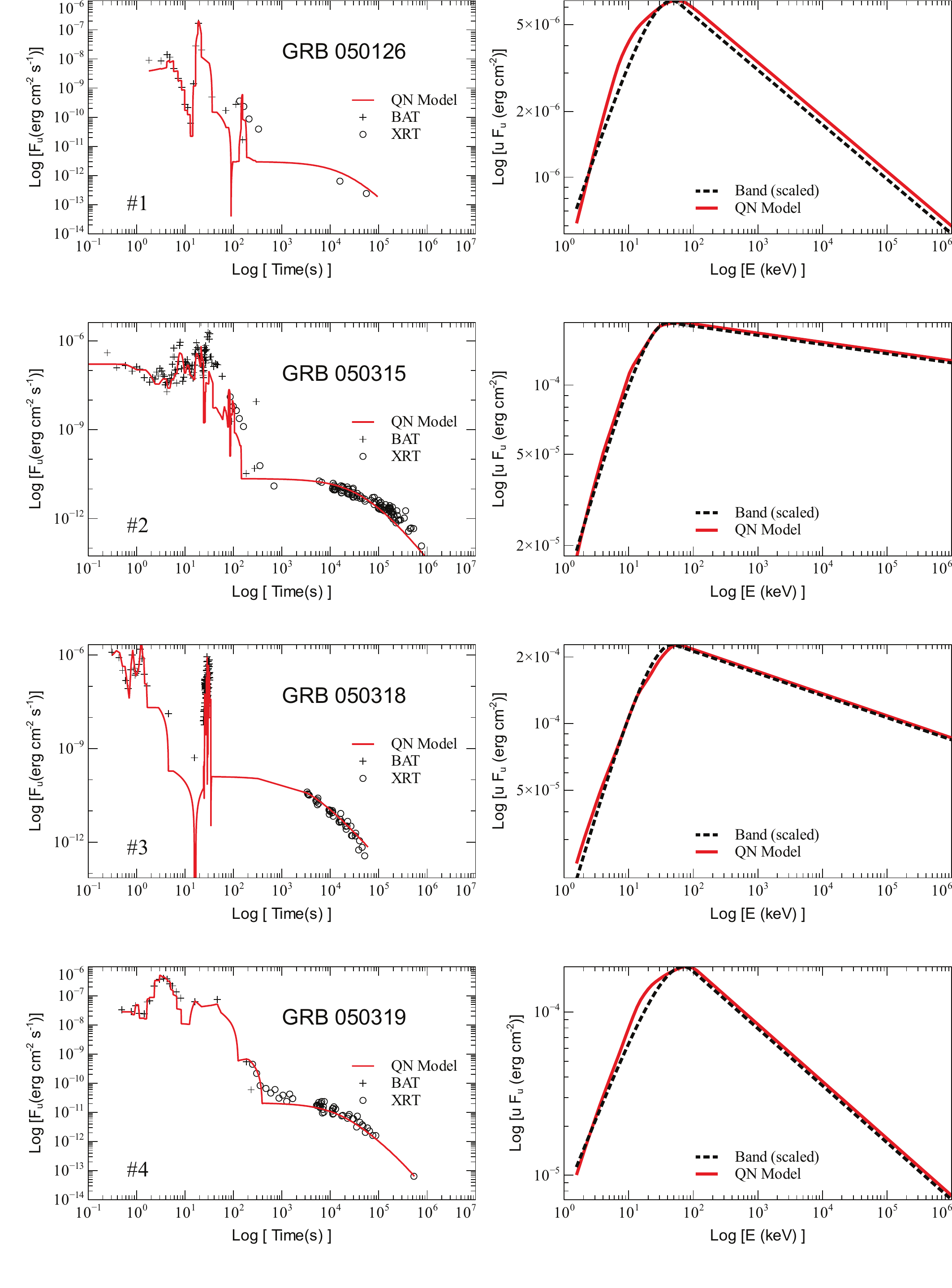}
\caption{{\bf GRB simultaneous light-curve and spectrum fits}: The  light-curve (left panel) and spectrum (right panel) fits for each of the 48 LGRBs listed in Table \ref{table:lcfits}.  For the light-curves, the BAT data is extrapolated to the XRT band and shown as black crosses.  The XRT data is shown as open circles.  The red line is the QN model.  For the spectra, the red line is the QN model whereas the black dashed line is the best-fit Band function to the observed spectrum from \citet{yonetoku_2010}.}
 \label{figure:lc0}
\end{figure}

\renewcommand{\thefigure}{\arabic{figure} (Cont.)}
\addtocounter{figure}{-1}
\begin{figure}[ht!] \label{figure:lc1}
\centering
\includegraphics[scale=0.65]{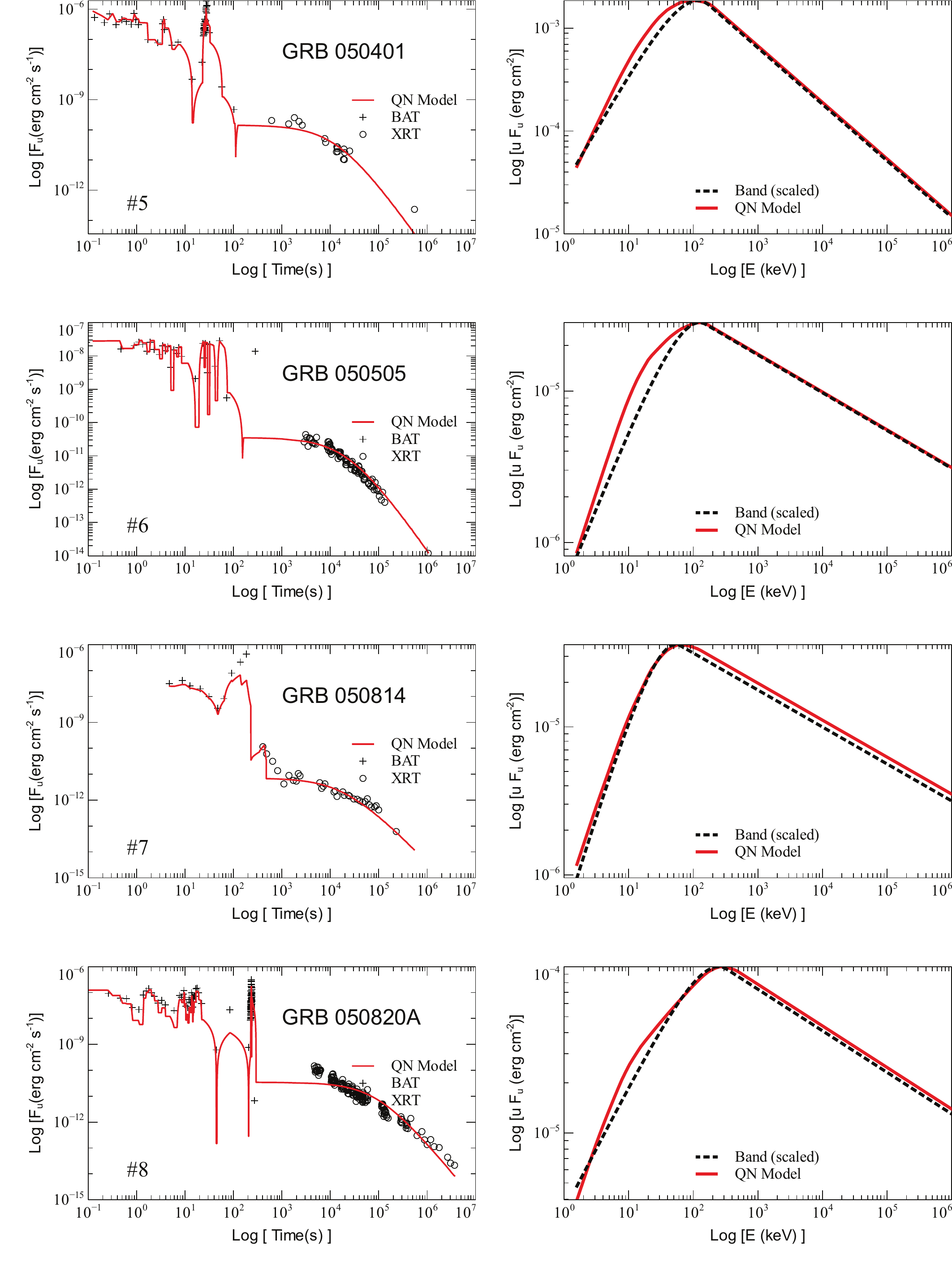}
\caption{}
\end{figure}
\renewcommand{\thefigure}{\arabic{figure}}

\renewcommand{\thefigure}{\arabic{figure} (Cont.)}
\addtocounter{figure}{-1}
\begin{figure}[ht!]
\centering
\includegraphics[scale=0.65]{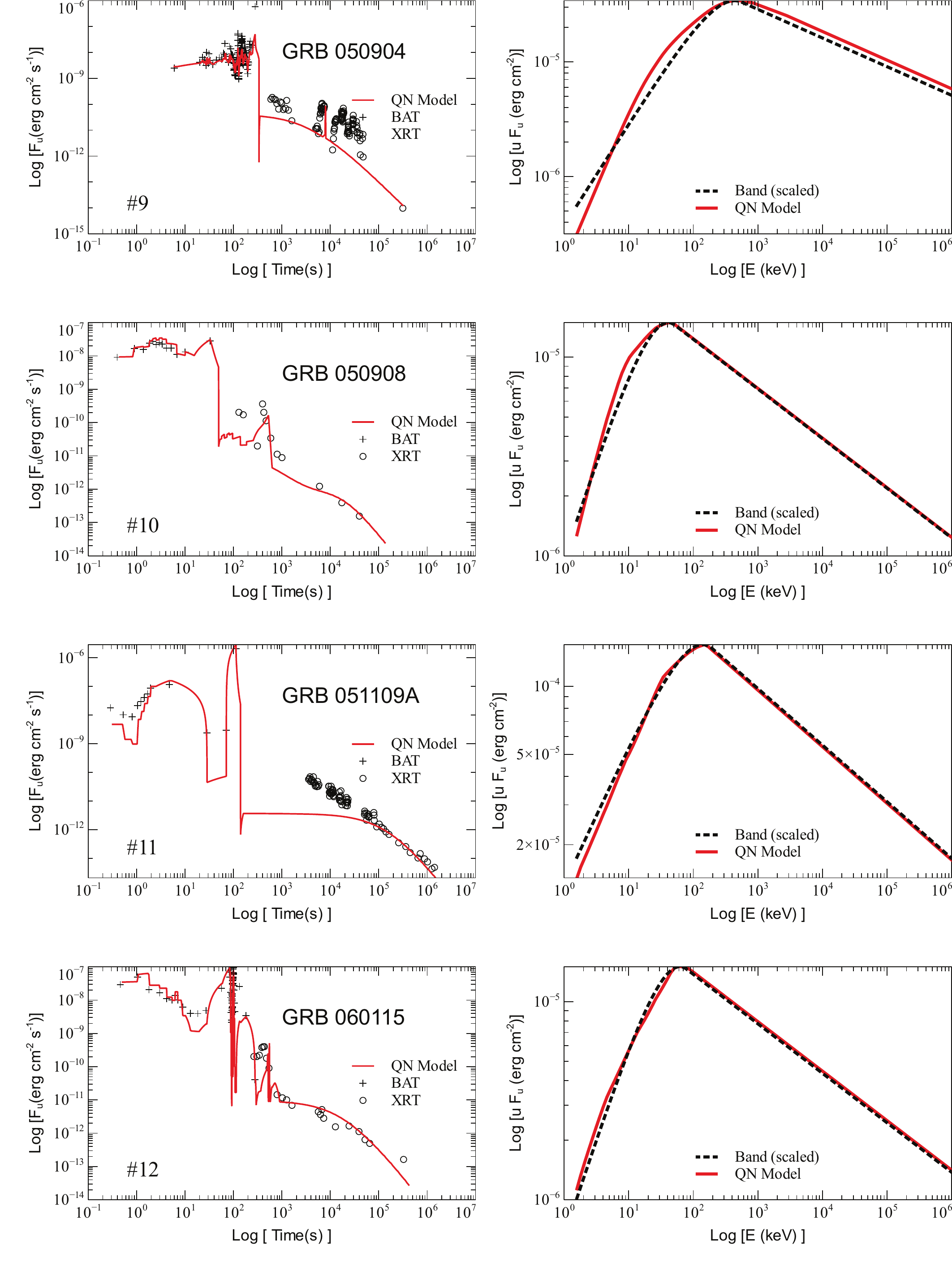}
\caption{}
 \label{figure:lc2}
\end{figure}
\renewcommand{\thefigure}{\arabic{figure}}

\renewcommand{\thefigure}{\arabic{figure} (Cont.)}
\addtocounter{figure}{-1}
\begin{figure}[ht!]
\centering
\includegraphics[scale=0.65]{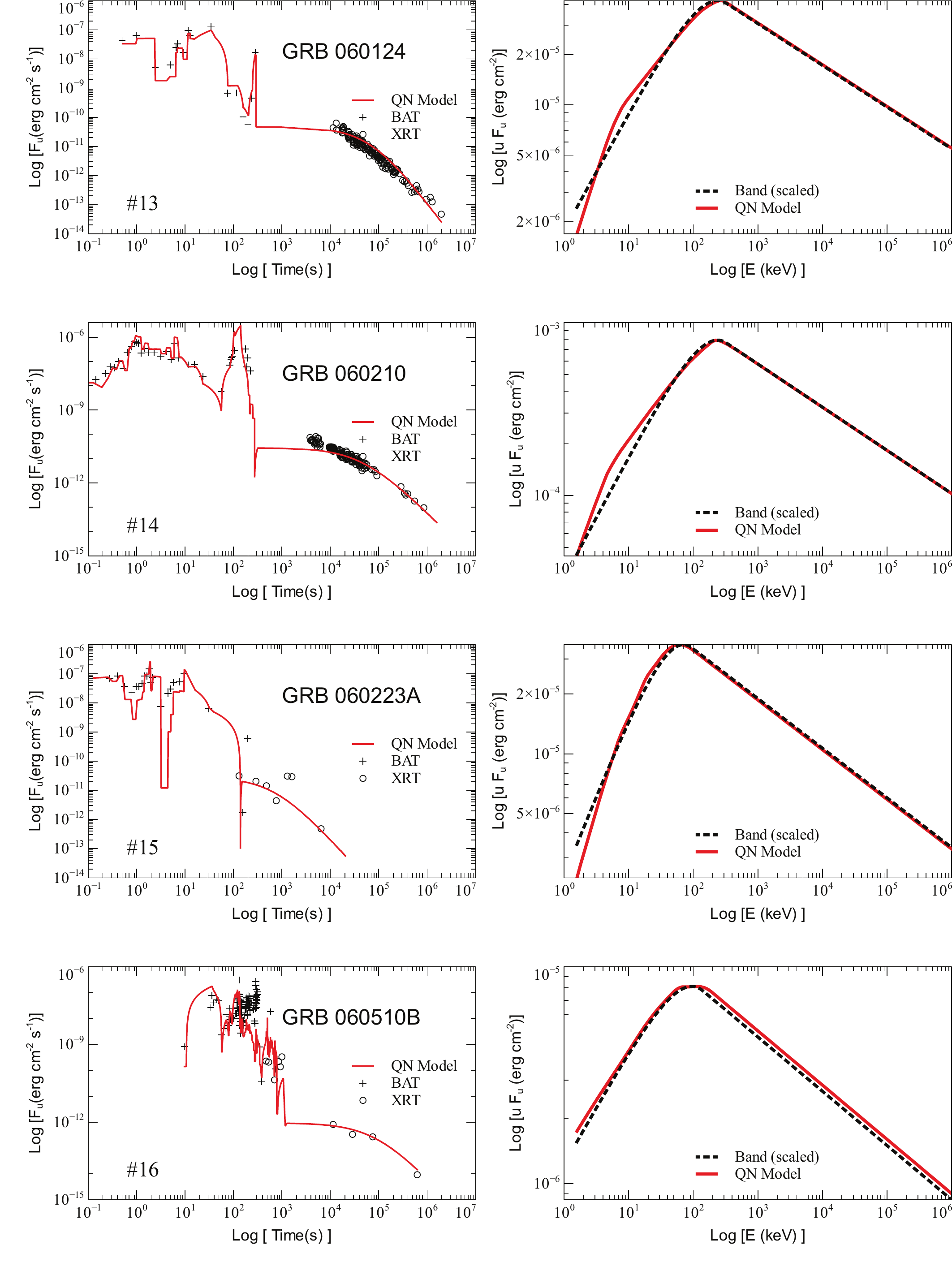}
\caption{}
 \label{figure:lc3}
\end{figure}
\renewcommand{\thefigure}{\arabic{figure}}

\renewcommand{\thefigure}{\arabic{figure} (Cont.)}
\addtocounter{figure}{-1}
\begin{figure}[ht!]
\centering
\includegraphics[scale=0.65]{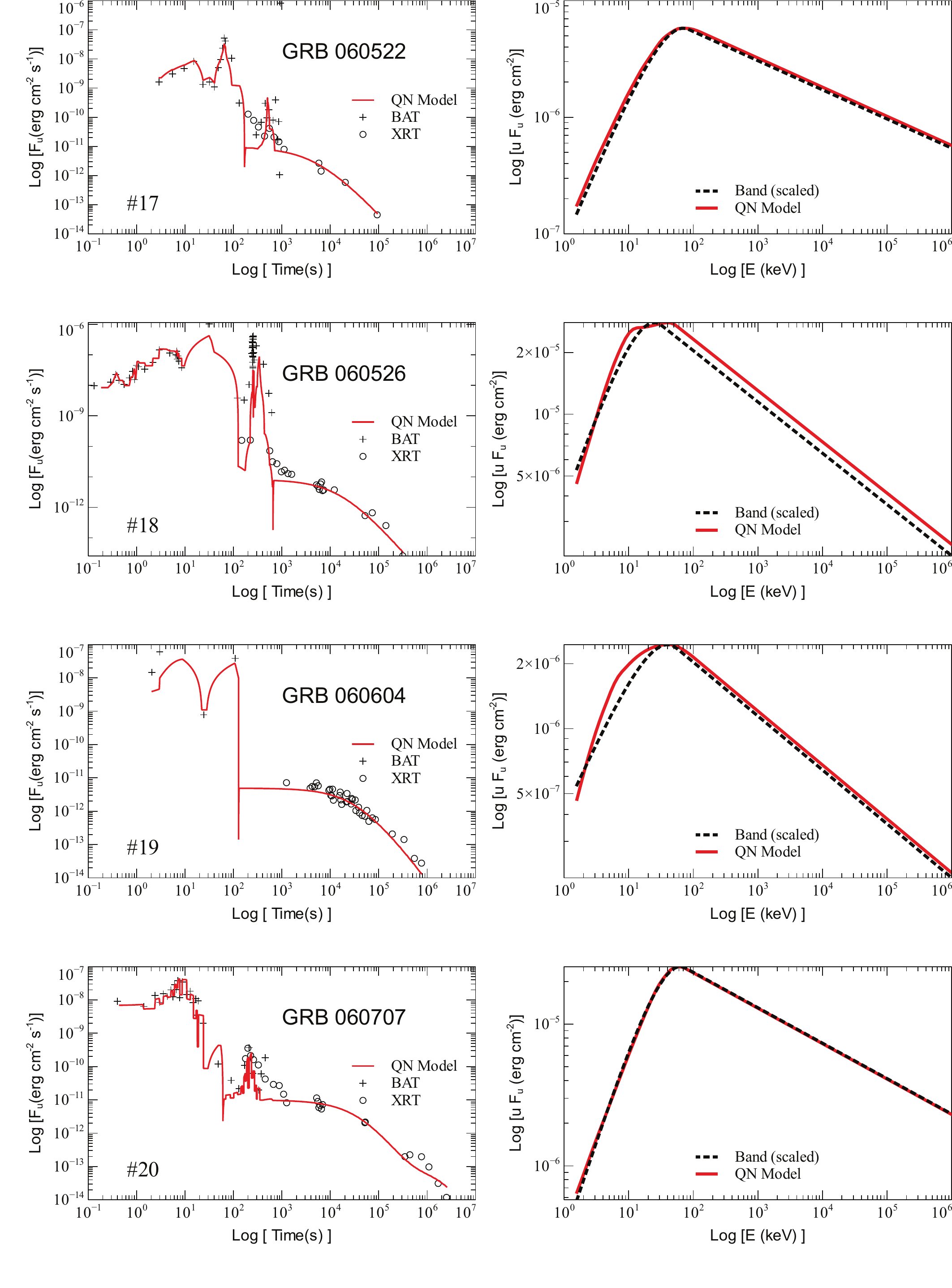}
\caption{}
 \label{figure:lc4}
\end{figure}
\renewcommand{\thefigure}{\arabic{figure}}

\renewcommand{\thefigure}{\arabic{figure} (Cont.)}
\addtocounter{figure}{-1}
\begin{figure}[ht!]
\centering
\includegraphics[scale=0.65]{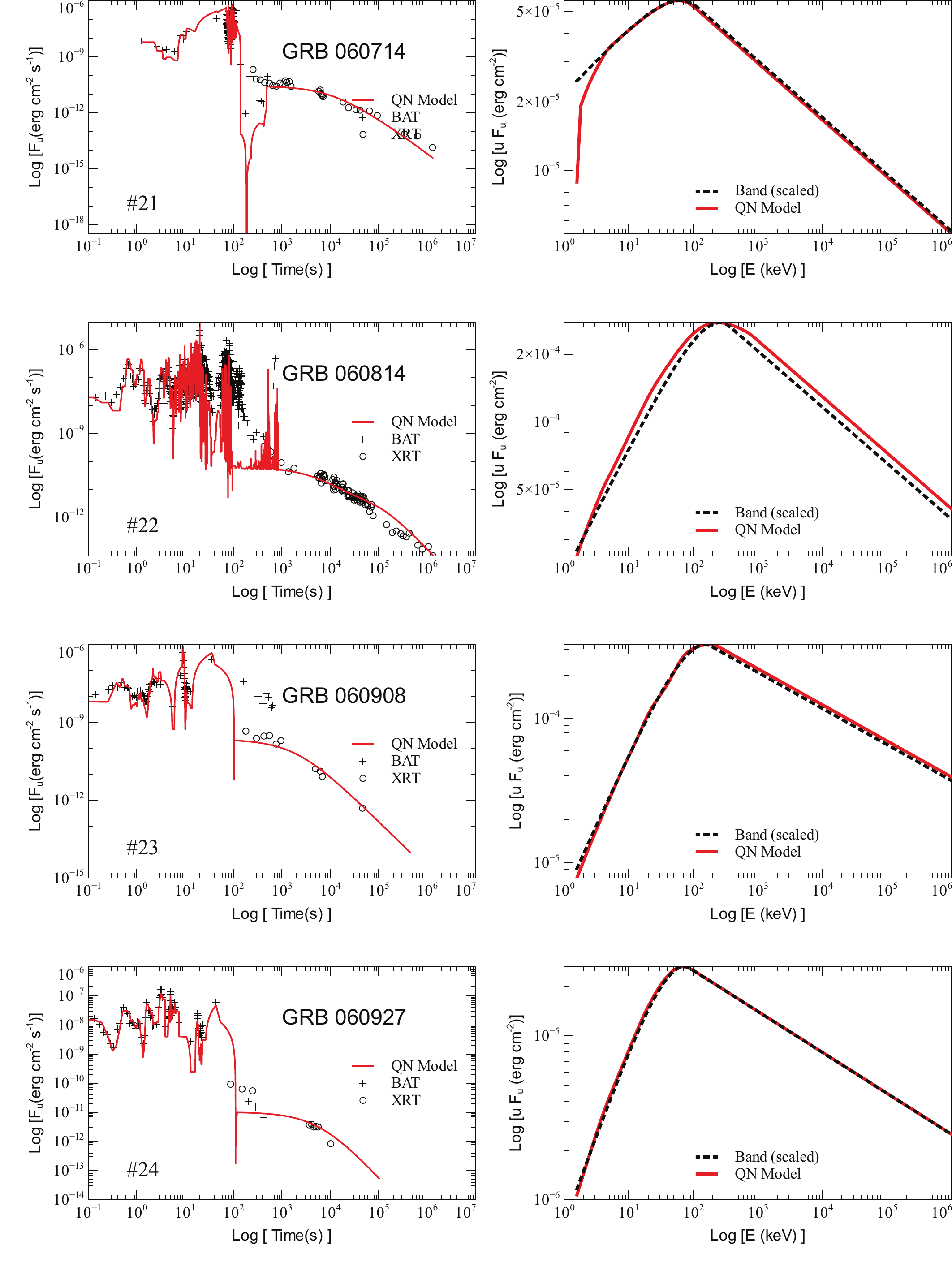}
\caption{}
 \label{figure:lc5}
\end{figure}
\renewcommand{\thefigure}{\arabic{figure}}

\renewcommand{\thefigure}{\arabic{figure} (Cont.)}
\addtocounter{figure}{-1}
\begin{figure}[ht!]
\centering
\includegraphics[scale=0.65]{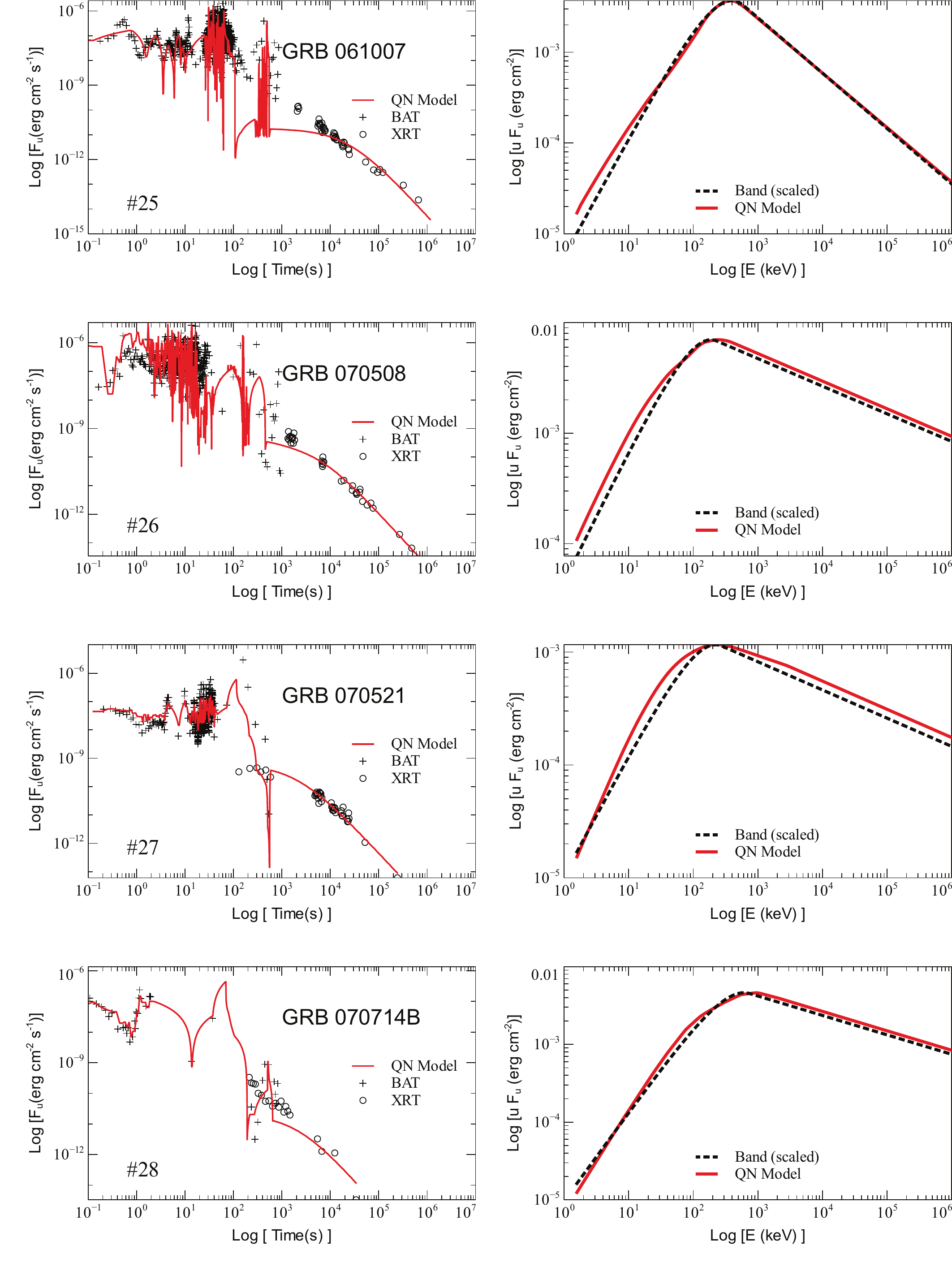}
\caption{}
 \label{figure:lc6}
\end{figure}
\renewcommand{\thefigure}{\arabic{figure}}

\renewcommand{\thefigure}{\arabic{figure} (Cont.)}
\addtocounter{figure}{-1}
\begin{figure}[ht!]
\centering
\includegraphics[scale=0.65]{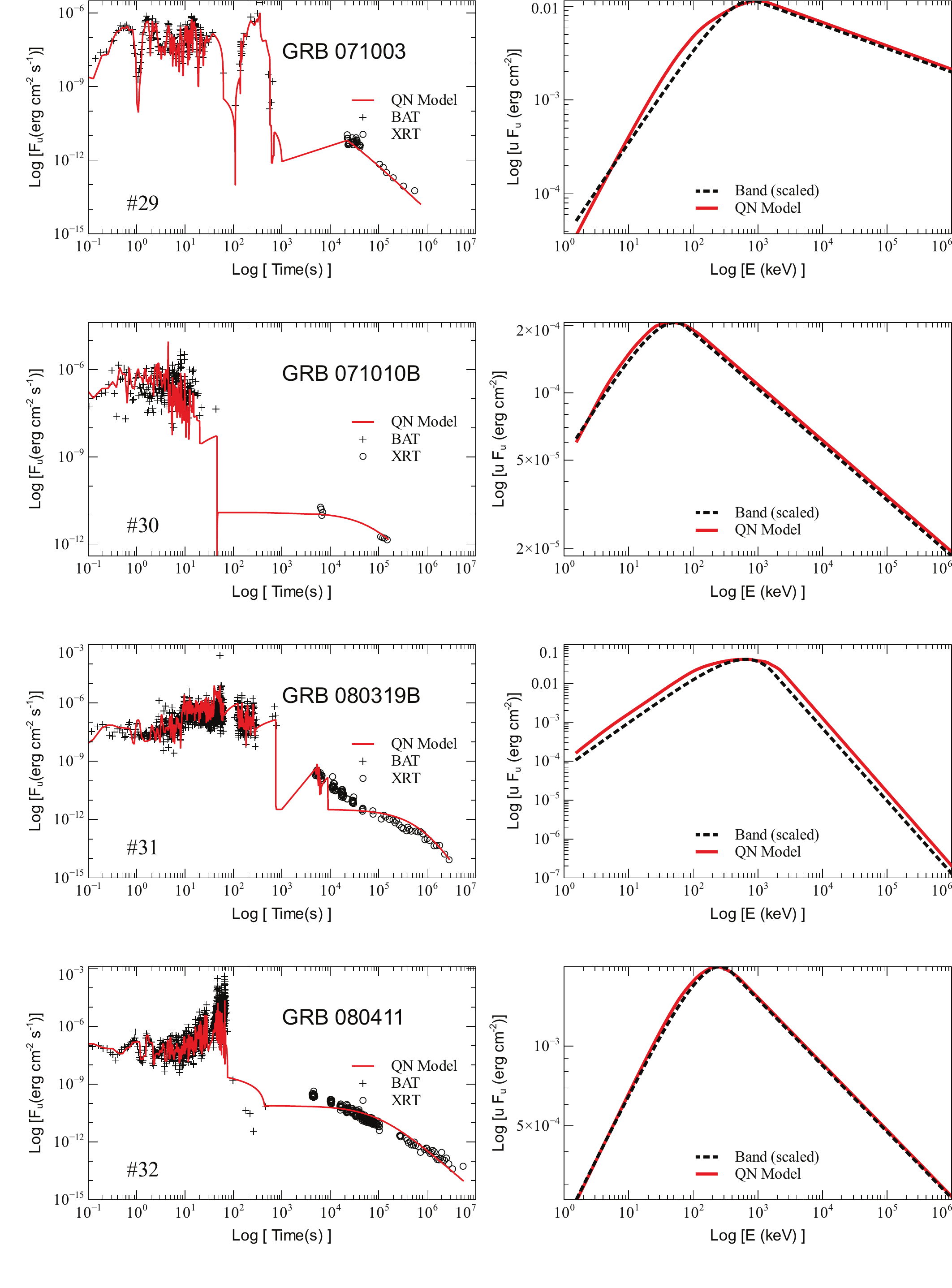}
\caption{}
 \label{figure:lc7}
\end{figure}
\renewcommand{\thefigure}{\arabic{figure}}

\renewcommand{\thefigure}{\arabic{figure} (Cont.)}
\addtocounter{figure}{-1}
\begin{figure}[ht!]
\centering
\includegraphics[scale=0.65]{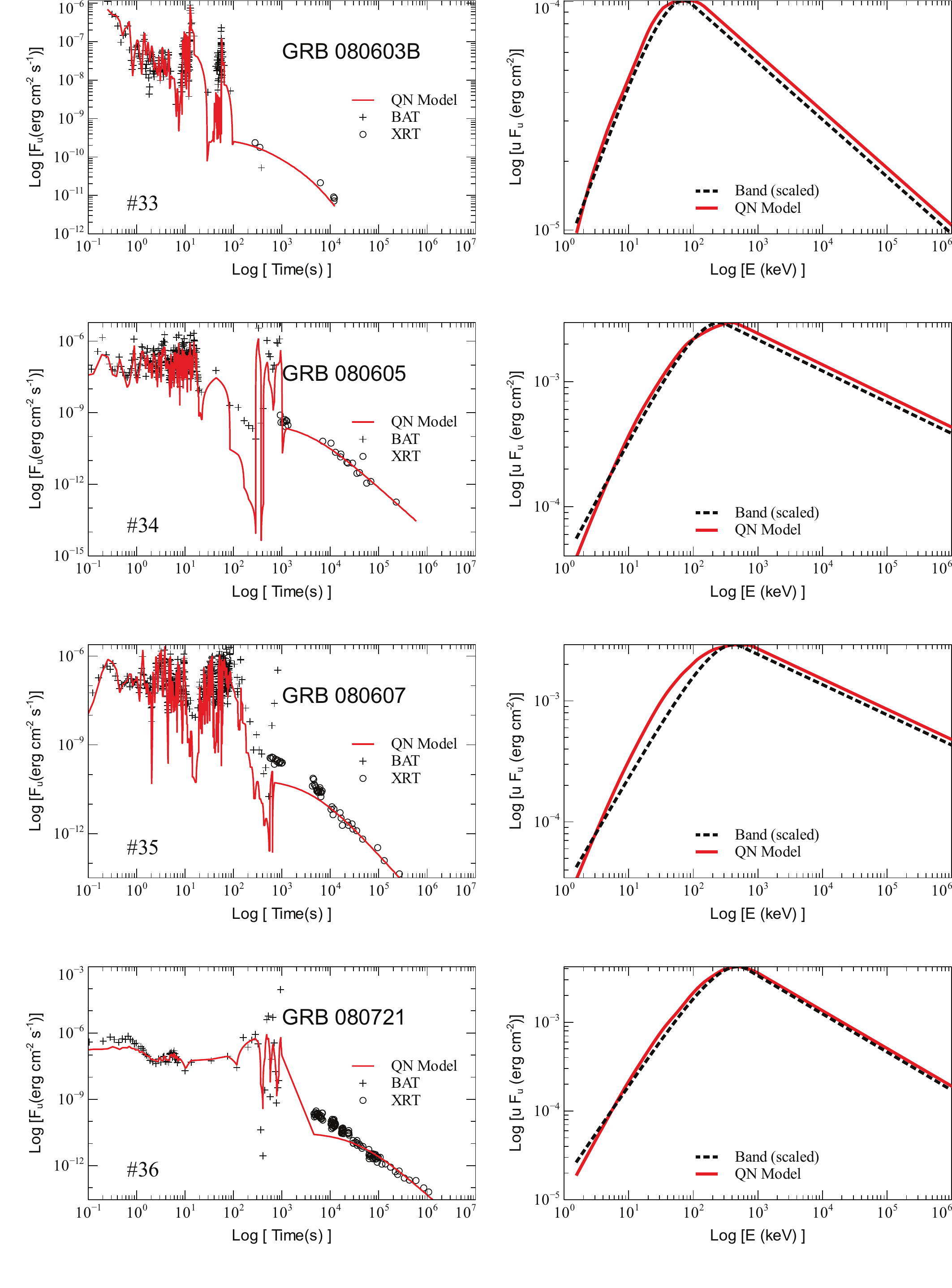}
\caption{}
 \label{figure:lc8}
\end{figure}
\renewcommand{\thefigure}{\arabic{figure}}

\renewcommand{\thefigure}{\arabic{figure} (Cont.)}
\addtocounter{figure}{-1}
\begin{figure}[ht!]
\centering
\includegraphics[scale=0.65]{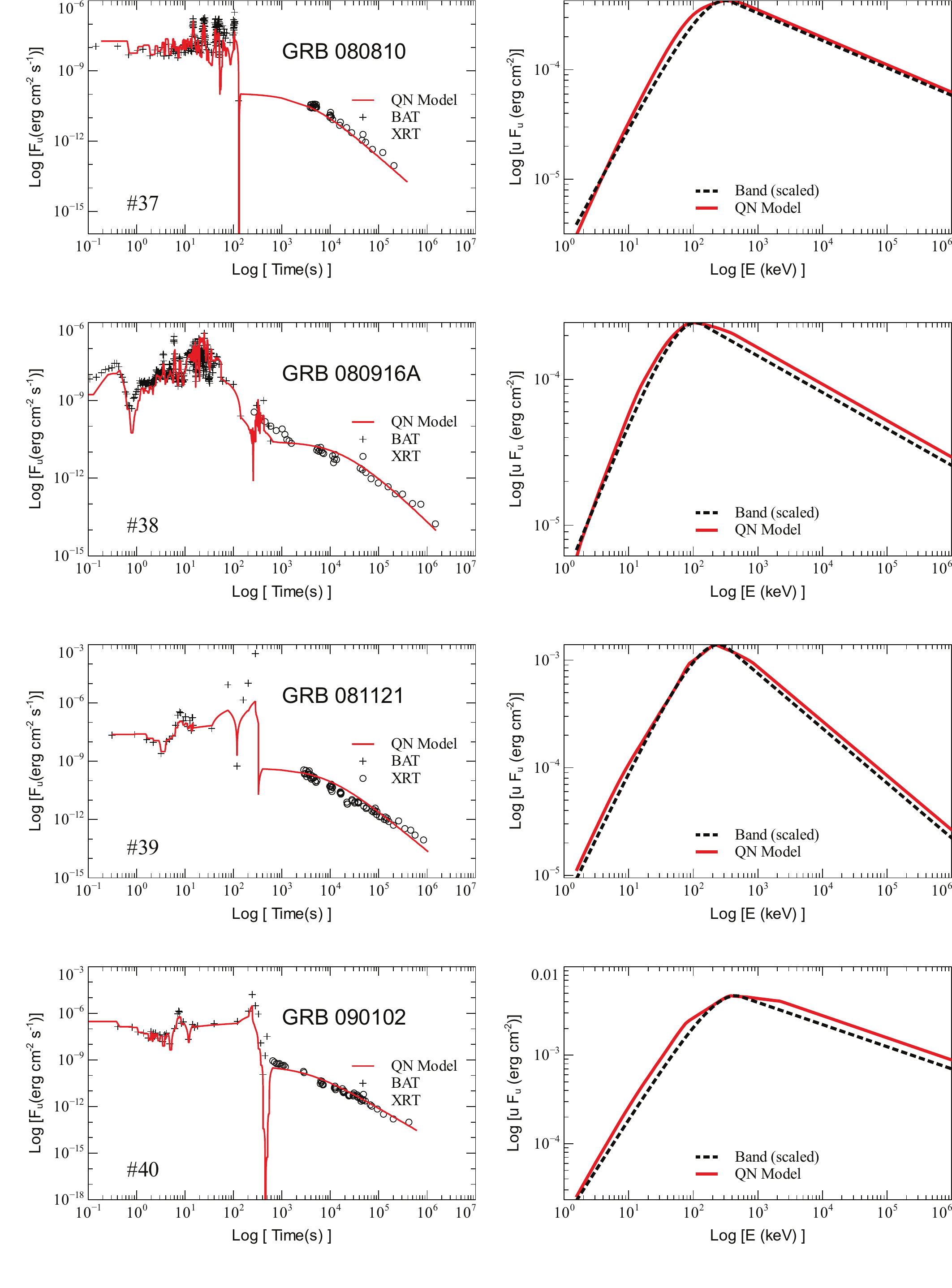}
\caption{}
 \label{figure:lc9}
\end{figure}
\renewcommand{\thefigure}{\arabic{figure}}

\renewcommand{\thefigure}{\arabic{figure} (Cont.)}
\addtocounter{figure}{-1}
\begin{figure}[ht!] 
\centering
\includegraphics[scale=0.65]{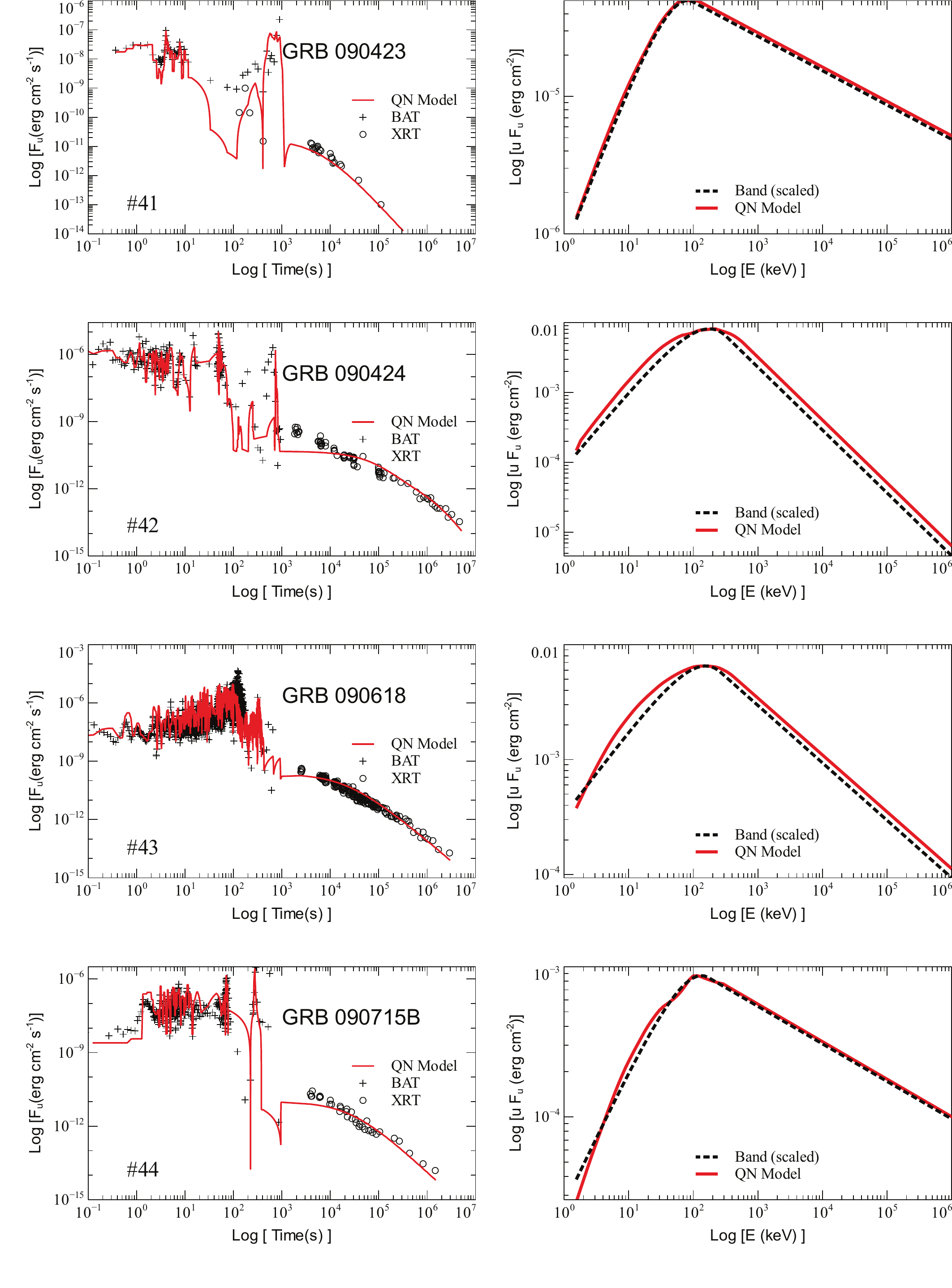}
\caption{}
\label{figure:lc10}
\end{figure}
\renewcommand{\thefigure}{\arabic{figure}}

\renewcommand{\thefigure}{\arabic{figure} (Cont.)}
\addtocounter{figure}{-1}
\begin{figure}[ht!] 
\centering
\includegraphics[scale=0.65]{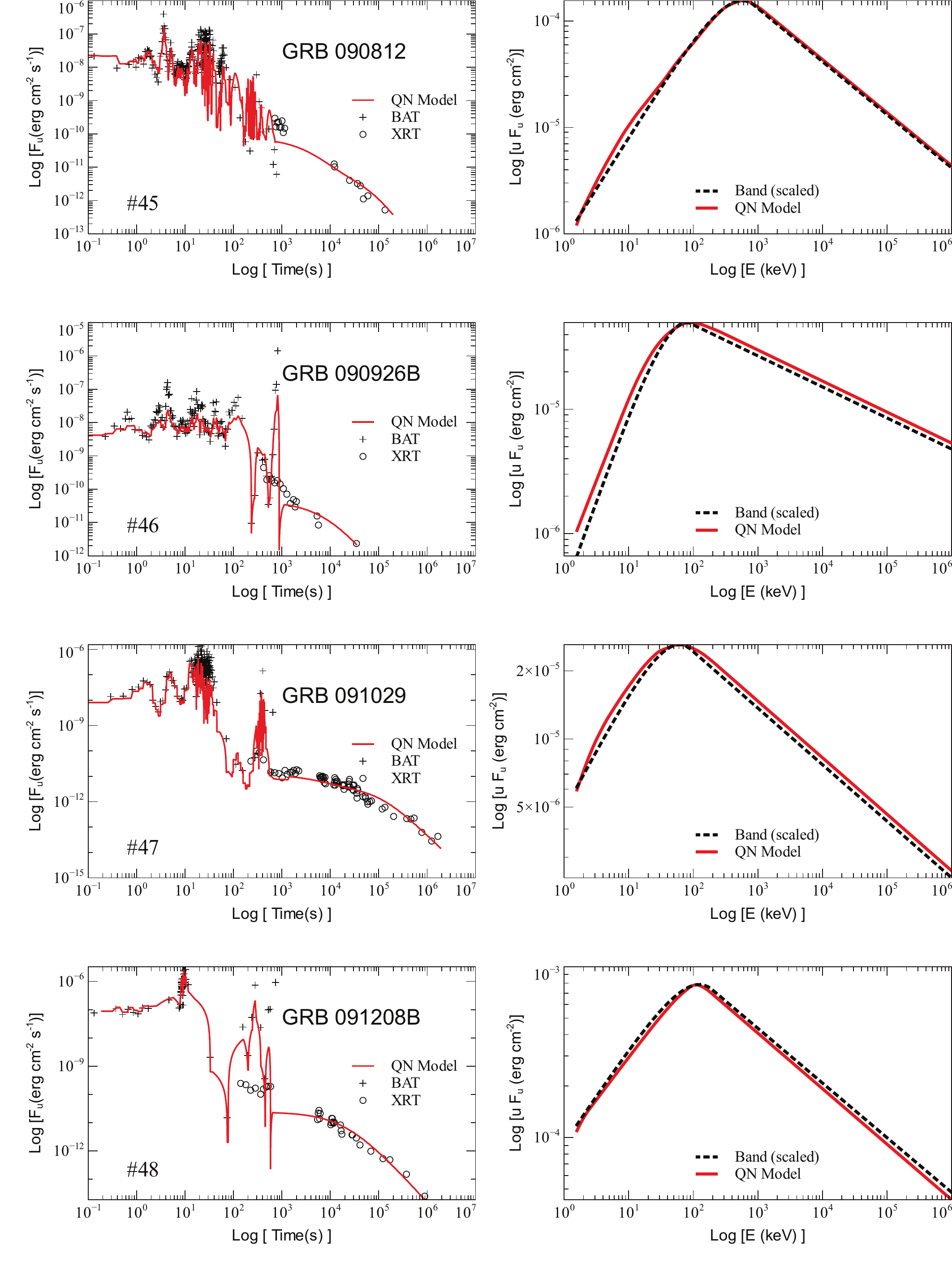}
\caption{}
\label{figure:lc11}
\end{figure}
\renewcommand{\thefigure}{\arabic{figure}}


\begin{figure}[ht!]
\centering
\includegraphics[scale=0.45]{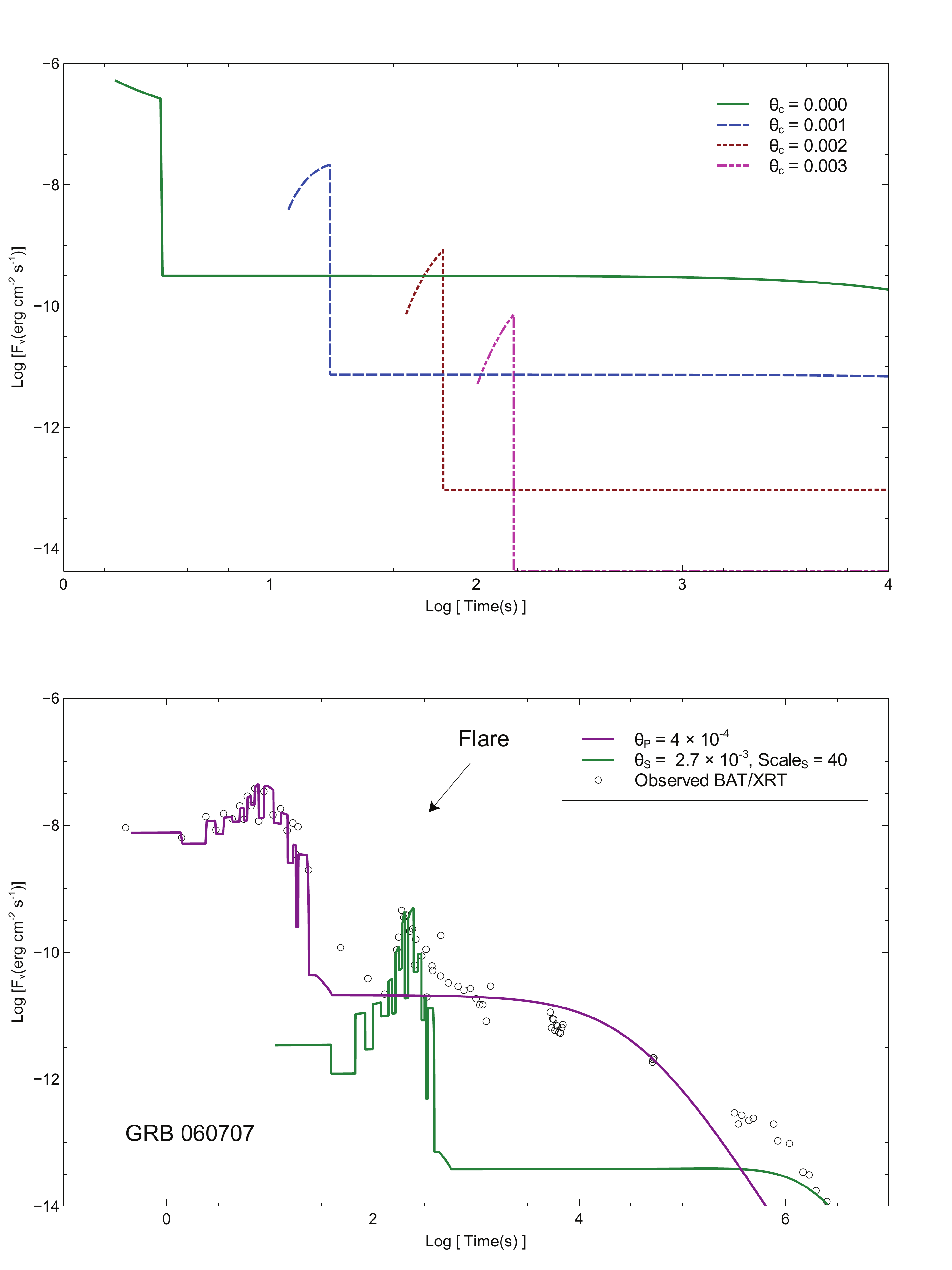}
\caption{{\bf Flares in our model}: {\bf Top panel}: Illustration of how flares are produced in our model.  A simulated GRB with a single filament and four chunks at $\theta_{\rm c} = 0.000, 0.001, 0.002, $ and $0.003$ rads.  This panel demonstrates how the emission gets shifted to longer times and lower flux as  $\theta_{\rm c}$ increases; the afterglow is the plateau for each $\theta_{\rm c}$. Here, $\theta_{\rm c}=0.000$ represents the
primary chunk responsible for the prompt emission.
To see a flare, the secondary chunk should be at a $\theta_{\rm c}$ large enough that it does not overlap with the primary
chunk peak, but not so large that it is fainter than the afterglow (like the chunk at $\theta_{\rm c} = 0.003$ rads).
 {\bf Bottom panel}: The data for GRB 060707 is represented by the open circles with a flare at $\sim 10^{2.5}$ s.  Using the simulation results from \S ~ \ref{section:results} we show, in purple, the light-curve produced by a single, primary chunk at $\theta_{\rm P}=4\times 10^{-4}$ rad.  The light-curve from  a secondary chunk at $\theta_{\rm S} = 2.7 \times 10^{-3}$ rad is shown in purple (scaled by 40).
}
\label{figure:genericflare}
\end{figure}

\begin{figure}[ht!]
\centering
\includegraphics[scale=0.4]{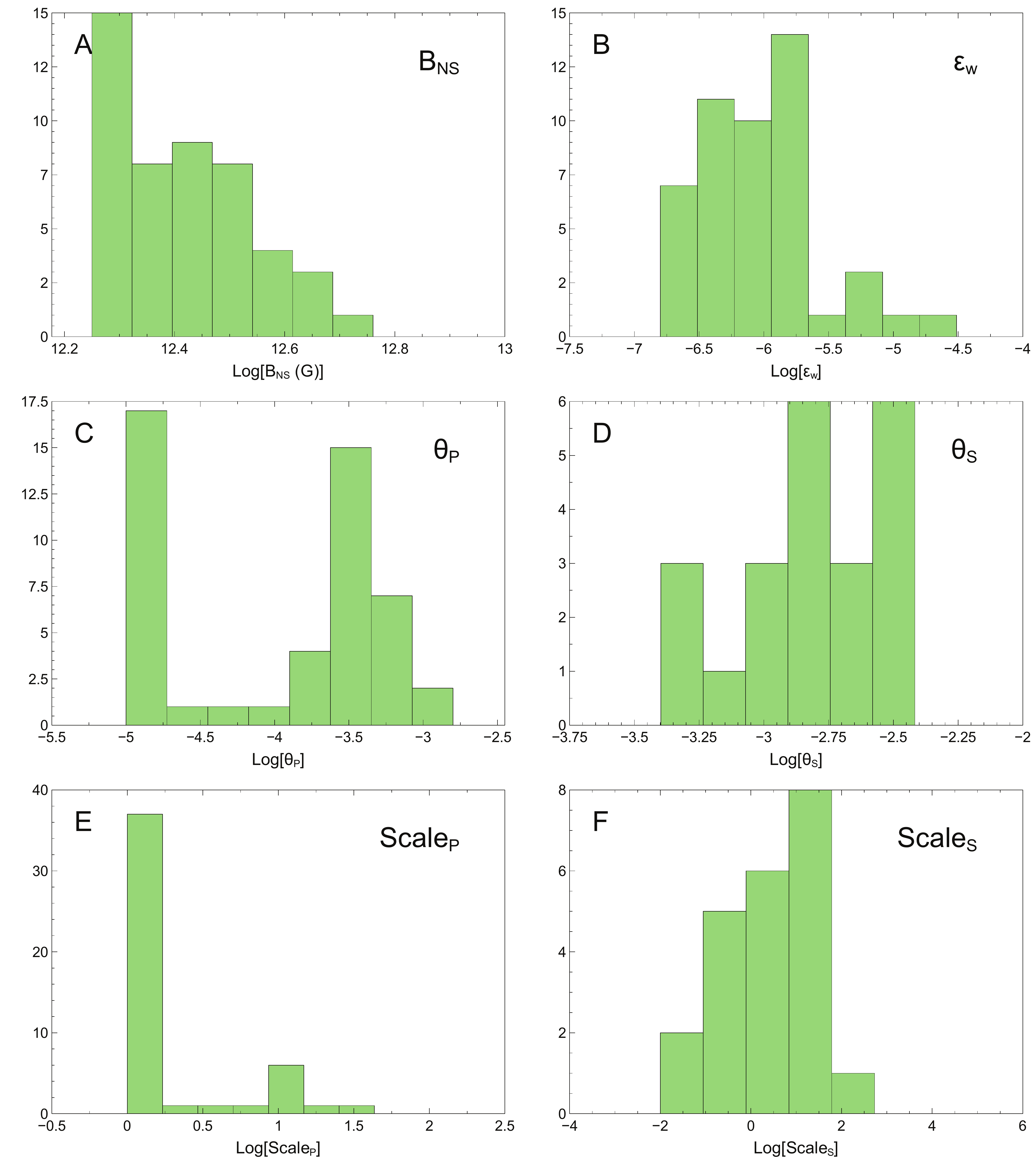}
\caption{{\bf Model parameters}: Distributions of  parameters from simultaneous fits to  light-curves and spectra of the 48 LGRBs listed in Table \ref{table:lcfits}. See \S ~ \ref{subsec:fitting}, Table \ref{table:parameters} and Table \ref{table:lcfits} for definitions of parameters.  We use Rice's rule for
binning.}
 \label{figure:fits-histograms}
\end{figure}

\renewcommand{\thefigure}{\arabic{figure} (Cont.)}
\addtocounter{figure}{-1}
\begin{figure}[ht!]
\centering
\includegraphics[scale=0.4]{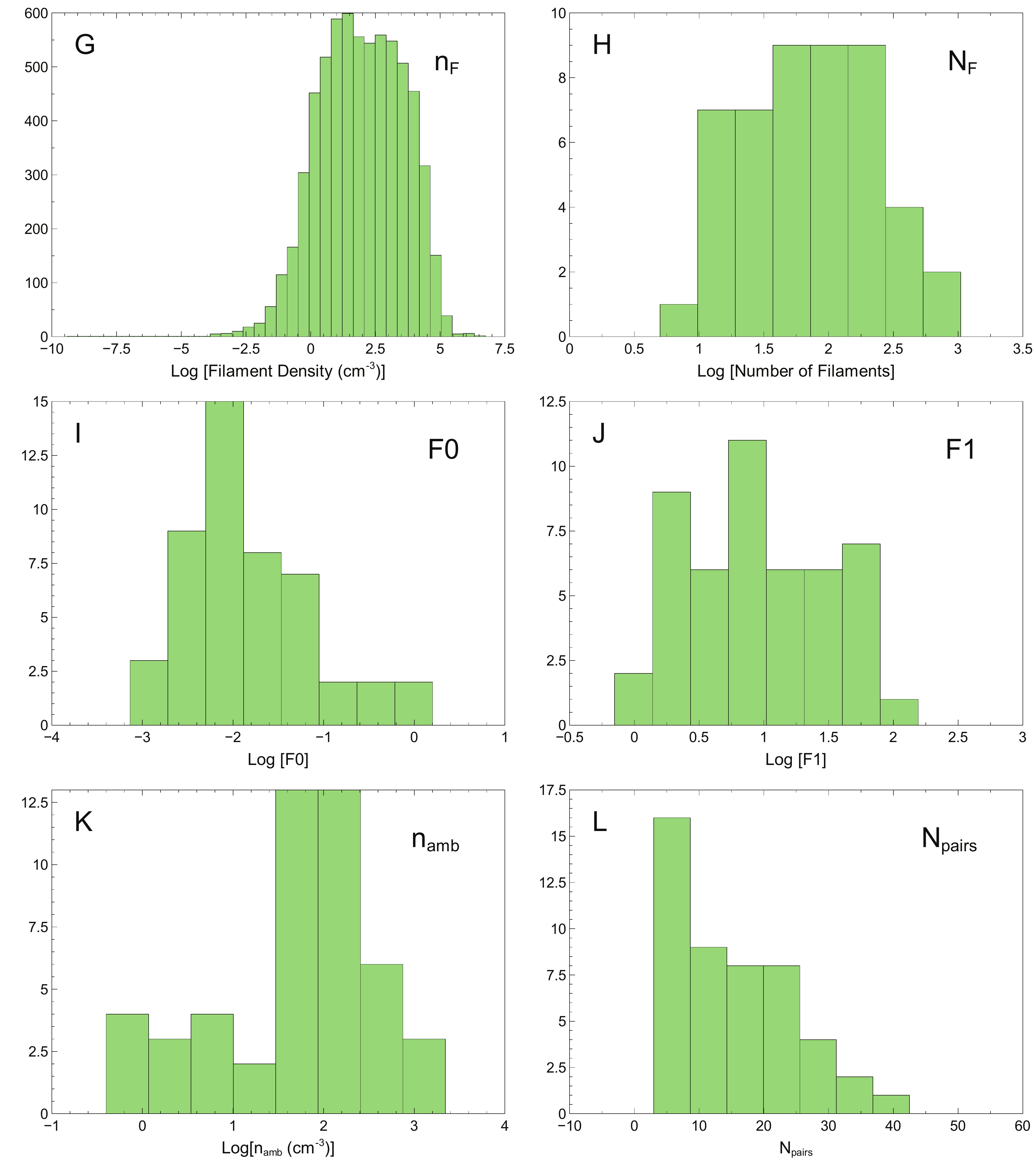}
\caption{}
\end{figure}
\renewcommand{\thefigure}{\arabic{figure}}

\begin{figure}[ht!]
\centering
\includegraphics[scale=0.4]{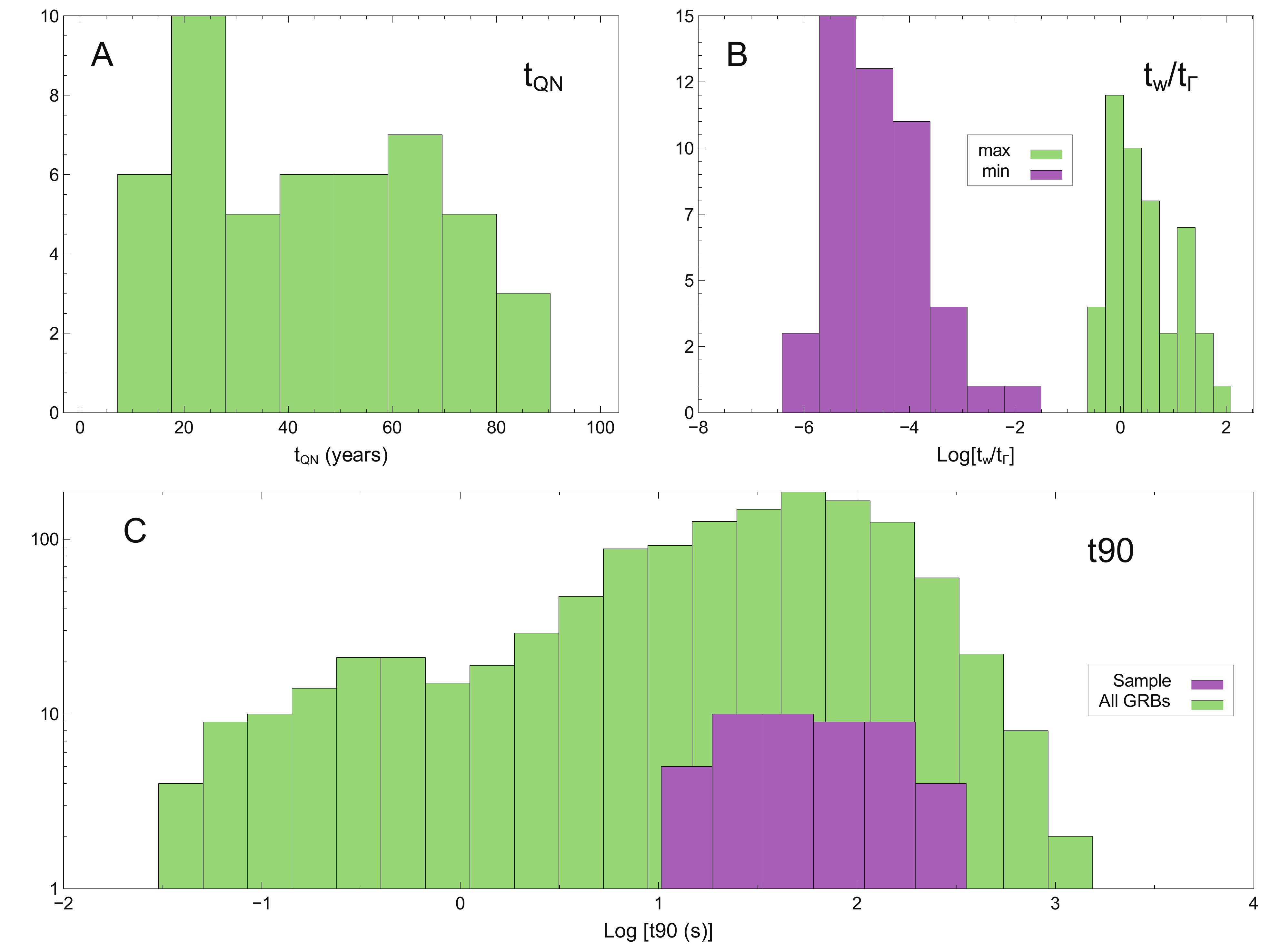}
\caption{{\bf Model timescales}: Distribution of timescales from simultaneous fits to  light-curves and spectra  of the 48 selected LGRBs.  {\bf Panel A}:  the distribution of $t_{\rm QN}=t_{\rm SpD}$ resulting
from the $B_{\rm NS}$ distribution (see panel A of Figure \ref{figure:fits-histograms}). {\bf Panel B}:  the distributions of the minimum and maximum values of the 
thickness parameter $t_{\rm w}/t_{\Gamma}$. This shows the wide variation of filaments' thickness within each GRB and from one GRB to another.
 {\bf Panel C}:  the distribution of  durations of the 48 fit LGRBs  compared to the $t_{90}$ of all
GRBs (data from \url{https://swift.gsfc.nasa.gov/archive/grb_table}). The fit LGRBs have durations representative of the LGRB population.}
\label{fig:distributions-t90}
\end{figure}

\begin{figure*}[t!]
\centering
\includegraphics[scale=0.4]{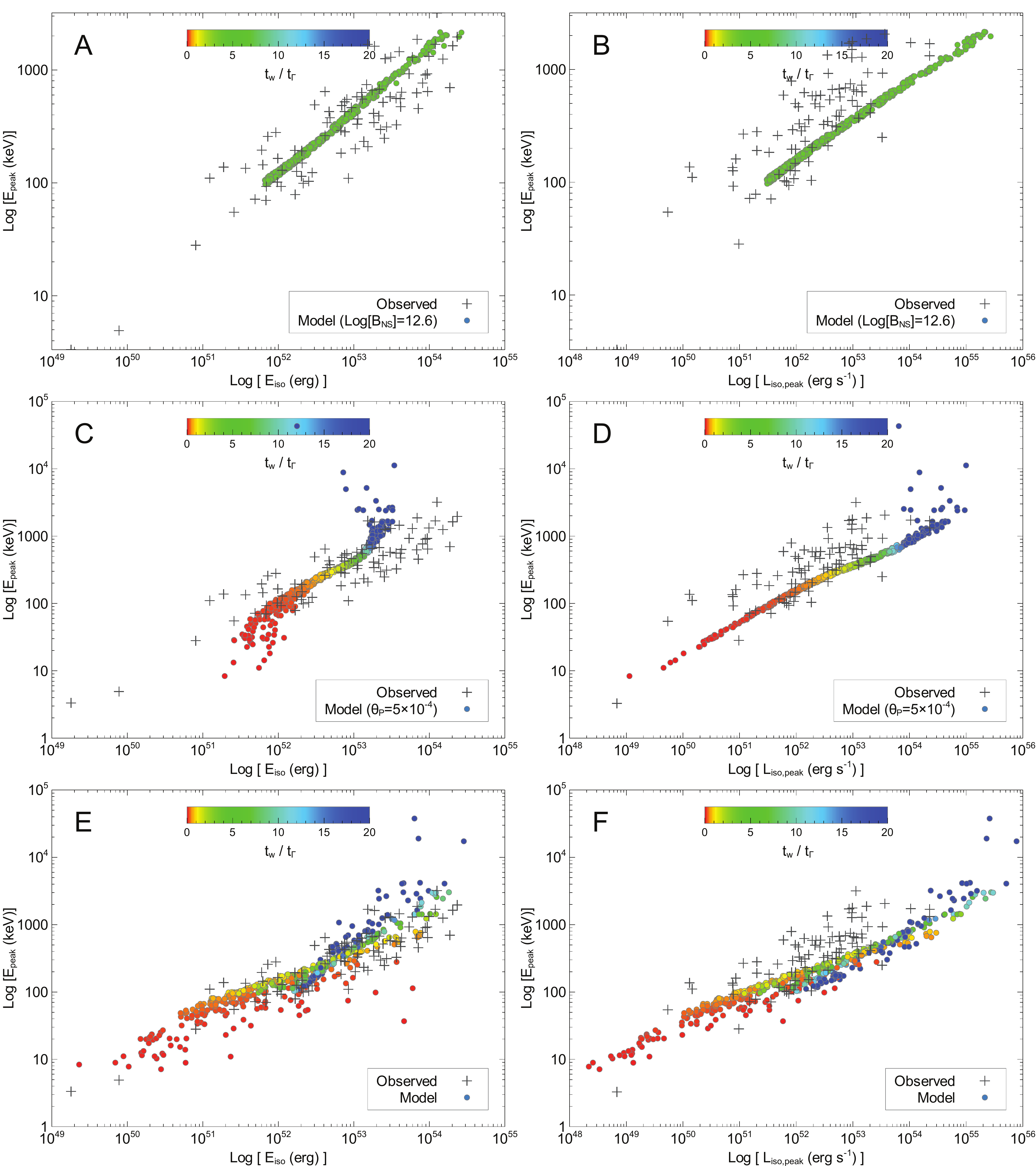}
\caption{{\bf Amati plot (left panels) and Yonetoku plot (right panels) for the single wall numerical simulations model (\S ~ \ref{sec:yonetoku-amati-simulations})}: 500 simulations (including deceleration of chunk for large values of $t_{\rm w}/t_{\Gamma}$)  are plotted against the observations.  For each simulation the primary chunk passes through a single  wall. Here all parameters  are kept to their fiducial values (see Table \ref{table:parameters}) except for
 the number of pairs which is $n_{\rm pairs}=15$  for best agreement with data in the left panels. The dots correspond to 500 simulations
  but for varied $t_{\rm QN}$ (i.e. $B_{\rm NS}$) and   $\theta_{\rm P}$ with ranges similar to those used in Figure \ref{fig:amati-single-theory2}.
  Also, binning into 64 ms time bins introduces scatter not present in the analytical case (see Figure \ref{fig:amati-single-theory2}).
 {\bf Top panels}: Effects of varying the viewing angle $\theta_{\rm P}$ for a fixed $B_{\rm NS}$. {\bf Middle panels}: Effects of varying $B_{\rm NS}$ for a fixed viewing angle $\theta_{\rm P}$. {\bf Bottom panels}: Effects of varying both the viewing angle $\theta_{\rm P}$ and the NS magnetic field $B_{\rm NS}$.
\\
 }
 \label{fig:amati-single-simulations}
\end{figure*}

\begin{figure*}[t!]
\centering
\includegraphics[scale=0.4]{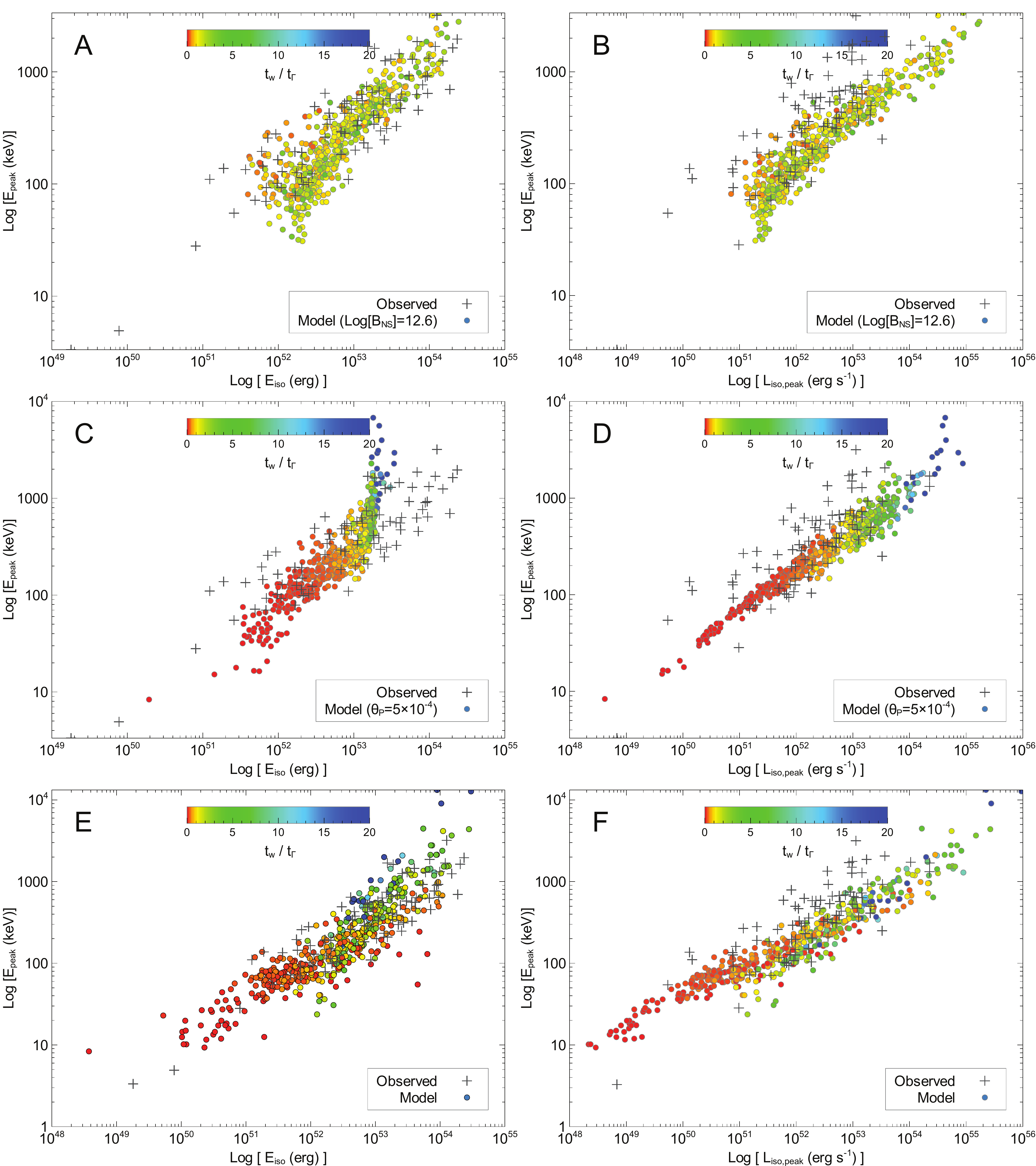}
\caption{{\bf Amati plot (left panels) and Yonetoku plot (right panels) for the multiple filaments numerical simulations model (\S ~ \ref{sec:yonetoku-amati-simulations})}: For each simulation a single chunk passes through multiple filaments of varied thickness. The palette
shows the highest thickness parameter (i.e. the thickest filament for each case).  The dots correspond to 500 simulations
  but for varied $t_{\rm QN}$ (i.e. $B_{\rm NS}$) and $\theta_{\rm P}$ with ranges similar to those used in Figure \ref{fig:amati-single-simulations}.
  Other parameters kept to their fiducial values (see Table \ref{table:parameters}) except for the number of pairs which is $n_{\rm pairs}=12$ for best agreement with data in the left panels.  {\bf Top panels}: Effects of varying the viewing angle $\theta_{\rm P}$ for a fixed $B_{\rm NS}$. {\bf Middle panels}: Effects of varying $B_{\rm NS}$ for a fixed viewing angle $\theta_{\rm P}$. {\bf Bottom panels}: Effects of varying both the viewing angle $\theta_{\rm P}$ and the NS magnetic field $B_{\rm NS}$.\\
}
\label{fig:amati-multiple-simulations}
\end{figure*}


\begin{figure*}[t!]
\centering
\includegraphics[scale=0.4]{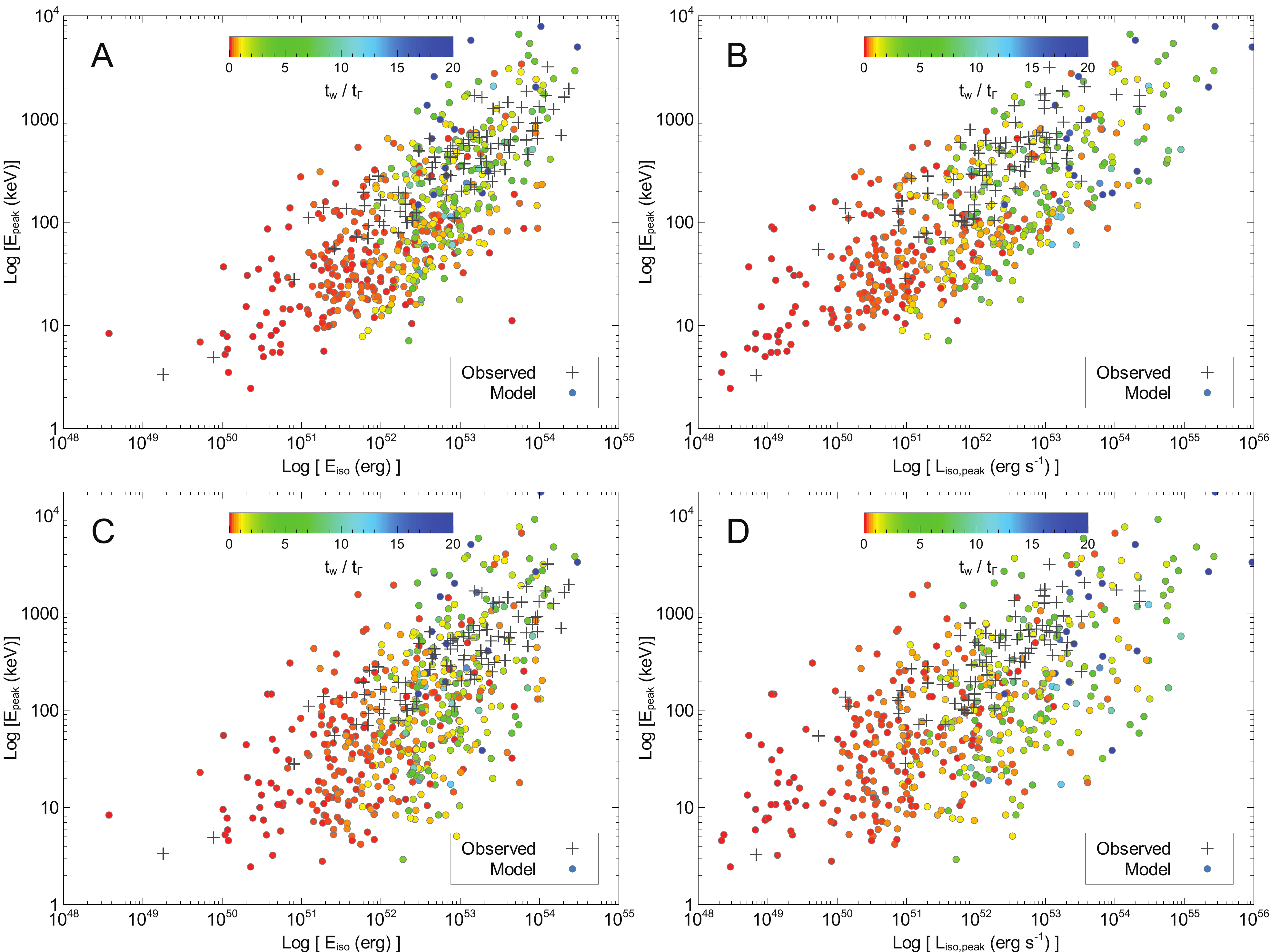}
\caption{{\bf Amati plot (left panels) and Yonetoku plot (right panels) for the multiple filaments case including scatter from other parameters}: These are the
same simulations as in Figure \ref{fig:amati-multiple-simulations}  but this time we include scatter from $n_{\rm pairs}$,  $\epsilon_{\rm w}$ and
 $p$  in the range of Table \ref{table:lcfits}. The two top panels correspond to  $5  \le n_{\rm pairs} \le 30$, while 
  in the two bottom panels  we vary simultaneously the three parameters  ($5  \le n_{\rm pairs} \le 30$, $-6.0 \le \log{\epsilon_{\rm w}}\le -4.5$ and $2 < p \le 3$).\\
}
\label{fig:amati-everything-simulations}
\end{figure*}

\begin{figure*}[t!]
\centering
\includegraphics[scale=0.1]{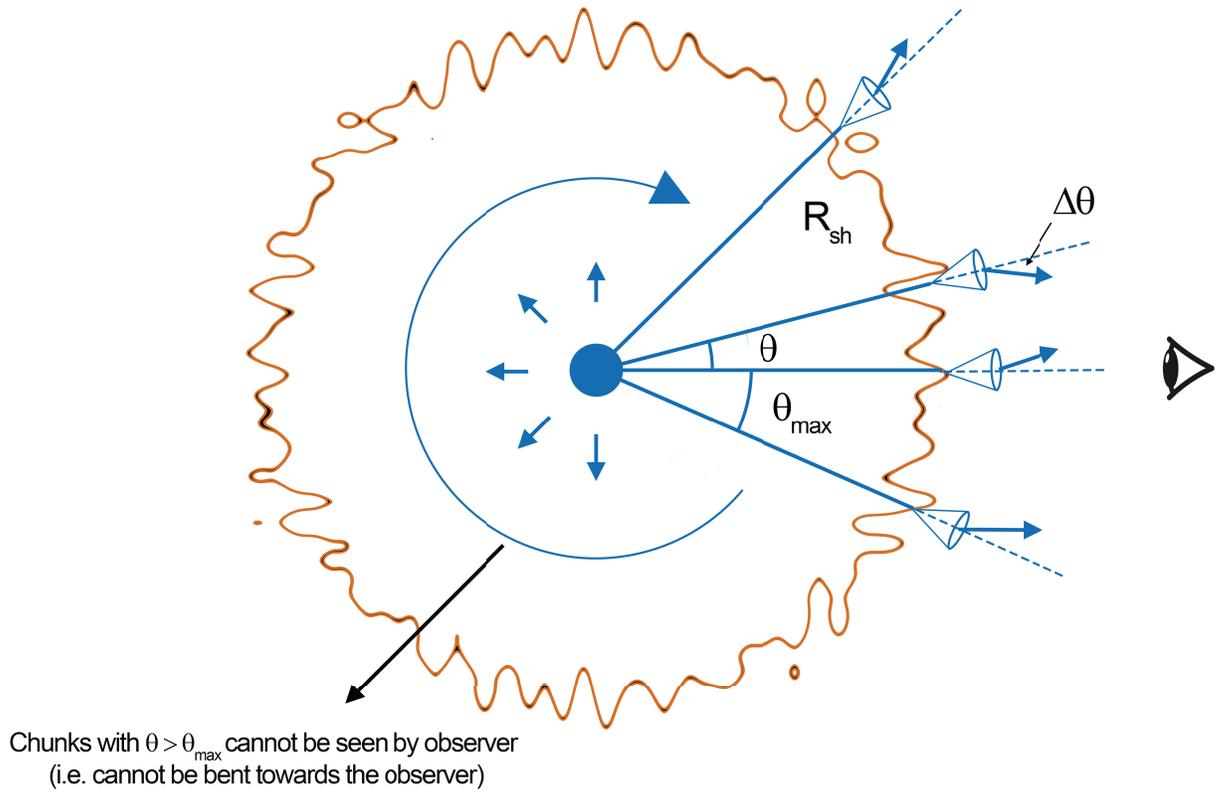}
\caption{{\bf Our model for repeating FRBs}: A  plasma shell (e.g. an HII region) at a distance $R_{\rm sh.}$ from the QN explosion 
 acts as a refractor. In this  simple geometry, the FRBs beams from the QN chunks are each bent by an angle $\Delta \theta \le \theta_{\rm max.}$, 
in random direction, by the refracting plasma. Repeating FRBs occur when multiple beams are bent towards the observer  by any inhomogeneities
 in the shell  (see Appendix \ref{appendix:RFRBs}).}
\label{fig:RFRBs}
\end{figure*}


\begin{figure}[t!]
\centering
\includegraphics[scale=0.12]{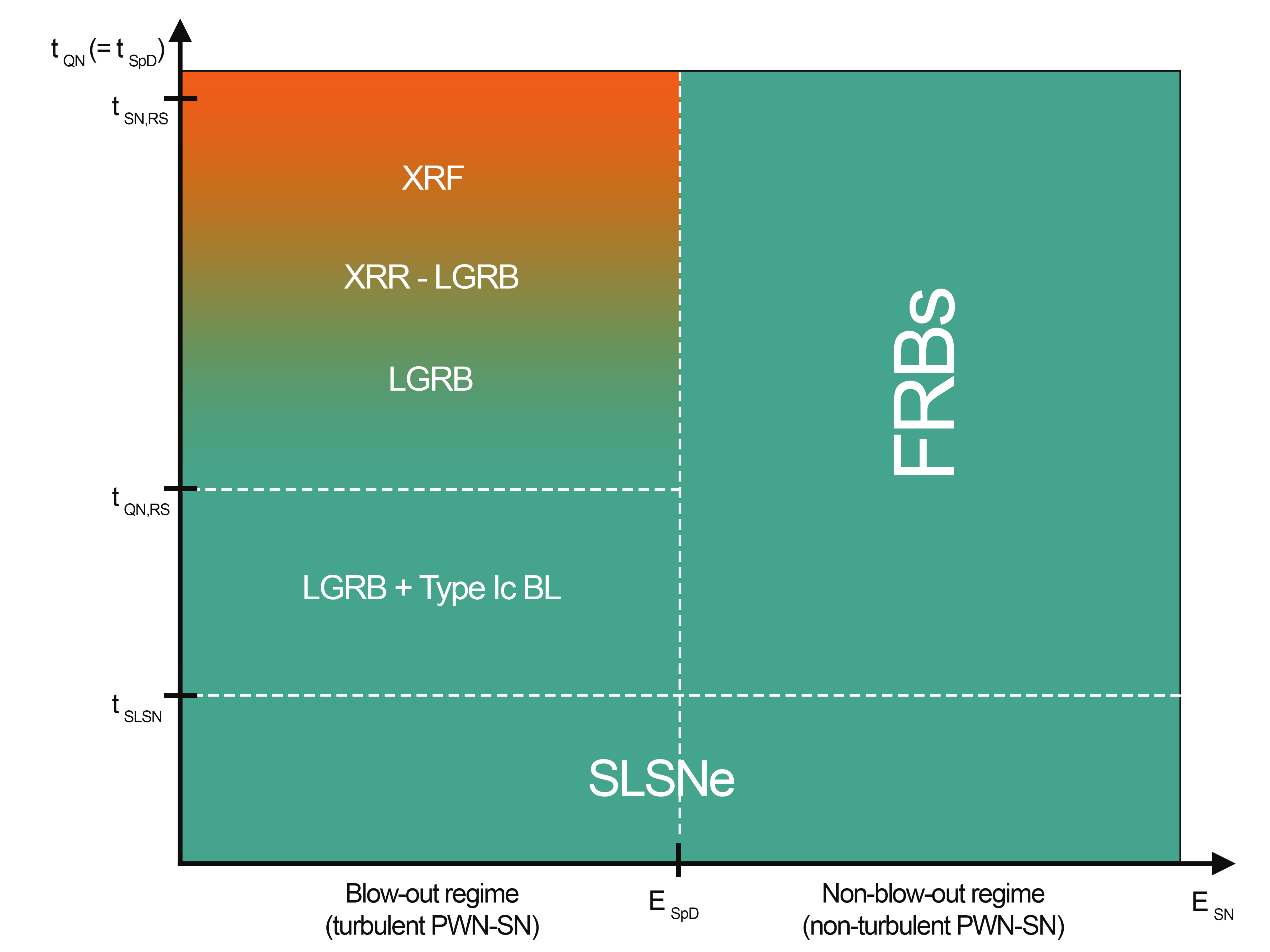}
\caption{{\bf The unification of bursts in our model}: The two regimes correspond to the blow-out regime ($E_{\rm SN}< E_{\rm SpD}$)
 and the non-blow-out regime ($E_{\rm SN}< E_{\rm SpD}$). The non-blow-out regime yields FRBs in our model
 since the non-turbulent, weakly magnetized, PWN-SN shell is prone to the Weibel instability, triggering coherent
 synchrotron emission (CSE) in the chunk's shock (see \S ~ \ref{sec:frbs}). The blow-out regime yields LGRBs, XRR-GRBs and XRFs (see \S ~ \ref{sec:unification}). For both
 regimes SLSNe result if the QN occurs on timescales $\le t_{\rm SLSN}$ when the PWN-SN shell is still optically thick.}
 \label{fig:unification}
\end{figure}

\begin{figure}[ht!]
\centering
\includegraphics[scale=0.4]{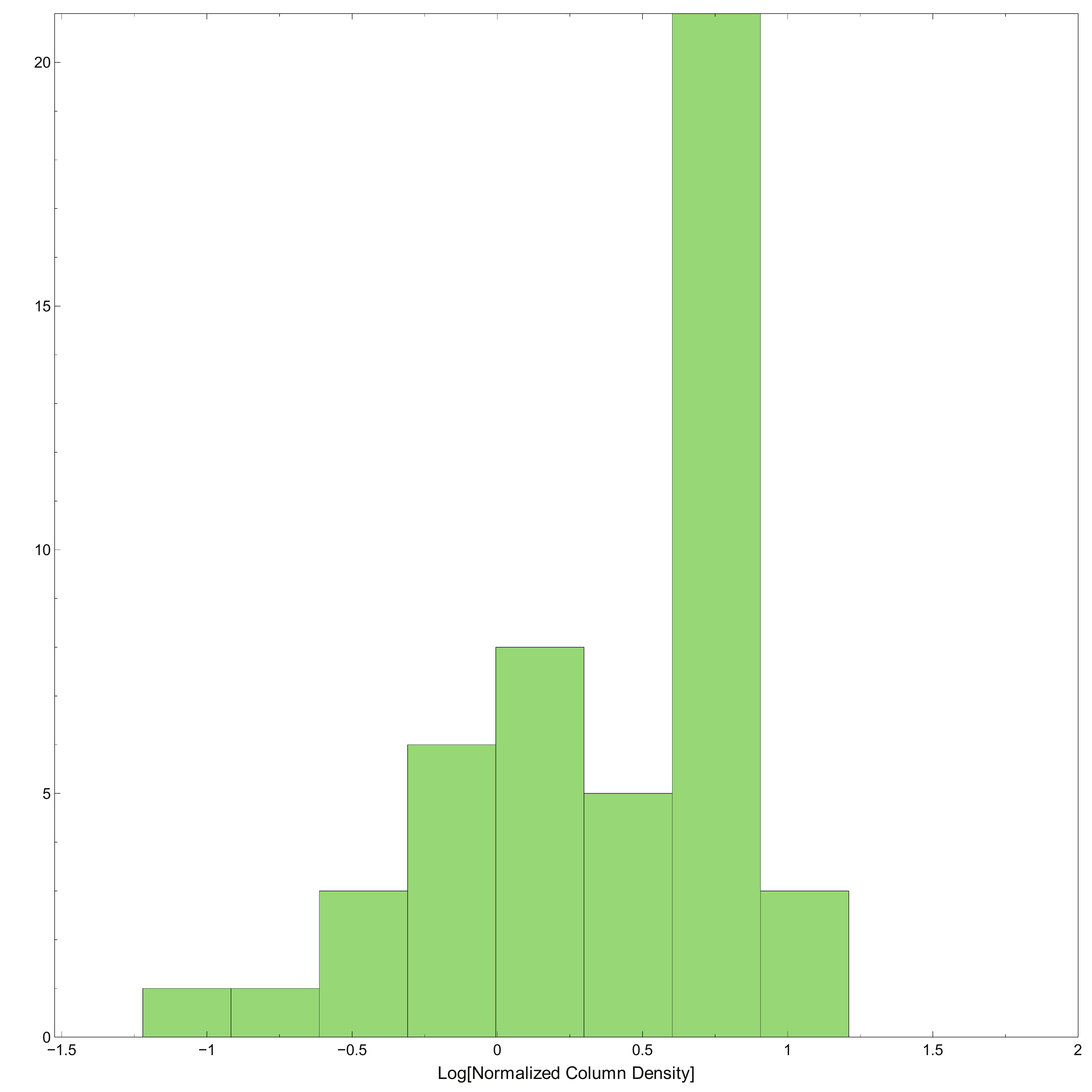}
\caption{{\bf PWN-SN shell (``the Wall") column density}: The relative  column density $\sum{n_{\rm F}\Delta R_{\rm F}}$
  normalized to the analytical value $n_{\rm Plat.}\times (R_{\rm w}/12$) (see \S ~ \ref{sec:QN-SN}). Each column density (one per fit LGRB)  is generated by adding up all the filaments along the line-of-sight for a single LGRB.}
\label{fig:distribution-nwdRw}
\end{figure}

\begin{figure}[ht!]
\centering
\includegraphics[scale=0.65]{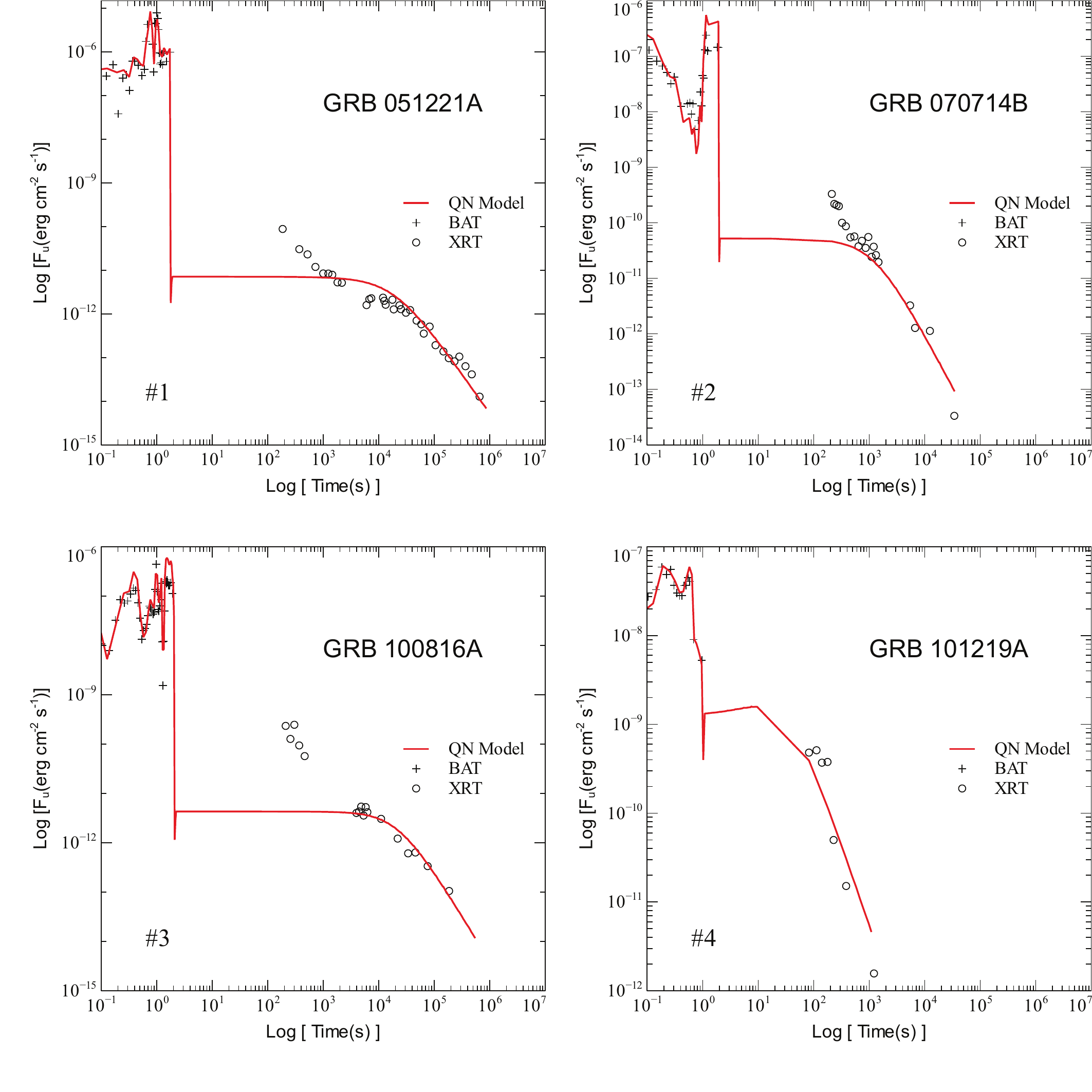}
\caption{{\bf SGRB  light-curve  fits}: The  XRT light-curve fit for 4 SGRBs (see \S ~ \ref{sec:SGRBs}).  The BAT data is extrapolated to the XRT band (see \citealt{evans_2010}) and shown as black crosses.  The XRT data is shown as open circles.  The red line is the QN model.}
 \label{figure:lcSGRB}
\end{figure}

 \label{lastpage}

\end{document}